\documentclass[10pt, journal]{IEEEtran}
\usepackage{cite}
\usepackage{amsmath,amssymb,amsfonts}
\usepackage[noend]{algorithmic}

\usepackage{textcomp}
\usepackage{xcolor}
\usepackage{amssymb}
\usepackage{indentfirst}
\usepackage{amsmath}
\usepackage{bm}
\usepackage{booktabs}
\usepackage{epsfig}
\usepackage{algorithm}
\usepackage{changepage}
\usepackage{amsthm} 
\usepackage{soul,color}
\usepackage{amsbsy}
\usepackage{hyperref}
\usepackage{setspace}
\usepackage{multirow}
\usepackage{xfrac}
\usepackage{diagbox}
\usepackage{graphicx}
\usepackage[font=small]{caption}
\usepackage[subrefformat=parens]{subcaption}

\DeclareMathOperator*{\argmax}{arg\,max}

\usepackage{tikz}
\usepackage{pgfplots}
\usetikzlibrary{plotmarks}
\usetikzlibrary{matrix,positioning,shapes,arrows,decorations,calc,fit}
\usetikzlibrary{decorations.pathreplacing}
\usetikzlibrary{spy,backgrounds}
\usetikzlibrary{external}
\usetikzlibrary{patterns}
\usetikzlibrary{shapes,arrows}
\usetikzlibrary{shadows.blur}
\usetikzlibrary{shapes.symbols}
\usetikzlibrary{
	pgfplots.groupplots,
	matrix
}

\newtheorem{theorem}{Theorem}
\newtheorem{lemma}{Lemma}

\newtheorem{corollary}{Corollary}
\newtheorem{remark}{Remark}
\theoremstyle{remark}

\ifCLASSINFOpdf
\else
\fi
\newcommand{\black}[1]{{\textcolor[rgb]{0, 0, 0}{#1}}}

\begin{document}

\title{The Guesswork of Ordered Statistics Decoding: \black{Guesswork Complexity} and \black{Decoder} Design}

\author{Chentao~Yue,~\IEEEmembership{Member,~IEEE,}
        Changyang~She,~\IEEEmembership{Senior Member,~IEEE}, 
        Branka~Vucetic,~\IEEEmembership{Life Fellow,~IEEE,}
        and~Yonghui~Li,~\IEEEmembership{Fellow,~IEEE}% <-this % stops a space
\thanks{C. Yue,  C. She, B. Vucetic, and Y. Li are with School of Electrical and Computer Engineering, University of Sydney, Sydney, NSW 2006, Australia (e-mail: chentao.yue@sydney.edu.au,  shechangyang@gmail.com, branka.vucetic@sydney.edu.au, yonghui.li@sydney.edu.au). }
\thanks{The work of Yonghui Li was supported by ARC under Grant DP210103410.}
}




\maketitle

\begin{abstract}
This paper investigates guesswork over ordered statistics and formulates the \black{achievable guesswork} complexity of ordered statistics decoding (OSD) in binary additive white Gaussian noise (AWGN) channels. \black{The achievable guesswork complexity is defined as the number of test error patterns (TEPs) processed by OSD immediately upon finding the correct codeword estimate.}   \black{The paper first develops a new upper bound for guesswork over independent sequences by partitioning them into Hamming shells and applying Hölder's inequality.} This upper bound is then extended to ordered statistics, by constructing the conditionally independent sequences within the ordered statistics sequences. \black{Next, we apply these bounds to characterize the statistical moments of the OSD guesswork complexity.} 
We show that the \black{achievable guesswork} complexity of OSD at maximum decoding order can be accurately approximated by the modified Bessel function, which increases exponentially with code dimension. We also identify a \black{guesswork complexity saturation threshold}, where increasing the OSD decoding order beyond this threshold improves error performance without further raising \black{the achievable guesswork complexity. Finally, the paper presents insights on applying these findings to enhance the design of OSD decoders.}

\end{abstract}



% \fi
\IEEEpeerreviewmaketitle

\section{Introduction}

One of the key requirements in 6G is the extreme ultra-reliable low-latency communications (xURLLC) \cite{Changyang2023xURLLC}, requiring a tenfold decrease in end-to-end transmission latency and a hundredfold \black{improvement} in network reliability compared to 5G URLLC services \cite{tataria20216g}. Moreover, networks are required to maintain this service quality among varying performance demands, diverse applications, and dynamic propagation environments. Providing xURLLC services far surpasses the capabilities of 5G, and one of the key bottlenecks is channel coding and decoding. Channel coding ensures reliable transmission by protecting messages against noise. Code blocklength
is the basic unit of communication latency, and decoding time dominates the receiver processing delay. Therefore, achieving xURLLC requires short blocklength codes with strong error-correction capabilities. However, the use of short blocklength codes presents a challenging trade-off between blocklength and reliability. The normal approximation (NA) bound for the finite blocklength regime \cite{erseghe2016coding} shows that the maximum ratio of the number of information to the number of coded bits for a given error probability over a noisy channel decreases as blocklength reduces. Thus, short blocklength codes typically have worse block error rate (BLER) performance compared to longer block codes at the same code rate. 

The diversity of 6G applications, each with vastly differing performance requirements and propagation environments, will also lead to variable channel conditions and block lengths. In these dynamic scenarios, codes with flexible rates/lengths and optimal rate-compatible (RC) codes are essential. Although several channel codes have been proposed for URLLC \cite{Mahyar2019ShortCode}, they were mainly designed based on fixed rates and blocklengths, and achieve the flexibility and RC capabilities through puncturing, shortening, and extending. Such approaches were shown to be suboptimal in both error performance and decoding complexity at short block lengths. For example, the successive cancellation list (SCL) decoding of shortened polar codes is actually performed on a longer mother code, leading to unnecessary decoding overhead \cite{3GPPRel16Coding}.

\black{Generic decoding techniques for linear block codes have recently gained interest as a potential solution to these challenges. Their capacity to decode any linear block code based solely on the generator matrix can significantly simplify transmitter and receiver design \cite{yue2023efficient}. These decoding techniques enable the use of best-known linear codes (BKLC), known for their superior error performance, at any blocklength and rate tailored to application requirements. This task is challenging for code-specific decoders, as BKLCs possess distinct structures at different lengths and rates \cite{Grasslcodetables}. Generic decoding approaches can simplify the design and application of optimal RC codes with bit-level granularity for incremental-redundancy hybrid automatic repeat request (IR-HARQ). These advantages also make such decoding techniques suitable for integration with machine learning \cite{cavarec2020learning}, or joint design with learning-based encoders as auto-encoders \cite{larue2022neural}, to further boost decoding performance and adaptability.}

%Ordered-statistics decoding (OSD) \cite{Fossorier1995OSD} and Guessing random additive noise decoding (GRAND) \cite{duffy2021guessing} are regarded as promising universal decoders for 6G. Consider a linear block code $\mathcal{C}(n, k)$ with block length $n$ and dimension $k$. OSD begins by permuting the received symbols and columns of the code generator matrix in descending order of symbol reliabilities. The permuted code generator matrix is then transformed into systematic form using Gaussian elimination (GE). After that, OSD flips the $k$ most reliable bits by XORing them with a test error pattern (TEP), and these $k$ bits are re-encoded to recover the remaining $n-k$ bits. OSD processes a specific group of TEPs to decode one block, where each TEP is in fact a guess of transmission errors over the $k$ most reliable bits. Compared to OSD, GRAND directly guesses transmission errors over all $n$ received symbols, with each guess termed a noise query. It then subtracts each noise query from the received symbols and verifies codeword membership. In a nutshell, OSD and GRAND share similarities while also exhibiting distinct characteristics. They both decode through guessing transmission errors, but differ in their guessing ranges within a codeword. While GRAND avoids the permutation and Gaussian elimination overhead of OSD, OSD typically requires fewer guesses due to its shorter guessing range. 

\black{Ordered-statistics decoding (OSD) \cite{Fossorier1995OSD} is regarded as a promising generic decoding technique for linear block codes in 6G. Consider a linear block code $\mathcal{C}(n, k)$ with block length $n$ and dimension $k$. OSD begins by permuting the received symbols and columns of the code generator matrix in descending order of symbol reliabilities. The permuted code generator matrix is then transformed into systematic form using Gaussian elimination (GE). After that, OSD flips the $k$ most reliable bits by XORing them with a test error pattern (TEP), and these $k$ bits are re-encoded to recover the remaining $n-k$ bits. OSD processes a specific group of TEPs to decode one block. From a guesswork perspective, OSD strategically uses TEPs to guess the $k$ most reliable bits. Correctly guessing these bits allows the entire codeword to be recovered through re-encoding. In this paper, we will focus on characterizing this guessing component of OSD, which is a fundamental factor to understanding its complexity behavior and performance.}

\black{In the context of generic decoding techniques, Guessing random additive noise decoding (GRAND) \cite{duffy2021guessing} represents another notable advancement. GRAND directly guesses transmission errors across all $n$ received symbols and achieves maximum-likelihood decoding when noise queries are performed in descending likelihood order \cite{duffy2018guessing}. Its complexity is well-characterized by the Rényi entropy of noise sequences \cite{duffy2021guessing}. Unlike OSD, GRAND does not require linear permutation of the generator matrix and GE. Thus, it can decode both linear and non-linear codes, making it a fully universal decoder. Additionally, the absence of GE enables highly efficient circuit- and chip-based implementations of GRAND. Several ASIC implementations have been proposed for GRAND decoders \cite{abbas2023guessing,riaz2023sub}, achieving wide recognition within the circuit design community for their efficiency. Although both OSD and GRAND involve guessing transmission errors, given their distinct characteristics and implementation approaches, we limit our discussion of GRAND to this brief overview.}

\black{The design of OSD has seen significant improvements in recent years. Key advances include introducing sufficient and necessary conditions for optimal decoding \cite{yue2021revisit, yue2021probability}, removing the need for GE \cite{yue2022ordered, choi2019fast}, and reducing the number of required TEPs \cite{wang2021efficient, yue2022linear, LCOSD2022}. These techniques effectively reduce the decoding complexity while maintaining error performance. The error performance of OSD was examined in \cite{Fossorier1995OSD}, where OSD was proved to approach MLD performance with the decoding order $m_e \!=\! \lceil d_{\min}/4\!-\!1\rceil$. Here, $d_{\min}$ is the minimum Hamming distance of the code, and the decoding order of OSD represents the maximum allowable Hamming weight for TEPs. This error performance analysis was revisited and simplified in \cite{dhakal2016error}. Recently, our work \cite{yue2021revisit} explored the distribution of distances between received sequences and codewords in OSD, which can be further used to guide efficient design of OSD decoders.}

%In terms of theoretical completeness, GRAND is proved to achieve maximum-likelihood decoding (MLD) if the noise queries are conducted in descending order of their likelihood \cite{duffy2018guessing}. Its complexity, in terms of the total number of noise queries, is theoretically characterized by the Rényi entropy of the noise sequences \cite{duffy2021guessing}. The error performance of OSD was examined in \cite{Fossorier1995OSD}, where OSD was also proved to be near MLD. The error performance analysis was revisited and simplified in \cite{dhakal2016error}, while \cite{yue2021revisit} further explored the distance distribution in OSD to inform efficient decoder design.

\black{The complexity of OSD comprises two components: Preprocessing and Reprocessing\footnote{\black{Despite the computational complexity challenges, circuit implementation remains crucial for real-world applications of OSD. While several research works have explored FPGA implementations of OSD \cite{scholl2014hardware,kim2021fpga}, the decoder still lacks efficient advanced ASIC designs. This hardware aspect lies outside the scope of this paper.}}\cite{Fossorier1995OSD}. The Preprocessing complexity includes sorting symbol reliabilities and performing GE. The Reprocessing complexity arises from the need to iteratively re-encoding multiple TEPs. We further define the guesswork complexity as the number of TEPs, or in other words, the number of attempts to guess the $k$ most reliable bits, required by OSD to achieve a target BLER. The overall complexity of OSD is approximately given as \cite{Fossorier1995OSD} 
\begin{equation} \notag
    C_{\mathrm{OSD}} = \underbrace{C_{\mathrm{Sorting}} + C_{\mathrm{GE}}}_{\text{Preprocessing}} + \underbrace{X_G \cdot C_{\mathrm{Re-encoding}}}_{\text{Reprocessing}},
\end{equation}}
\black{where $C_{\mathrm{Sorting}}$ and $C_{\mathrm{GE}}$ represent the complexity for sorting symbols and GE, respectively. $C_{\mathrm{Re-encoding}}$ is the complexity to re-encode a single TEP. $X_G$ is the number of TEPs, i.e., the guesswork complexity as defined earlier, which will be the main focus of this paper.}

\black{
The preprocessing complexity of OSD is typically fixed for a given code dimension and becomes non-negligible only at very high signal-to-noise ratios (SNRs). In contrast, guesswork complexity (i.e., the number of TEPs) is dominates the overall OSD complexity in most case \cite{yue2022ordered}, because it can be very large for even short codes. For instance, order-4 OSD (near-optimal) for a $(128,64)$ BCH code requires up to 679,121 TEPs.}

Despite the \black{theoretical research} efforts in \cite{Fossorier1995OSD, dhakal2016error, yue2021revisit}, \black{the guesswork complexity} of OSD \black{is yet to be fully characterized}, with current understanding being largely intuitive. An order-$m$ OSD may execute up to $\sum_{i=0}^m\binom{k}{i}$ guesses (i.e., TEPs), \black{which represent the worst-case scenario. However, an OSD decoder could terminate earlier upon identifying the correct codeword, without enumerating all TEPs.}  Such early termination is achievable through techniques reported in \cite{yue2021probability, wu2007preprocessing_and_diversification, jin2006probabilisticConditions, LCOSD2022}.  \black{In this early-termination case, the guesswork complexity is a random variable due to the randomness of channel noise, making its characterization particularly challenging.} The challenge stems from the correlated ordered statistics in OSD, which complicates the probability analysis and renders the conventional guesswork theory unsuitable. Consequently, a theoretical analysis of this guesswork complexity of OSD is still lacking. \black{Characterizing this guesswork complexity will answer a fundamental question: \textit{how many guess attempts are typically required in OSD to achieve optimal performance without reaching the worst-case number of guesses?}}

\subsection*{Main Contributions:}

\black{This paper focuses on the \textit{achievable guesswork complexity}} of an order-$m$ OSD \black{and its statistical moments}. \black{Achievable guesswork complexity is defined as the minimum number of TEPs (or guesses) processed, ensuring no loss in error performance compared to decoding that exhaustively enumerates all $\sum_{i=0}^m \binom{k}{i}$ TEPs}. This achievable guesswork complexity is realized by an OSD decoder that terminates early upon accurately identifying the correct TEP or correctly guessing \black{transmission} errors in the $k$ most reliable bits.

Such a decoder is feasible as existing OSD stopping methods proposed in \cite{yue2021probability, wu2007preprocessing_and_diversification, jin2006probabilisticConditions, LCOSD2022}, can identify the correct OSD output with satisfactory accuracy. \black{These methods} can be further enhanced by combining cyclic redundancy check (CRC) to achieve a negligible false alarm rate. 

It is worth noting that one can always achieve a lower complexity than the achievable guesswork complexity by allowing \black{some loss in error performance}. This occurs when \black{certain TEPs are discarded without being processed}, \black{as in the approaches proposed in } \cite{Chentao2019SDD, wang2021self, yue2021probability}. However, this trade-off between complexity and error performance loss is beyond the scope of this paper \black{and merits future investigation}. 

\black{The main contributions of this paper are outlined as follows.}

\subsubsection{\black{The average} achievable \black{guesswork} complexity of OSD} When considering the highest order, i.e., $m=k$, OSD is strictly equivalent to MLD since the largest decoding effort allows examining all codewords from $\mathcal{C}(n,k)$. Our result shows that the \black{average} achievable guesswork complexity \black{(first moment)} of order-$k$ OSD is tightly approximated by \black{(see Theorem \ref{The::Newbound:XnYn:ordered:1k:expapp})}
 \begin{equation}  \notag
        e^{-k p_e}I_0(2k\sqrt{p_e}) \approx \frac{1}{\sqrt{4\pi k\sqrt{p_e}}} \ 2^{k(2\sqrt{p_e}-p_e) \log_2(e)},
\end{equation}
where $p_e \in [0,\frac{1}{2}]$ is determined by the code rate and SNR, and $I_0$ is the modified Bessel function. \black{This reveals that the number of TEPs in order-$k$ OSD (with the full codebook examined) to achieve MLD grows exponentially, i.e., $O(2^{k(2\sqrt{p_e}-p_e)})$, with information length $k$.} Compared to brute-force MLD that examines $2^k$ codewords, OSD substantially reduces the complexity by an exponential factor of $(2\sqrt{p_e}-p_e) \log_2(e)$. For instance, when $k = 64$ and $p_e = 0.1$, the achievable guesswork complexity of an order-$k$ OSD is only about $1/(2.7\times 10^4)$ of that of brute-force MLD.

For a more practical order-$m$ OSD with $m<k$, we also provide an approximation of the \black{average} achievable guesswork complexity:
\begin{equation} \notag
    e^{-kp_e} \left( \sum_{j=0}^{m} \left(\frac{k^j}{j!}\right)^{2}p_e^{j} + \frac{k^m}{m!}\frac{(kp_e) ^{m+1}}{(m+1)!}\right),
\end{equation}
\black{which shows that order-$m$ OSD exhibits  polynomial complexity growth $O(k^m)$ for any fixed $m$ (when $m < k$)}. This instant evaluation of guesswork complexity, along with the error rate provided in \cite{dhakal2016error}, helps quickly assess the performance-complexity trade-offs when \black{selecting OSD parameters for specific codes}. \black{When considering both the preprocessing overhead (Gaussian elimination) and the re-encoding overhead for each TEP, our results indicate that the overall complexity of order-$m$ OSD grows as approximately $O(k^{m+2})$. This compounding polynomial growth effectively restricts OSD's application to short block codes.}

\subsubsection{Guesswork for ordered statistics} The above results of guesswork complexity are obtained by developing the guesswork theory for ordered statistics. 
\black{Consider sequences $(X^n, Y^n)$ with $n$ i.i.d. pairs. After ordering these pairs by descending $\max_{X_i}\mathbb{P}(X_i|Y_i)$, let $(\widetilde{X}_i, \widetilde{Y}_i)$ denote the $i_\mathrm{th}$ ordered pair. For any segment from position $a$ to position $b$ ($1 \leq a \leq b \leq n$), we use $(\widetilde{X}_a^b, \widetilde{Y}_a^b)$ to denote the subsequence of ordered pairs from the $a_\mathrm{th}$ to the $b_\mathrm{th}$ position.}
The aim is to find \black{upper bounds on the statistical moments} of guesses needed to accurately identify $\widetilde{X}_a^b$ when given $\widetilde{Y}_a^b$. We tackle this problem by first developing a guesswork upper bound for \black{the i.i.d. sequence $(X^n, Y^n)$} of length $n$. \black{This upper bound is characterized by applying Hölder's inequality to the subspaces of guess error patterns, where these subspaces are partitioned according to their Hamming weights.} Then, the bound is extended to ordered statistics \black{$(\widetilde{X}_a^b, \widetilde{Y}_a^b)$} by leveraging the conditional independence between ordered statistics variables. \black{Finally, we introduce the asymptotic approximation of the bound for OSD guesswork, which is a special case of $a=1$ and $b=k$.} Comparisons between the derived bounds and simulation results validate their tightness.

\subsubsection{Guesswork complexity saturation threshold of OSD} Our results on the achievable \black{guesswork} complexity of OSD provide new insights into this decoding technique. \black{We show that for given} for given $k$, $n$, and SNR, there exist a guesswork complexity saturation threshold, $m_s = \lceil k\sqrt{p_e} \rceil$. Increasing the OSD decoding order $m$ beyond $m_s$ will not further increase the average \textit{achievable guesswork complexity}. This result echoes the findings in \cite{Fossorier1995OSD}, which proved that an OSD decoder of order $m_e \!=\! \lceil d_{\min}/4\!-\!1\rceil$ \black{achieves near-MLD performance}, suggesting increasing decoding order $m$ beyond $m_e$ will not further \black{reduce} the error probability. A widely accepted view from existing research is that OSD with early termination is efficient for both low-rate and high-rate codes, but less so \black{for codes with a rate near $\frac{1}{2}$}. Our discovery provides a straightforward rationale: low-rate codes usually have $m_s < m_e$, despite their relatively large $d_{\min}$, and therefore their achievable guesswork complexity with OSD is mainly governed by $m_s$. In contrast, high-rate codes have relatively small $d_{\min}$ and a small decoding order $m = m_e$ suffices for MLD. \black{Codes with a rate close to $\frac{1}{2}$ require more computational effort for OSD because neither $m_s$ nor $m_e$ is small.}

\black{
\subsubsection{Insights for OSD Algorithm Design}
Beyond the aforementioned theoretical analysis, this paper discusses how these results can guide the design of efficient OSD algorithms. Specifically, we investigate the development of an OSD decoder that can achieve the derived average guesswork complexity bounds by utilizing the sufficient condition for optimal decoding and CRC. We also explore an efficient IR-HARQ system design using OSD, which optimizes retransmission length to reduce overall decoding complexity. Furthermore, we propose a complexity cutoff criterion (CCC) for OSD. CCC employs second-order moment estimation of \textit{achievable guesswork complexity} to determine a cutoff threshold. When the number of processed TEPs surpasses this threshold, the decoding process is terminated immediately. Simulations demonstrate that CCC effectively reduces both the average and worst-case number of TEPs required in OSD algorithms, while maintaining the near optimal BLER performance.
}
\\ 

The rest of this paper is organized as follows. Section \ref{Sec::related_works} reviews existing guesswork theories. Section \ref{sec::guess::upper} derived a new non-asymptotic upper bound of guesswork over i.i.d. sequences. Then. this bound is extended to ordered statistics in \ref{sec::OSguess}. Section \ref{sec::OSD:bound} uses the non-asymptotic guesswork bound to characterize the achievable \black{guesswork} complexity of OSD, \black{and provides the asymptotic approximation. Section \ref{sec::property} discusses properties of the guesswork complexity, including the asymptotic behavior and the saturation threshold.}
Section \ref{sec::discussion} provides discussions on \black{OSD algorithm design based on our theoretical findings.}. Finally, Section \ref{sec::conclusion} concludes the paper.

\emph{Notation}: In this paper, we use $a^n$ or $A^n$ to denote a sequence of $n$ scalars $a^n = [a_1,\ldots,a_n]$ or $n$ random variables $A^n = [A_1,\ldots,A_n]$, respectively. A contiguous subsequence of $a^n$ is represented as $a_i^j = [a_i,\ldots,a_j]$ for $1\leq i \leq j \leq n$. We use $\mathbb{P}(\cdot)$ to denote the probability of an event. $\mathbb{P}_{A}(\cdot)$ denotes the probability mass function (pmf) or probability density function (pdf) of $A$, with the subscript usually omitted when there is no ambiguity. \black{$\mathbb{E}[A^\omega]$ denotes the $\omega_{\rm{th}}$ order moment of random variable $A$.} We use $\phi(x)$ to denote the pdf of the standard normal distribution $\mathcal{N}(0,1)$.

\section{Previous Works} \label{Sec::related_works}
 We consider the guesswork in the context of transmission over a channel with uncertainty. Let $(X,Y) \in (\mathcal{X,\mathcal{Y}})$ be a pair of discrete random variables with the joint pmf $\mathbb{P}_{X,Y}$. Assume that $X$ has $M$ possible values. The guesswork $G(x|y)$ is defined as the number of attempts required to correctly guess $X = x$ with given $Y = y$ according to some guessing strategy. \black{Of particular interest is} the $\omega_{\rm{th}}$ moment of $G(X|Y)$; that is
 \begin{equation}
     \mathbb{E} [G(X|Y)^\omega] =  \sum_{x} \sum_{y} G(x|y)^{\omega} \mathbb{P}_{X,Y}(x,y).
 \end{equation}
 For the $n$-tuples $(X^n,Y^n)$, the corresponding moment of guesswork is
 \begin{equation} \label{equ::def:guess:n}
     \mathbb{E} [G(X^n|Y^n)^\omega] =  \sum_{x^n} \sum_{y^n} G(x^n|y^n)^{\omega} \mathbb{P}(x^n,y^n).
 \end{equation}

The moment $\mathbb{E} [G(X^n|Y^n)^\omega]$ is minimized with the optimal guessing strategy \cite{arikan1996inequality}, which guesses possible values of $X^n$ in decreasing order of the \textit{a posteriori} probability $\mathbb{P}(x^n|y^n)$, given $Y^n = y^n$. We denote the optimal guesswork as $G^*(x^n|y^n)$, and $\mathbb{E} [G^*(X^n|Y^n)^\omega] \leq \mathbb{E} [G(X^n|Y^n)^\omega]$ for any guessing strategies $G(X^n|Y^n)$. However, it is challenging to directly compute the moments of $G^*(X^n|Y^n)$ or $G(X^n|Y^n)$, due to the expansive space of $(X^n,Y^n)$. Instead, these moments are typically \black{characterized} using bounds provided in the literature.

\subsubsection{Arikan's bounds}
A lower bound \black{for} $\mathbb{E} [G(X^n|Y^n)^\omega]$ was given by Arikan \cite{arikan1996inequality}. Specifically,
\begin{align} \label{equ::Arikan::lowerbound}
    \mathbb{E}[G(X^n & |Y^n)^{\omega}] \geq \notag\\
    & \left(1 + \ln(M_1\cdots M_n)\right)^{-\omega} \exp E_\omega(X^n|Y^n),
\end{align}
where 
\begin{align}
    E_\omega(X^n|Y^n) = \ln \sum_{y^n}\left[\sum_{x^n}\mathbb{P}(x^n, y^n)^{\frac{1}{\omega+1}} \right]^{1+\omega}.
\end{align}
Observing the relationship between $E_\omega$ and the Rényi entropy, i.e., $E_{\omega}(X|Y)=\omega H_{\frac{1}{1+\omega}}(X|Y)$, the following bound is derived for i.i.d. pairs $(X^n,Y^n)$ of length $n$.
\begin{theorem} [Arikan's lower bound \cite{arikan1996inequality}] \label{The::Arikan:lower}
    \begin{align} \label{equ::Arikan:lower}
         \ln\mathbb{E}[G(X^n & |Y^n)^{\omega}] ^{\frac{1}{\omega}} \geq \notag\\
         &nH_{\frac{1}{1+\omega}}(X|Y) -\ln\left[1 + \ln(M_1\cdots M_n)\right],
    \end{align}
    where $H_{\alpha}(X|Y)$ is the Rényi entropy at rate $\alpha$, given by
    \begin{equation}
        H_{\alpha}(X|Y) = \frac{\alpha}{1-\alpha} \ln\sum_{y}\Big[\sum_{x} \mathbb{P}(x,y)^{\alpha}\Big]^{1/\alpha}.
    \end{equation}
\end{theorem}
As an extension of Theorem \ref{The::Arikan:lower}, the optimal guesswork has an upper bound \cite{arikan1996inequality}, given as
\begin{equation}   \label{Equ::Arikan:upper:optimal}
    \ln\mathbb{E}[G^*(X^n |Y^n)^{\omega}] ^{\frac{1}{\omega}}  \leq nH_{\frac{1}{1+\omega}}(X|Y),
\end{equation}
and \black{it follows that}
\begin{equation} \label{equ::Arikan::relationRenyi}
    \lim_{n\to \infty}\frac{1}{n}\ln\left(\mathbb{E}[G^*(X^n|Y^n)^{\omega}]\right)^{1/\omega} = H_{\frac{1}{1+\omega}}(X|Y),
\end{equation}
which provides \black{an asymptotic characterization} of $\mathbb{E}[G^*(X^n|Y^n)^{\omega}]$ only when the blocklength reaches infinity.

\subsubsection{Bounds for Markov Source}
Let $P$ be an irreducible Markov chain on $\mathcal{A}$ with the stochastic matrix $\mathbf{U} = [U_{ab}]$ and invariant probability $\mathbf{u} = [u_a]$ satisfying $\mathbf{u}\mathbf{U} = \mathbf{u}$, so that for $\nu = {\nu_1,\ldots,\nu_{n+1}} \in A^{n+1}$
\begin{equation}
P_{n+1}(\nu) = u_{\nu_1} \prod_{i=1}^{n} U_{\nu_i\nu_{i+1}},
\end{equation}
where $P_{n+1}$ is the restricted Markov chain on $\mathcal{A}_{n+1}$.
\begin{theorem} [Markov Source \cite{malone2004guesswork}]
For such a Markov chain $P$, its guesswork moments are characterized by
\begin{align}
\lim{n\to \infty} \frac{1}{n} \ln \mathbb{E}[G(P_n)^{\omega}] = (1+\omega)\ln(\lambda),
\end{align}
where $\lambda$ is the Perron–Frobenius eigenvalue of the matrix with entries $U_{ab}^{1/(1+\omega)}$.
\end{theorem}

\subsubsection{Relation to Compression}

The equivalence between the optimal guesswork and the optimal length function was shown in \cite{hanawal2010guessing}. Given the optimal length function $L^*(X)$ for the random variable $X\in \mathcal{X}$, there is 
\begin{align} \label{equ::manjesh:relation:GandL}
    \left|\log_2\mathbb{E}[G^*(X)^\omega] - \log_2\mathbb{E}[\exp_2(\omega L^*(X))]  \right| \leq \omega+\log_2 c,
\end{align}
for $c = \sum_{i = 1}^{\mathcal{X}}\frac{1}{i} \leq 1+ \ln |\mathcal{X}|$ and $\omega>0$. 

For the $n$-tuples $(X^n, Y^n)$, let $n\to \infty$, and then the right side of \eqref{equ::manjesh:relation:GandL} \black{vanishes as} $O(\log_{2}n/n)$. Thus, the limit 
\begin{equation}
    \lim_{n \to \infty} \frac{1}{n} \ln \mathbb{E}[G^*(X^n)^\omega]
\end{equation}
exists if and only if 
\begin{equation}
    \lim_{n \to \infty}\inf_{L} \frac{1}{n} \ln \mathbb{E}[\exp_2\{\omega L(X^n)\}]
\end{equation}
exists. Furthermore, these two limits are equal.

\subsubsection{Relation to Large Deviation Principle (LDP)}

As shown in \cite{malone2004guesswork,hanawal2010guessing}, for $\omega > 0$, $\lim_{n\to \infty}\frac{1}{n}\ln\left(\mathbb{E}[G^*(X^n)^{\omega}]\right)$ exists if and only if the Rényi entropy rate 
\begin{equation}
    \lim_{n \to \infty} \frac{1}{n} H_{\alpha}(X^n)
\end{equation}
exists. Let $v_n$ denote the distribution of the information spectrum $-\frac{1}{n} \ln \mathbb{P}(X^n)$. The large deviation principle immediately yields a sufficient condition of existence. 
\begin{theorem} [Existence of Rényi entropy rate \cite{hanawal2010guessing}]

    Let the sequence of distributions of the information spectrum $(v_n:n\in\mathbb{N})$ satisfy the LDP with rate function $I$. Then the limiting Rényi entropy rate of order $1/(1+\omega)$ exists for all $\omega>0$, and equals
    \begin{equation}
        \frac{1+\omega}{\omega} \sup_{t\in\mathbb{R}}\left\{\frac{\omega}{1+\omega} t - I(t)\right\},
    \end{equation}
\end{theorem}
According to \eqref{equ::Arikan::relationRenyi},
\begin{align}   \label{equ::Manjesh::LDP}
    \lim_{n\to \infty}\frac{1}{n}\ln\mathbb{E}[G^*(X^n)^{\omega}] = (1+\omega) \sup_{t\in\mathbb{R}}\left\{\frac{\omega}{1+\omega} t - I(t)\right\},
\end{align}
which is a scalar multiple of the Legendre-Fenchel dual of the rate function $I$.

The limiting guesswork itself satisfies an LDP as well.
\begin{theorem} [LDP of guesswork \cite{christiansen2012guesswork}] \label{The::Ken:LDP}
    The sequence $\frac{1}{n}\log G^*(X^n)$ satisfies an LDP with rate function $\Lambda^*$, where
    \begin{equation}
        \Lambda^*(x) :=
        \begin{cases}
        -x-g_1,   \ \ \ & \textup{for}\ \  x\in[0,\gamma],\\
        \sup_{\alpha\in\mathbb{R}}\{x\alpha - \Lambda(\alpha)\},  \ \ \ & \textup{for}\ \ x\in(\gamma,\ln{M}]\\
        \infty,   \ \ \ & \textup{for}\ \ x\notin[0,\ln{M}]\\
        \end{cases} 
\end{equation}
with
\begin{equation}
    \Lambda(\alpha) := \lim_{n\to\infty} \frac{1}{n} \log \mathbb{E}\left[e^{\alpha\log G(X^n)}\right],   \notag
\end{equation}
\begin{equation}
    g_1 := \lim_{n\to \infty} \frac{1}{n} \log \mathbb{P}(G^*(X^n)=1), \notag
\end{equation}
\begin{equation}
    \gamma := \lim_{\alpha\downarrow -1} \frac{d \Lambda(\alpha)}{d\alpha}    \notag
\end{equation}
$\Lambda(\alpha) $ is proved to exist for every $\alpha \in \mathbb{R}$ \textup{\cite{christiansen2012guesswork}}.
\end{theorem}
This LDP of guesswork can result in an approximation of the probability, $\mathbb{P}(G^*(X^n)=i) \approx i^{-1} \exp (-n \Lambda^*(n^{-1}\ln i))$, which \black{corresponds to} the probability of the $i_{\rm{th}}$ most likely sequence of $X^n$.

Theorem \ref{The::Ken:LDP} lays the foundation for analyzing the GRAND algorithm \black{\cite{duffy2021guessing}}. For blocklength $n$, define the random variable $N^n$ of the noise sequence. Then, $1/n \log G(N^n)$ satisfies the LDP with the rate function \cite{duffy2021guessing}
\begin{align}
    I^N(x) := \sup_{\alpha\in \mathbb{R}} \{x\alpha - \Lambda^{N}(\alpha)\},
\end{align}
where $\Lambda^{N}(\alpha)$ is given by \cite[Eq. (4)]{duffy2018guessing}. Combining the LDP from Theorem \ref{The::Ken:LDP} and the probability of guessing a non-transmitted codeword from \cite[Theorem 2]{duffy2018guessing} \black{enables the determination of} the average number of guesses in GRAND to find the MLD codeword at blocklength $n\to \infty$. \black{As GRAND is out of scope of this paper, we refer readers to \cite{duffy2021guessing,duffy2018guessing} for further details.} 

\begin{figure}  [t]
     \centering
\definecolor{mycolor1}{rgb}{0.46600,0.67400,0.18800}%
\definecolor{mycolor2}{rgb}{0.92900,0.69400,0.12500}%
\definecolor{mycolor3}{rgb}{0.63500,0.07800,0.18400}%
\begin{tikzpicture}

\begin{axis}[%
width=2.8in,
height=2in,
at={(0.758in,0.481in)},
scale only axis,
xmin=8,
xmax=60,
xlabel style={at={(0.5,1ex)},font=\color{white!15!black}, font=\small},
xlabel={Sequence length $n$},
ymode=log,
ymin=1,
ymax=10000000000,
yminorticks=true,
ylabel style={at={(1.5ex,0.5)},font=\color{white!15!black}, font=\small},
ylabel={$\mathbb{E} [G^*(X^n|Y^n)]$},
axis background/.style={fill=white},
tick label style={font=\footnotesize},
xmajorgrids,
ymajorgrids,
yminorgrids,
legend style={at={(0,1)}, anchor=north west , fill opacity=0.5, text opacity=1, font = \scriptsize	 , legend cell align=left, align=left, draw=white!15!black}
]

\addplot [color=red]
  table[row sep=crcr]{%
8	18.070456001469\\
12	76.816350094069\\
16	326.541380101027\\
20	1388.10647456831\\
24	5900.75164177457\\
28	25083.716974041\\
32	106629.272918283\\
36	453274.203932708\\
40	1926839.58473847\\
44	8190871.55876689\\
48	34818869.9379053\\
52	148012784.116381\\
56	629192857.233791\\
60	2674658503.02999\\
64	11369801843.0818\\
};
\addlegendentry{Arikan's upper bound \eqref{Equ::Arikan:upper:optimal}}

\addplot [color= black, mark=triangle, mark size = 1.5pt, dotted, line width=0.5pt, mark options={solid}]
  table[row sep=crcr]{%
8	3.7434\\
12	12.4548\\
16	44.987\\
20	169.532633333333\\
24	645.629833333333\\
28	2688.5404\\
32	11081.5566333333\\
36	39713.0891\\
40	196539.9829\\
};
\addlegendentry{Simulation, optimal guesswork}

\addplot [color=mycolor2]
  table[row sep=crcr]{%
8	2.76088099287671\\
12	8.24407360300983\\
16	27.0084197775885\\
20	93.3937792458827\\
24	334.594472693414\\
28	1229.10467385049\\
32	4599.91406388503\\
36	17464.9940519101\\
40	67076.7649340383\\
44	260040.249984264\\
48	1015984.48359725\\
52	3995631.38553349\\
56	15802416.8999697\\
60	62801876.7051961\\
64	250649162.976181\\
};
\addlegendentry{Arikan's lower bound \eqref{equ::Arikan::lowerbound}}

\end{axis}

\end{tikzpicture}%
	\vspace{-0.5em}
    \caption{The average number of guesses at different blocklength $n$ with binary symmetric channel with error probability $0.05$. }
    \vspace{-0.25em}
	\label{Fig::Arikan:exmp}
        
    \end{figure}
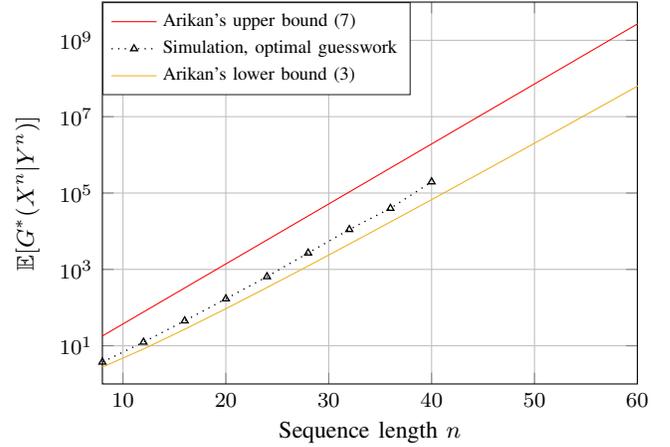

The guesswork approaches presented in Theorems 1-4 were established using asymptotic analysis, i.e., as $n\to \infty$. For example, Arikan's bounds, given in \eqref{equ::Arikan::lowerbound} and \eqref{Equ::Arikan:upper:optimal}, are found to be loose for short blocklengths, as demonstrated by the example in Fig. \ref{Fig::Arikan:exmp}. On the other hand, these results were mainly developed for i.i.d. pairs $(X^n,Y^n)$. Therefore, \black{to analyze the guesswork complexity of OSD, we need to develop new methods suitable for ordered statistics.}

\section{A new upper bound of guesswork} \label{sec::guess::upper}

Consider an i.i.d sequence $X^n$ with each element following the distribution of $X \in \mathcal{X}$. The $\omega_{\rm{th}}$ moment of its guesswork can be expressed as
\begin{equation}  \label{equ::mydefine::guesswork}
\begin{split}
    \mathbb{E}[G(X^n)^\omega] = \sum_{g=1}^{|M^n|}  \mathbb{P}\left(X^n = x^n_{(g)}\right) g^\omega, 
\end{split}
\end{equation}
where $x^n_{(g)}$ is the realization of $X^n$ satisfying $G(x^n_{(g)}) = g$. Then, we have the following simple upper bound of $\mathbb{E}[G(X^n)^\omega]$ from Hölder's inequality.

\begin{lemma} \label{lem::holder}
For i.i.d sequence $X^n$, there is

  \begin{equation} \label{equ::Holder}
    \mathbb{E}[G(X^n)^\omega] \leq   \left(\frac{  M^{n\omega  p+n}}{\omega p+1} \right)^{\frac{1}{p}}\left(\sum_{x\in\mathcal{X}} \mathbb{P}(x)^{q}\right)^{\frac{n}{q}} ,
\end{equation}
for $\frac{1}{p} + \frac{1}{q} = 1$, $p > 1$, and $q>1$.

\end{lemma}

\begin{IEEEproof}
   Choose \black{$a_g = \mathbb{P}(X^n = x^n_{(g)})$} and $b_g = g^\omega$ in Holder's inequality, and obtain
     \begin{equation} \label{equ::Holder:p1}
        \sum_{g=1}^{M^n} a_g b_g \leq  \left(\sum_{g=1}^{M^n}  \mathbb{P}\left(X^n = x^n_{(g)}\right)^{q}\right)^{\frac{1}{q}}  \left(\sum_{g=1}^{M^n}   g^{\omega p}\right)^{\frac{1}{p}}.
    \end{equation}  
    Also, we have 
    \begin{equation} \label{equ::Holder:p2}
        \sum_{g=1}^{M^n} g^{\omega p} \leq \int_{0}^{M^n} g^{\omega p} dg = \frac{1}{\omega  p+1}M^{n(\omega  p+1)},
    \end{equation}
    and
    \begin{equation}  \label{equ::Holder:p3}
         \sum_{g=1}^{M^n}  \mathbb{P}\left(X^n \!=\! x^n_{(g)}\right)^{q} = \sum_{x^n \in \mathcal{X}^n} \!\! \mathbb{P}(x^n)^{q} =  \left(\sum_{x \in \mathcal{X}} \mathbb{P}(x)^{q}\right)^{n}
    \end{equation}
    Substituting \eqref{equ::Holder:p2} and \eqref{equ::Holder:p3} into \eqref{equ::Holder:p1} completes the proof.
\end{IEEEproof}

The bound given in \eqref{equ::Holder} applies to general guesswork $G(X^n)$, and might be loose for the optimal guesswork $G^*$. \black{This can be demonstrated through two cases:} 
As $q$ approaches 1 from above, $p \to \infty$ and the bound will be no tighter than 
\begin{equation} \label{equ::bound::insight1}
\lim_{p \to \infty} \left(\frac{M^{n\omega p+n}}{\omega p+1}\right)^{\frac{1}{p}} = M^{\omega n}.
\end{equation}
On the other hand, as $q\to \infty$, we have
\begin{equation}
\lim_{q \to \infty} \left(\sum_{x} P(x)^{q}\right)^{\frac{n}{q}} = \max_{x\in\mathcal{X}}\{\mathbb{P}(x)\}^n,
\end{equation}
and the upper bound becomes  
\begin{equation} \label{equ::bound::insight2}
    \frac{1}{\omega + 1}M^{n(\omega+1)}\max_{x\in\mathcal{X}}\{\mathbb{P}(x)\}^n.
\end{equation}
In contrast, for any $X^n$ and $\omega = 1$, a trivial bound is $\mathbb{E}[G^*(X^n)] \leq \frac{1}{2}({M^n}+1)$, derived from assuming equiprobable $X^n$. This trivial bound is already tighter than \eqref{equ::bound::insight1} and \eqref{equ::bound::insight2}, since $\max_{x}\{\mathbb{P}(x)\} \geq \frac{1}{M^n}$ for any $X^n$.

However, \eqref{equ::Holder} is tighter than Arikan's bound \eqref{equ::Arikan::relationRenyi} for some distributions. For example, with $n = 1$ and a uniform distribution $X$, \eqref{equ::Arikan::relationRenyi} provides 
\begin{equation} \label{equ::Arikon:Uniform}
    \mathbb{E}[G^*(X)^\omega] = \mathbb{E}[G(X)^\omega] \leq   M^{\omega},
\end{equation}
while \eqref{equ::Holder}, with $q = \omega + 1$, yields 
\begin{equation}  \label{equ::Holder:Uniform}
    \mathbb{E}[G(X)^\omega] \leq \left(\omega + 2\right)^{-\frac{\omega}{\omega + 1}} M^{\omega}
\end{equation} 
In this case, \eqref{equ::Holder:Uniform} is tighter than \eqref{equ::Arikon:Uniform}, as $\left(\omega + 2\right)^{-\frac{\omega}{\omega + 1}} < 1$ for $\omega \geq 1$. 

To refine  \eqref{equ::Holder:Uniform}, we can divide $\mathcal{X}^n$ into $n+1$ subsets $\{\mathcal{X}_0,\ldots,\mathcal{X}_{n}\}$. Let $\mathbb{E}[G(X^n)^\omega \mid X^n \in \mathcal{X}_j]$ denote the $\omega_{\rm{th}}$ moment of the guesswork for $X^n \in \mathcal{X}_j$. Then, according to the law of total expectation, there is 
\begin{equation} \label{equ::lawofEx1}
\mathbb{E}[G(X^n)^\omega] = \sum_{j = 0}^{n} \mathbb{P}(X^n \in \mathcal{X}_j) \mathbb{E}[G(X^n)^\omega \mid X^n \in \mathcal{X}_j].
\end{equation}
Applying \black{Holder's inequality} \eqref{equ::Holder:p1} to $\mathbb{E}[G(X^n)^\omega \mid X^n \in \mathcal{X}_j]$ obtains that 
\begin{align}  \label{equ::lawofEx2}
    \mathbb{E}[G(X^n)^\omega] &\leq  \sum_{j = 0}^{n}\!\! \left(\sum_{x^n \in \mathcal{X}_j}\!\! G(x^n)^{\omega p}\right)^{\frac{1}{p}} \!\!\! \left(\sum_{x^n \in \mathcal{X}_j}\mathbb{P}(x^n)^{q} \right)^{\frac{1}{q}} 
\end{align}

Then, selecting subsets $\{\mathcal{X}_0,\ldots,\mathcal{X}_{n}\}$ according to the Hamming shells results in the following tighter bound with respect to $G^*(X^n)$.
\begin{theorem} [Hamming subset bound] \label{The::Newbound:Xn}
    Let $\bar{x}^n = [\bar{x}, \ldots, \bar{x}]$ denote the most likely sequence of i.i.d. $X^n\in \mathcal{X}^n$. Then, $\mathbb{E}[G^*(X^n)^\omega]$ is upper bounded by 
    \begin{align} \label{equ::Newbound:Xn}
        \mathbb{E}[G^*(X^n)^\omega] \leq \sum_{j=0}^{n}(\gamma_j)^{\frac{1}{p}} \binom{n}{j}^{\frac{1}{q}}\cdot \mathbb{P}(\bar{x})^{n-j} \Big(\sum_{x\in\mathcal{X}\backslash\bar{x}}\mathbb{P}(x) ^q \Big)^{\frac{j}{q}},
    \end{align}
    where $p > 1$, $q > 1$, $\frac{1}{p} + \frac{1}{q} = 1$, 
    \begin{equation}  \label{equ::gamma_j}
        \gamma_j = \frac{\beta_j^{\omega p +1} - \beta_{j-1}^{\omega p +1}}{\omega p +1} ,
    \end{equation}
    and
    \begin{equation} \label{equ::beta_j}
        \beta_j := \sum_{i = 0}^{j} (M-1)^{i}\binom{n}{i}. 
    \end{equation}
\end{theorem}
\begin{IEEEproof}
    Let $\bar{x}^n$ denote the most likely sequence in $\mathcal{X}^n$. Then, \black{let us construct} $\{\mathcal{X}_0,\ldots,\mathcal{X}_n\}$ with
    \begin{equation}
        \mathcal{X}_j := \left\{x^n \in \mathcal{X}^n \mid d_{\mathrm{H}} (x^n,\bar{x}^n) = j \right\},
    \end{equation} 
    where $d_{\mathrm{H}} (x^n,\bar{x}^n) = |\{i \mid x_i \neq \bar{x}_i, 1 \leq i \leq n\}|$ is the Hamming distance between $x^n = [x_1,\ldots,x_n]$ and $\bar{x}^n$. In other words, $\mathcal{X}_j$ represents the Hamming shell with radius $j$ to $\bar{x}^n$.  Then, for $\mathcal{X}_j$, we have
    \begin{equation} \label{equ::subset:proof1}
    \begin{split}
             \left(\sum_{x^n \in \mathcal{X}_j}\mathbb{P}(x^n)^{q} \right)^{\frac{1}{q}} &= \binom{n}{j}^{\frac{1}{q}}\mathbb{P}(\bar{x})^{n-j} \Big(\sum_{x\in\mathcal{X}\backslash\bar{x}}\mathbb{P}(x) ^q \Big)^{\frac{j}{q}} .
    \end{split}
    \end{equation}
    Consider a suboptimal guess strategy, \black{denoted by $G^\dagger(X^n)$}, that always guesses sequences with lower Hamming distance to $\bar{x}^n$. Let $\beta_j$ \black{represents the total number of sequences in $\mathcal{X}_0 \cup \mathcal{X}_1 \cup \ldots \cup \mathcal{X}_j$; that is, 
    \begin{equation}
        \beta_j = \sum_{i=0}^{j} |\mathcal{X}_i| = \sum_{i = 0}^{j} (M-1)^{i}\binom{n}{i}.
    \end{equation}
    } Then, we have
    \black{
    \begin{align}  
        \sum_{x^n \in \mathcal{X}_j}\!\! G^*(x^n)^{\omega p} &\leq \sum_{x^n \in \mathcal{X}_j}\!\! G^\dagger(x^n)^{\omega p}  \label{equ::subset:proof3} \\
        &\leq \int_{\beta_{j-1}}^{\beta_{j}} g^{\omega p} dg  \notag\\
        &\leq \frac{\beta_j^{\omega p +1} - \beta_{j-1}^{\omega p +1}}{\omega p +1},  \label{equ::subset:proof2}
    \end{align}
    which is obtained similarly to \eqref{equ::Holder:p2}. The expression in \eqref{equ::subset:proof2} is denoted as $\gamma_j$.}
    Substituting \eqref{equ::subset:proof1} and \eqref{equ::subset:proof2} into \eqref{equ::lawofEx2} completes the proof.
\end{IEEEproof}
\black{Note that \eqref{equ::subset:proof2} shows that the bound in \eqref{equ::Newbound:Xn} applies to both optimal and suboptimal guesswork moments, i.e., $\mathbb{E}[G^*(X^n)^\omega]$ and $\mathbb{E}[G^\dagger(X^n)^\omega]$.}

\black{We now extend the bound to the $n$-fold i.i.d. pair $(X^n,Y^n)$. In the context of wireless transmission, $X^n$ represents the transmitted data sequence and $Y^n$ represents the received sequence from the channel. The channel transmission probability is characterized by $\mathbb{P}_{X^n|Y^n}(x^n|y^n)$, which we denote as $\mathbb{P}(x^n|y^n)$ for simplicity.}

\begin{corollary}  \label{Cor::HSB}
For i.i.d. pair $(X^n,Y^n)$ with each following $(X,Y)$, the guesswork $\mathbb{E}[G^*(X^n|Y^n)^\omega]$ is upper bounded by 

\begin{align} \label{equ::Newbound:XnYn}
    \mathbb{E}&[G^*(X^n|Y^n)^\omega] \notag\\
    &\leq \sum_{j=0}^{n}\black{(}\gamma_j\black{)^{\frac{1}{p}}}\black{\binom{n}{j}^\frac{1}{q}} 
     \mathbb{E}[\mathbb{P}(\bar{x}|Y)^q]^{\frac{n-j}{q}}\Big(\sum_{x\in\mathcal{X}\backslash\bar{x}} \mathbb{E}[\mathbb{P}(x|Y)^q] \Big)^{\frac{j}{q}},
\end{align}
    where $p,q>1$, $\frac{1}{p} + \frac{1}{q} = 1$, and $\gamma_j$ is given by \eqref{equ::gamma_j}. Here, $\bar{x}$ is the most likely value of $X$ given $Y=y$, i.e., $\bar{x} := \argmax_{x} \mathbb{P}(x|y)$.
\end{corollary}
\begin{IEEEproof}
    Eq. \eqref{equ::lawofEx2} yields 
    \begin{align}  \label{equ::lawofEx:XY}
        \mathbb{E}[\black{G}(X^n|Y^n)^\omega] \leq  & \sum_{j = 0}^{n}\!\! \left(\sum_{x^n \in \mathcal{X}_j}\!\! G(x^n)^{\omega p}\right)^{\frac{1}{p}} \notag\\
        \cdot  & \sum_{y^n} \mathbb{P}(y^n) \left(\sum_{x^n \in \mathcal{X}_j}\mathbb{P}(x^n|y^n)^{q} \right)^{\frac{1}{q}} . 
    \end{align}  
    \black{Let $f(x) = x^{\frac{1}{p}}$. Since $p > 1$, $f(x)$ is a concave function. A}ccording to Jensen's inequality, \black{there is $\mathbb{E}\left[f(X)\right] \leq f(\mathbb{E}[X])$}  for the concave function $f(x) = x^{\frac{1}{p}}$. \black{Applying Jensen's inequality to \eqref{equ::lawofEx:XY}, we obtain}
    \begin{align}   \label{equ::cor1::Jensen}
        \mathbb{E}[G(X^n|Y^n)^\omega] \leq   \sum_{j = 0}^{n} &\!\! \left(\sum_{x^n \in \mathcal{X}_j}\!\! G(x^n)^{\omega p}\right)^{\frac{1}{p}} \notag \\
        &\cdot  \left(\sum_{y^n} \sum_{x^n \in \mathcal{X}_j}\mathbb{P}(x^n|y^n)^{q} \mathbb{P}(y^n) \right)^{\frac{1}{q}}   \notag \\
        =  \sum_{j = 0}^{n}&\!\! \left(\sum_{x^n \in \mathcal{X}_j}\!\! G(x^n)^{\omega p}\right)^{\frac{1}{p}} \notag \\
        &\cdot  \left(\sum_{x^n \in \mathcal{X}_j} \prod_{i=1}^{n} \mathbb{E}[\mathbb{P}(x_i|Y_i)^{q} ] \right)^{\frac{1}{q}},  
    \end{align}
    \black{Then, \eqref{equ::Newbound:XnYn} is obtained by following similar steps as in \eqref{equ::subset:proof1} and \eqref{equ::subset:proof2}. Specifically, considering the suboptimal guesswork $G(x^n) = G^\dagger(x^n)$, we have
    \begin{align}   \label{equ::cor1::1}
        \left(\sum_{x^n \in \mathcal{X}_j}\!\! G^*(x^n)^{\omega p}\right)^{\frac{1}{p}} &\leq \left( \sum_{x^n \in \mathcal{X}_j}\!\! G^\dagger(x^n)^{\omega p} \right)^{\frac{1}{p}} \leq (\gamma_j)^{\frac{1}{p}}.
    \end{align}
    Moreover, we obtain
    \begin{align}   \label{equ::cor1::2}
        &\left(\sum_{x^n \in \mathcal{X}_j} \prod_{i=1}^{n} \mathbb{E}[\mathbb{P}(x_i|Y_i)^{q} ] \right)^{\frac{1}{q}} = \notag\\
        & \hspace{1.5cm} \binom{n}{j}^\frac{1}{q}
     \mathbb{E}[\mathbb{P}(\bar{x}|Y)^q]^{\frac{n-j}{q}}\Big(\sum_{x\in\mathcal{X}\backslash\bar{x}} \mathbb{E}[\mathbb{P}(x|Y)^q] \Big)^{\frac{j}{q}}
    \end{align}
    Substituting  \eqref{equ::cor1::1} and  \eqref{equ::cor1::2} into \eqref{equ::cor1::Jensen} obtains \eqref{equ::Newbound:XnYn}.
    }
\end{IEEEproof}

For binary $X^n$, \eqref{equ::Newbound:XnYn} is simply reduced to 
    \begin{align} \label{equ::Newbound:XnYn:binary}
        \mathbb{E}&[G^*(X^n|Y^n)^\omega] \notag\\
        &\leq \sum_{j=0}^{n}\black{(}\gamma_j\black{)^{\frac{1}{p}}} \cdot \black{\binom{n}{j}^\frac{1}{q}}  \cdot \mathbb{E}[\mathbb{P}(\bar{x}|Y)^q]^{\frac{n-j}{q}} \cdot \mathbb{E}[\mathbb{P}(1-\bar{x}|Y)^q]^{\frac{j}{q}},
    \end{align}
    \black{since for binary $X$, the complement set $\mathcal{X}\backslash\bar{x}$ contains only the single element $1-\bar{x}$.}

\begin{figure}  [t]
     \centering
\definecolor{mycolor1}{rgb}{0.00000,0.44706,0.74118}%
\definecolor{mycolor2}{rgb}{0.92900,0.69400,0.12500}%
\definecolor{mycolor3}{rgb}{0.46600,0.67400,0.18800}%
\begin{tikzpicture}

\begin{axis}[%
width=2.8in,
height=2in,
at={(1.01in,0.685in)},
scale only axis,
xmin=5,
xmax=50,
xlabel style={at={(0.5,1ex)},font=\color{white!15!black}, font=\small},
xlabel={Sequence length $n$},
ymode=log,
ymin=1,
ymax=1e11,
yminorticks=true,
ylabel style={at={(1.5ex,0.5)},font=\color{white!15!black}, font=\small},
ylabel={The average number of guesses},
axis background/.style={fill=white},
tick label style={font=\footnotesize},
xmajorgrids,
ymajorgrids,
yminorgrids,
legend style={at={(0,1)}, anchor=north west , fill opacity=0.5, text opacity=1, font = \scriptsize	 , legend cell align=left, align=left, draw=white!15!black}
]
\addplot [color=red, line width=0.5pt]
  table[row sep=crcr]{%
8	44.3723209678601\\
12	295.575354210657\\
16	1968.9028680748\\
20	13115.3644872242\\
24	87364.7900167555\\
28	581959.162637588\\
32	3876578.50391318\\
36	25822878.7547421\\
40	172012785.632743\\
44	1145821064.41956\\
48	7632606534.67178\\
};
\addlegendentry{Arikan's upper bound, Eq. \eqref{Equ::Arikan:upper:optimal}}

\addplot [color=black, line width=0.5pt]
  table[row sep=crcr]{%
8	24.0972744487952\\
12	191.47986480107\\
16	1551.27857944387\\
20	12805.4333678437\\
24	106806.188285465\\
28	895585.447673489\\
32	7573116.493138\\
36	64458817.9820575\\
40	551520931.822786\\
44	4739166305.74067\\
48	40868999374.7302\\
};
\addlegendentry{Hamming subset bound, Eq. \eqref{equ::Newbound:XnYn:binary}}

\addplot [color= black, mark=square, mark size = 1pt, dotted, line width=0.5pt, mark options={solid}]
  table[row sep=crcr]{%
8	17.733\\
12	136.89\\
16	1053.502\\
20	8491.173\\
24	83521.7385\\
28	663693.649\\
};
\addlegendentry{Simulation, suboptimal guesswork}

\addplot [color= black, mark=triangle, mark size = 1.5pt, dotted, line width=0.5pt, mark options={solid}]
  table[row sep=crcr]{%
8	9.157\\
12	50.8242\\
16	299.3594\\
20	1785.109\\
24  1.1624e+04\\
};
\addlegendentry{Simulation, optimal guesswork}

\addplot [color=mycolor2]
  table[row sep=crcr]{%
8	6.77939159698189\\
12	31.7216969091128\\
16	162.849054982894\\
20	882.420389278883\\
24	4953.90716678868\\
28	28516.057948199\\
32	167232.951063732\\
36	994974.829676656\\
40	5988075.64413965\\
44	36377031.9057176\\
48	222712851.464128\\
};
\addlegendentry{Arikan's lower bound, Eq. \eqref{equ::Arikan::lowerbound}}

\end{axis}
\end{tikzpicture}%
	\vspace{-0.5em}
    \caption{The average number of guesses for various block lengths $n$ in a BI-AWGN channel for the pair $(X^n,Y^n)$ at \black{SNR = 0 dB}. The suboptimal guesswork is performed in ascending order of the Hamming distance from the most likely $X^n$ sequence given $Y^n$.}
    \vspace{-0.25em}
	\label{Fig::DSB:exmp}
        
    \end{figure}
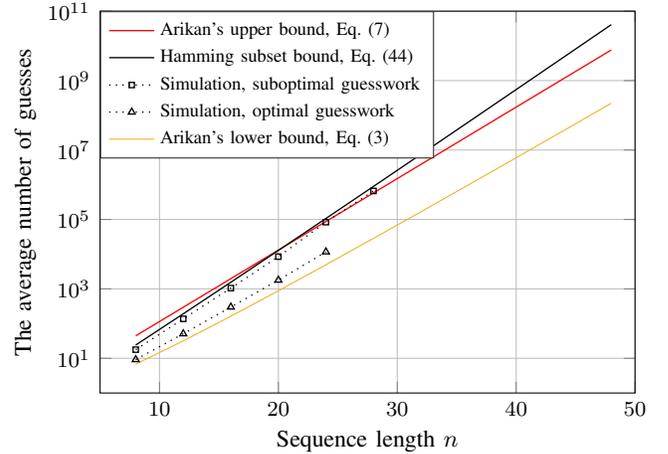

\begin{figure}  [t]
     \centering
        \input{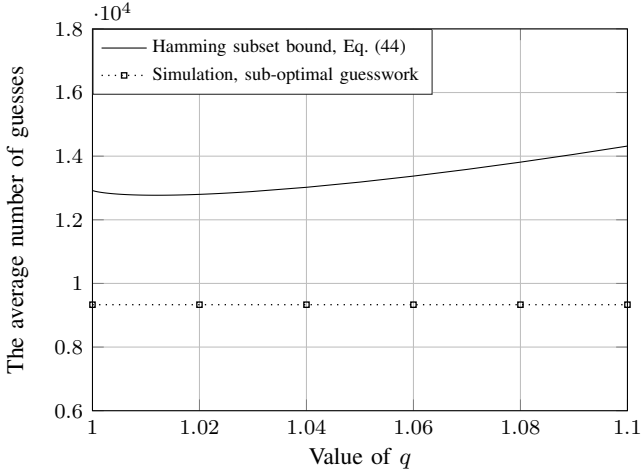}
	\vspace{-0.5em}
    \caption{\black{The performance of the proposed Hamming subset bound with various values of parameter $q$ for blocklengths $n = 20$ in a BI-AWGN channel at SNR = 0 dB.}}
    \vspace{-0.25em}
	\label{Fig::DSB:qValues}      
    \end{figure}

    We verify the performance of the Hamming subset bound \eqref{equ::Newbound:Xn} in a binary AWGN (BI-AWGN) channel, as depicted in Fig. \ref{Fig::DSB:exmp}. The optimal guesswork and suboptimal guesswork obtained from simulations are included for comparison, where the suboptimal guesswork guesses sequences in the order of increasing Hamming distance from the most likely sequence \black{of} $X^n$, given the received sequence $Y^n = y^n$. As shown, the Hamming subset bound is tighter than Arikan's upper bound for optimal guesswork at very short blocklengths. Moreover, it provides a tight upper bound for suboptimal guesswork, \black{since} it is derived by upscaling the required guess number in each Hamming shell, following \eqref{equ::subset:proof2}. \black{Note that the Hamming subset bound \eqref{equ::Newbound:XnYn:binary} is built based on Hölder's inequality. Thus, the choice of parameter $q$ and the corresponding  $p = \frac{q}{q-1}$ is crucial to the bound tightness. The Hamming subset bound in Fig. \ref{Fig::DSB:exmp} is produced by using the tightest results of \eqref{equ::Newbound:XnYn:binary} obtained across values of $q$ in the range $(1,3]$. For example, Fig. \ref{Fig::DSB:qValues} illustrates how the tightness of bound \eqref{equ::Newbound:XnYn:binary} varies with different values of $q$ when $n = 20$ and SNR = 0 dB. The tightest bound with the value $12,771$ is achieved at $q = 1.012$, compared to $\mathbb{E}[G^\dagger(X^n|Y^n)] = 9,332$ obtained through simulation of suboptimal guesswork.}
    
    \black{Based on the results in Fig. \ref{Fig::DSB:exmp},} the Hamming subset bound will be suitable for characterizing the OSD complexity, as \black{OSD algorithms} typically process TEPs in ascending order of Hamming weight \cite{Fossorier1995OSD}.

    \section{Guesswork for Ordered symmetric channel} \label{sec::OSguess}

    \black{This section establishes the theoretical framework for analyzing the guesswork for ordered symbol sequences, which provides the mathematical foundation for analyzing the guesswork complexity of OSD in Section \ref{sec::OSD:bound}. We first introduce the binary ordered symmetric channel (BI-OSC), which formalizes how transmitted symbols $X^n$ are ordered based on their reliabilities. Then, we demonstrate that under certain conditions, guesswork over ordered sequences can be related to guesswork over independent sequences through  a \textit{conditional independence} relationship. Based on this relationship, we develop bounds for guesswork moments for ordered sequences in BI-OSC.}

    \subsection{Binary Ordered Symmetric Channel \black{(BI-OSC)}}
    \black{
        We begin by introducing the BI-OSC, which defines the ordering process of transmitted signal $X^n$ and received sequence $Y^n$, and characterizes its statistical properties.
    }
    
    Consider the $n$-tuple pair $(X^n,Y^n)$, with the \black{independent and identical} transition probability $\mathbb{P}(X_i|Y_i) = \mathbb{P}(X|Y)$. Let $y^n$ be a specific realization of $Y^n$. Then, $(X^n,y^n)$ are ordered in descending order of $\max\mathbb{P}(X_i|y_i)$. Let $(\widetilde{X}_i, \tilde{y}_i)$ denote the $i_{\rm{th}}$ \black{ordered pair}, and  $(\widetilde{X}^n, \tilde{y}^n)$ satisfies
    \black{
    \begin{equation} \label{equ::ordering::XnYn}
        \max_{\widetilde{X}_1 \in \mathcal{X}}\mathbb{P}(\widetilde{X}_1|\tilde{y}_1) \geq \max_{\widetilde{X}_2 \in \mathcal{X}}\mathbb{P}(\widetilde{X}_2|\tilde{y}_2)\geq \ldots \geq  \max_{\widetilde{X}_n \in \mathcal{X}}\mathbb{P}(\widetilde{X}_n|\tilde{y}_n).
    \end{equation}
    }
    For binary $X_i\in\{-1,+1\}$, the ordering is equivalent to 
    \begin{equation} \label{equ::ordering::BI-continuous}
        \infty \geq |\tilde{\ell}_1| \geq |\tilde{\ell}_2| \geq \ldots \geq  |\tilde{\ell}_n| \geq 0
    \end{equation}
    where $\tilde{\ell}_i$ is the log-likelihood ratio (LLR), defined as $\log\mathbb{P}_{X|Y}(1|\tilde{y}_i) - \log \mathbb{P}_{X|Y}(-1|\tilde{y}_i) $. \black{For notational convenience, we define two auxiliary variables $|\tilde{\ell}_0| = \infty$ and $|\tilde{\ell}_{n+1}| = 0$.}

    Let us consider a binary continuous channel, and denote the random variable of $\tilde{y}_i$ as $\widetilde{Y}_i$. Also, we define $\ell_i := \log\mathbb{P}_{X|Y}(1|y_i) - \log \mathbb{P}_{X|Y}(-1|y_i) $ as the LLR of the unordered pair $(X_i,y_i)$, with its random variable $L_i$. Then, according to the ordered statistics \cite{balakrishnan2014order}, the distribution of $\widetilde{Y}_i$ will be 
    \begin{equation}
        \mathbb{P}_{\widetilde{Y}_i}(y) = f_{i}(y) \cdot \mathbb{P}_{Y}(y),
    \end{equation}
    with 
    \begin{equation} \label{equ::order:coefficient1}
        f_{i}(y) = \frac{1}{\mathrm{B}(i,n-i+1)} \black{[1 - p(y)]^{n-i}} p(y)^{i},
    \end{equation}
    \begin{equation}
        p(y) = F_{|L|}\left(\left|\log\mathbb{P}_{X|Y}(+1|y) - \log \mathbb{P}_{X|Y}(-1|y) \right|\right),
    \end{equation}
    and
    \begin{equation}
        \frac{1}{\mathrm{B}(i,n-i+1)} = \frac{n!}{(i-1)!(n-i)!} .
    \end{equation}
    where $F_{|L|}(\ell)$ is the $\mathrm{cdf}$ of $|L_i|$. The subscript $i$ is omitted since both $\{(X_i,Y_i)\}$ and $\{L_i\}$ are i.i.d.

    \black{If $\mathbb{P}(y|-x) = \mathbb{P}(-y|x)$}, there will be $\mathbb{P}(\tilde{y}_i|-\tilde{x}_i ) = \mathbb{P}(-\widetilde{y}_i|\widetilde{x}_i )$ for $1\leq i \leq n$, because $f_{i}(y) = f_{i}(-y)$. We \black{refer to} the channel $(\widetilde{X}^n, \widetilde{Y}^n)$ with $\widetilde{X}_i \in \{-1,+1\}$ described by $\mathbb{P}_{\widetilde{Y}_i|\widetilde{X}_i}$ as the length-$n$ binary ordered symmetric channel (BI-OSC)\footnote{In \cite{fossorier1996first}, such a channel was also referred to as the ordered binary symmetric channel.}. In BI-OSC, $(\widetilde{X}_i,\widetilde{Y}_i)$ and $(\widetilde{X}_j,\widetilde{Y}_j)$ are dependent for $i\neq j$, due to the ordering \eqref{equ::ordering::BI-continuous}. Before the ordering, the original pairs $(X^n,Y^n)$ are referred to as the original channel. 
    
    %We are interested in the guesswork over a contiguous subsequence of $(\widetilde{X}^n, \widetilde{Y}^n)$ from the $a_{\rm{th}}$ ordered pair to the $b_{\rm{th}}$ ordered pair, denoted by $(\widetilde{X}_a^b, \widetilde{Y}_a^b)$. \black{However}, due to the dependency between ordered variables, bounds in Theorem \ref{The::Arikan:lower} and Theorem \ref{The::Newbound:Xn} are not directly applicable here.

    \black{OSD needs to process the first $k$ most reliable bits among $n$ received symbols. This motivates our analysis of guesswork over contiguous subsequences of $(\widetilde{X}^n, \widetilde{Y}^n)$. Specifically, for any segment from the $a_{\rm{th}}$ to $b_{\rm{th}}$ ordered positions ($1 \leq a \leq b \leq n$), we denote the corresponding subsequence by $(\widetilde{X}_a^b, \widetilde{Y}_a^b)$. By developing a general analysis for arbitrary $a$ and $b$, we can then characterize guesswork complexity of OSD as a special case where $a = 1$ and $b = k$. However, due to the dependency introduced by the ordering process \eqref{equ::ordering::BI-continuous}, bounds in Theorem \ref{The::Arikan:lower} and Theorem \ref{The::Newbound:Xn} are not directly applicable to these ordered subsequences. Therefore, we next establish a relationship between ordered sequences and independent sequences under certain conditions.}

    \subsection{Conditional Independence} \label{sec::conditional::indep}

    \black{
    To analyze guesswork over ordered sequences, we construct a new pair of random variables $(\hat{X}, \hat{Y})$ from the original channel variables $(X,Y)$. This construction allows us to transform the ordered sequence guesswork problem into an equivalent independent sequence guesswork problem under certain boundary conditions. We first define sets based on LLR boundaries, then use these sets to derive the distribution of $(\hat{X}, \hat{Y})$ from the original channel distribution.
    }

    Let us define the sets
    $$\black{\mathcal{Y}_{a,b}^{(+)}} :=  \left\{y: e^{-|\tilde{\ell}_{a-1}|} + 1 \leq \mathbb{P}_{X|Y}^{-1}(1|y) \leq   e^{-|\tilde{\ell}_{b+1}|}+1 \right\},$$
    $$\black{\mathcal{Y}_{a,b}^{(-)}} :=  \left\{y: e^{|\tilde{\ell}_{b+1}|} + 1 \leq \mathbb{P}_{X|Y}^{-1}(1|y) \leq   e^{|\tilde{\ell}_{a-1}|}+1 \right\},$$
    and $\mathcal{Y}_{a,b} := \mathcal{Y}_{a,b}^{(+)} \cup \mathcal{Y}_{a,b}^{(-)}$\black{, where $ \mathbb{P}_{X|Y}^{-1}(1|y)$ is the reciprocal of the density function $ \mathbb{P}_{X|Y}(1|y)$. In fact, $\mathcal{Y}_{a,b}$ defines a set where the received symbols $y$ have reliability levels (measured by their LLR magnitudes) between those of the $(a-1)_{\rm{th}}$ and $(b+1)_{\rm{th}}$ ordered positions. The subsets $\mathcal{Y}_{a,b}^{(+)}$ and $\mathcal{Y}_{a,b}^{(-)}$ correspond to positive and negative LLR regions, respectively.}

    \black{Using the set $\mathcal{Y}_{a,b}$, we now construct a pair of random variables $(\hat{X}, \hat{Y})$ with the following joint density}
    \begin{equation} \label{equ::auxiliary:iidvariable}
        \mathbb{P}_{\hat{X},\hat{Y}}(x,y) := \frac{\mathbb{P}_{X,Y}(x,y)}{\int_{\mathcal{Y}_{a,b}} \mathbb{P}_{X,Y}(x,y') dy'}
    \end{equation}
    for $\hat{X} \in \{-1,1\}$ and $\mathrm{supp}(\hat{Y}) = \mathcal{Y}_{a,b}$. Similarly, $\mathbb{P}_{\hat{Y}}$  can be accordingly defined as 
    \begin{equation}  \label{equ::auxiliary:hatY}
        \mathbb{P}_{\hat{Y}}(y) := \black{\frac{1}{2}}\sum_{x\in\{-1,1\}}\frac{\mathbb{P}_{X,Y}(x,y)}{\int_{\mathcal{Y}_{a,b}}\mathbb{P}_{X,Y}(x,y') dy'},
    \end{equation}
    and $\mathbb{P}_{\hat{X}|\hat{Y}}(x,y) = \mathbb{P}_{\hat{X},\hat{Y}}(x,y)/\mathbb{P}_{\hat{Y}}(y)$. \black{In \eqref{equ::auxiliary:hatY}, the coefficient $\frac{1}{2}$ comes from that $x\in\{-1,1\}$ is equiprobable.}
    
    %Let $(\hat{X}_a^b, \hat{Y}_a^b)$ be a sequence of i.i.d. pairs following $(\hat{X}, \hat{Y})$. We then demonstrate that a given guesswork for the pair $(\widetilde{X}_a^b, \widetilde{Y}_a^b)$ is equivalent to that of $(\hat{X}_a^b, \hat{Y}_a^b)$ under certain conditions, and thus the bound for $\mathbb{E}[G(\hat{X}_a^b| \hat{Y}_a^b)^\omega]$ suffices to bound $\mathbb{E}[G(\widetilde{X}_a^b| \widetilde{Y}_a^b)^\omega]$.

    \black{
        Let $(\hat{X}_a^b, \hat{Y}_a^b)$ be a sequence of i.i.d. pairs following $(\hat{X}, \hat{Y})$. By this construction, these pairs are independent but their LLRs are confined within the reliability boundaries $|\tilde{\ell}_{a-1}|$ and $|\tilde{\ell}_{b+1}|$. We then demonstrate that a given guesswork for the ordered sequence $(\widetilde{X}_a^b, \widetilde{Y}_a^b)$ is equivalent to that of $(\hat{X}_a^b, \hat{Y}_a^b)$ under these reliability boundaries. Thus, any bound derived for $\mathbb{E}[G(\hat{X}_a^b| \hat{Y}_a^b)^\omega]$ suffices to bound $\mathbb{E}[G(\widetilde{X}_a^b| \widetilde{Y}_a^b)^\omega]$.
    }
    \begin{lemma} \label{lem::independent}
        Given $|\widetilde{L}_{a-1}| = |\tilde{\ell}_{a-1}|$ and $|\widetilde{L}_{b+1}| = |\tilde{\ell}_{b+1}|$, \black{we have}
        \begin{equation}  \label{equ::Gequiv:independent}
           \mathbb{E}[G(\hat{X}_a^b| \hat{Y}_a^b)^\omega] = \mathbb{E}[G(\widetilde{X}_a^b| \widetilde{Y}_a^b)^\omega],
        \end{equation}
        for a specific guess strategy $G$, where $\widetilde{L}_{a}$ and $\widetilde{L}_{b}$ are the random variable of LLRs of $\widetilde{Y}_a$ and $\widetilde{Y}_b$, respectively.
    \end{lemma}
    \begin{IEEEproof}
        When $|\widetilde{L}_{a-1}| = |\tilde{\ell}_{a-1}|$ and $|\widetilde{L}_{b+1}| = |\tilde{\ell}_{b+1}|$, the pairs $(\widetilde{X}_a^b, \widetilde{Y}_a^b)$ satisfy
        \begin{equation} 
            |\tilde{\ell}_{a-1}| \geq  |\widetilde{L}_a| \geq \ldots |\widetilde{L}_b| \geq |\tilde{\ell}_{b+1}|.
        \end{equation}

        Since $\mathrm{supp}(\hat{Y}) = \mathcal{Y}_{a,b}$\black{,} the LLR, denoted by $\hat{L}$, of $\hat{Y}$ satisfies 
        \begin{equation}
           \tilde{\ell}_{a-1}  \geq  |\hat{L}| \geq \tilde{\ell}_{b+1}.
        \end{equation}
        Thus, for $\hat{Y}_a^b$, their LLRs, denoted by $\hat{L}_a^b$, satisfy 
        \begin{equation}  \label{equ::Lem:independent:proof1}
           |\tilde{\ell}_{a-1}|  \geq  \max\{|\hat{L}_a^b|\} \geq \min\{|\hat{L}_a^b|\} \geq |\tilde{\ell}_{b+1}|.
        \end{equation}

        For a given guesswork $G$, we observe that $G(\hat{X}_a^b| \hat{Y}_a^b)$ is unchanged for an arbitrary permutation $\pi$ that randomly interchanges the indices of a length $b-a+1$ sequence. That is,
        \begin{equation}
            G(\hat{X}_a^b| \hat{Y}_a^b) = G(\pi(\hat{X}_a^b)\mid \pi(\hat{Y}_a^b)).
        \end{equation}
        Based on \eqref{equ::Lem:independent:proof1}, there exist\black{s} one permutation $\pi'$ that can make
        \begin{equation} 
            \tilde{\ell}_{a-1} \geq  \pi'(\hat{L}_a) \geq \ldots \pi'(\hat{L}_b) \geq \tilde{\ell}_{b+1}.
        \end{equation}
        \black{This shows that} $(\pi'(\hat{X}_a^b), \pi'(\hat{Y}_a^b))$ and $(\widetilde{X}_a^b, \widetilde{Y}_a^b)$ are identically distributed. Therefore, 
        \begin{equation}
            G(\hat{X}_a^b| \hat{Y}_a^b) = G(\pi'(\hat{X}_a^b)\mid \pi'(\hat{Y}_a^b)) = G(\widetilde{X}_a^b| \widetilde{Y}_a^b).
        \end{equation}
        This proves \eqref{equ::Gequiv:independent}.
    \end{IEEEproof}

    %We note that Lemma \ref{lem::independent} is based on conditions $\{|\widetilde{L}_{a-1}| = |\tilde{\ell}_{a-1}|\}$ and $\{|\widetilde{L}_{b+1}| = |\tilde{\ell}_{b+1}|\}$. They mean that the LLR levels of the $(a-1)_{\rm{th}}$ and $(b+1)_{\rm{th}}$ outputs of BI-OSC are known. In essence, with these two conditions, pairs of $(\widetilde{X}_a^b, \widetilde{Y}_a^b)$ exhibit a degree of independence, since they result from permuting independent pairs in $(\hat{X}_a^b, \hat{Y}_a^b)$.

    \black{
    We note that Lemma \ref{lem::independent} requires the conditions ${|\widetilde{L}_{a-1}| = |\tilde{\ell}_{a-1}|}$ and ${|\widetilde{L}_{b+1}| = |\tilde{\ell}_{b+1}|}$, which fix the LLR levels at the boundary positions $(a-1)_{\rm{th}}$ and $(b+1)_{\rm{th}}$ of the BI-OSC outputs. Under these boundary conditions, the pairs in $(\widetilde{X}_a^b, \widetilde{Y}_a^b)$ become equivalent to a permutation of independent pairs from $(\hat{X}_a^b, \hat{Y}_a^b)$, thus exhibiting a \textit{conditional independence.}
    }

    \subsection{An upper bound for the guesswork on BI-OSC}

    With Lemma \ref{lem::independent}, guesswork bounds derived for i.i.d. random variable pairs can be readily used for BI-OSC. 
    \begin{corollary}   \label{Cor::Newbound:XnYn:ordered:condition}
        Let ($\widetilde{X}^n, \widetilde{Y}^n)$ be the input and output of \black{a} length-$n$ BI-OSC channel. Given $|\widetilde{L}_{a-1}| = |\tilde{\ell}_{a-1}|$ and $|\widetilde{L}_{b+1}| = |\tilde{\ell}_{b+1}|$, $\mathbb{E}[G^*(\widetilde{X}_a^b \mid \widetilde{Y}_a^b)^\omega]$ is upper bounded by 
        \begin{align} \label{equ::Newbound:XnYn:ordered:condition}
            \mathbb{E}[G^*(\widetilde{X}_a^b | \widetilde{Y}_a^b)^\omega] & \leq  \sum_{j=0}^{b-a+1} (\gamma_j)^{\frac{1}{p}} \binom{b-a+1}{j}^{\frac{1}{q}} \notag \\
            & \cdot \mathbb{E}[\mathbb{P}_{\hat{X}|\hat{Y}}(\bar{x}|Y)^q]^{\frac{n-j}{q}}\cdot\mathbb{E}[\mathbb{P}_{\hat{X}|\hat{Y}}(1-\bar{x}|Y)^q]^{\frac{j}{q}},
        \end{align}
        where 
        \begin{equation} \label{equ::gamma_j2}
            \gamma_j = \frac{(\beta_j)^{\omega p +1} \!-\! (\beta_{j-1})^{\omega p +1}}{\omega p +1},
        \end{equation}
        \begin{equation}
            \beta_j = \sum_{i=0}^{j} \binom{b-a+1}{i},
        \end{equation}
        $\mathbb{P}_{\hat{X}|\hat{Y}}$ is defined as \eqref{equ::auxiliary:iidvariable}, and $\bar{x} := \argmax_{x} \mathbb{P}_{\hat{X}|\hat{Y}}(x|y)$.
        
    \end{corollary}
    \begin{IEEEproof}
        According to Lemma \ref{lem::independent}, $\mathbb{E}[G(\hat{X}_a^b| \hat{Y}_a^b)^\omega] = \mathbb{E}[G(\widetilde{X}_a^b| \widetilde{Y}_a^b)^\omega]$. Then, \eqref{equ::Newbound:XnYn:binary} is \black{directly applied to} $(\hat{X}_a^b, \hat{Y}_a^b)$.
    \end{IEEEproof}

    We note that \eqref{equ::Newbound:XnYn:ordered:condition} still presumes conditions $\{|\widetilde{L}_{a-1}| = |\tilde{\ell}_{a-1}|\}$ and $\{|\widetilde{L}_{b+1}| = |\tilde{\ell}_{b+1}|\}$. \black{Conditions $|\tilde{\ell}_{a-1}|$ and $|\tilde{\ell}_{b+1}|$ are implicitly included in $\mathbb{E}[\mathbb{P}_{\hat{X}|\hat{Y}}(\bar{x}|Y)^q]$ and $\mathbb{E}[\mathbb{P}_{\hat{X}|\hat{Y}}(1-\bar{x}|Y)^q]$ through the definition \eqref{equ::auxiliary:iidvariable}. To derive a general result, we need to remove these conditions} to obtain the unconditional upper bound of $\mathbb{E}[G^*(\widetilde{X}_a^b \mid \widetilde{Y}_a^b)^\omega]$.

    \begin{theorem} \label{The::Newbound:XnYn:ordered}
        Let ($\widetilde{X}^n, \widetilde{Y}^n)$ be the input and output of a length-$n$ BI-OSC channel. Then, $\mathbb{E}[G^*(\widetilde{X}_a^b \mid \widetilde{Y}_a^b)^\omega]$ is upper bounded by 
        %\begin{align}  \label{equ::Newbound:XnYn:ordered}
        %    \mathbb{E}&[G^*(\widetilde{X}_a^b | \widetilde{Y}_a^b)^\omega] \notag \\
        %    &\leq \sum_{j=0}^{b-a+1} (\gamma_j)^{\frac{1}{p}}  \bigg[ 
        %        \iint_{(\mathbb{R}^+)^2} \mathbb{E}\left[\mathbb{P}_{\hat{X}|\hat{Y}}(\bar{x}|Y)^q\right]^{\frac{b-a+1-j}{q}} \notag \\
        %    &\hspace{0.5cm} \cdot  \mathbb{E}\left[\mathbb{P}_{\hat{X}|\hat{Y}}(1-\bar{x}|Y)^q\right]^{\frac{j}{q}} \notag \\
        %    &\hspace{0.5cm} \cdot \mathbb{P}_{|\widetilde{L}_{a\!-\!1,b\!+\!1}|}\left(|\tilde{\ell}_{a-1}|,|\tilde{\ell}_{b+1}|\right) \ d|\tilde{\ell}_{a-1}| \  d|\tilde{\ell}_{b+1}| 
        %    \bigg], 
        %\end{align}

        \black{
        \begin{align}  \label{equ::Newbound:XnYn:ordered}
            \mathbb{E}&[G^*(\widetilde{X}_a^b | \widetilde{Y}_a^b)^\omega] \notag \\
            &\leq \sum_{j=0}^{b-a+1} (\gamma_j)^{\frac{1}{p}} \binom{b-a+1}{j}^{\frac{1}{q}} \\
            &\hspace{0.5cm}\bigg[ 
                \iint_{(\mathbb{R}^+)^2} \mathbb{E}\left[\mathbb{P}_{\hat{X}|\hat{Y}}(\bar{x}|Y)^q\right]^{\frac{b-a+1-j}{q}}  \mathbb{E}\left[\mathbb{P}_{\hat{X}|\hat{Y}}(1-\bar{x}|Y)^q\right]^{\frac{j}{q}} \notag \\
            &\hspace{0.5cm} \cdot \mathbb{P}_{a-1,b+1}\left(|\tilde{\ell}_{a-1}|,|\tilde{\ell}_{b+1}|\right) \ d|\tilde{\ell}_{a-1}|\ d|\tilde{\ell}_{b+1}|
            \bigg], 
        \end{align}
        }
        where $\gamma_j$ is given by \eqref{equ::gamma_j2}, $\bar{x} := \argmax_{x} \mathbb{P}_{\hat{X}|\hat{Y}}(x|y)$, and \black{$ \mathbb{P}_{a-1,b+1}$} is the joint distribution of $|\widetilde{L}_{a-1}|$ and $|\widetilde{L}_{b+1}|$.
    \end{theorem}
    \begin{IEEEproof}
        Theorem \ref{The::Newbound:XnYn:ordered} is obtained by removing conditions $\{|\widetilde{L}_{a-1}| = |\tilde{\ell}_{a-1}|\}$ and $\{|\widetilde{L}_{b+1}| = |\tilde{\ell}_{b+1}|\}$ from Corollary \ref{Cor::Newbound:XnYn:ordered:condition}. This is achieved by integrating $|\widetilde{L}_{a-1}|$ and $|\widetilde{L}_{b+1}|$ alongside their joint distribution.
    \end{IEEEproof}
    %We note that $|\tilde{\ell}_{a-1}|$ and $|\tilde{\ell}_{b-1}|$ are implicitly included in $\mathbb{E}\left[\mathbb{P}_{\hat{X}|\hat{Y}}(\bar{x}|Y)^q\right]$ and $\mathbb{E}\left[\mathbb{P}_{1 - \hat{X}|\hat{Y}}(\bar{x}|Y)^q\right]$ according to the definition \eqref{equ::auxiliary:iidvariable}. Despite the apparent computational complexity of \eqref{equ::Newbound:XnYn:ordered}, it can be significantly simplified, as will be discussed in Section \ref{sec::OSD:bound}.

    \black{While \eqref{equ::Newbound:XnYn:ordered} appears computationally complex, we will show in Section \ref{sec::OSD:bound} that it can be significantly simplified for practical channels. In the next subsection, we demonstrate the application of this bound to a BI-OSC derived from a BI-AWGN channel, and provide explicit expressions of the joint distribution of $|\widetilde{L}_{a-1}|$ and $|\widetilde{L}_{b+1}|$.}
    
    \subsection{\black{An Example with the BI-OSC Derived from BI-AWGN}}

    \black{
    Given the known LLR distribution of the original channel $(X^n,Y^n)$, we can deduce the joint distribution of $|\widetilde{L}_{a-1}|$ and $|\widetilde{L}_{b+1}|$. As an example for the BI-OSC derived from BI-AWGN, we start with BI-AWGN $(X^n,Y^n)$ with noise power $\sigma^2$, where the distribution of $Y_i$ conditioned on $X_i$ is given by
    }
    \begin{equation}
        \mathbb{P}_{Y|X} (y|x) \sim \mathcal{N}(x,\sigma^2)
    \end{equation}
    for $X_i\in\{-1,1\}$. For this scenario, the LLR of each $Y_i$ from $Y^n$ is described by
    \begin{equation}
        L_i = \frac{2 Y_i}{\sigma^2}.
    \end{equation}

    Assume that $X_i$ is equiprobable to be -1 or 1, %we have $L_i\sim 0.5\mathcal{N}(\mu_{\ell},2\mu_{\ell}) + 0.5\mathcal{N}(-\mu_{\ell},2\mu_{\ell})$, 
    \black{the distribution of $L_i$ follows a mixture of Gaussian distributions:
    \begin{equation}
        L_i = \begin{cases}
            \mathcal{N}(\mu_{\ell},2\mu_{\ell}) & \text{w.p. } \frac{1}{2} \\
            \mathcal{N}(-\mu_{\ell},2\mu_{\ell}) & \text{w.p. } \frac{1}{2}
        \end{cases}
    \end{equation}}
    where $\mu_{\ell} = 2/\sigma^2$. 

    \black{
    Given this mixture Gaussian distribution, the probability density function of $|L_i|$ can be derived as
    \begin{equation}
        \mathbb{P}_{|L|}(\ell) = \phi\left(\frac{\ell-\mu_{\ell}}{\sqrt{2\mu_{\ell}}}\right) + \phi\left(\frac{\ell+\mu_{\ell}}{\sqrt{2\mu_{\ell}}}\right),
    \end{equation}
    for $\ell \geq 0$, where $\phi(\cdot)$ denotes the standard normal $\mathrm{pdf}$. The corresponding cumulative distribution function is
    \begin{equation} \label{equ::cdf::L}
        F_{|L|}(\ell) = 1 - Q\left(\frac{\ell-\mu_{\ell}}{\sqrt{2\mu_{\ell}}}\right) - Q\left(\frac{\ell+\mu_{\ell}}{\sqrt{2\mu_{\ell}}}\right),
    \end{equation}
    where $Q(\cdot)$ is the standard Q-function.
    }

    \black{
    Then, following ordered statistics theory \cite{balakrishnan2014order}, the distribution of $|\widetilde{L}_i|$ is given by\footnote{For notation simplicity, we use $\mathbb{P}_{i}(\ell) $ to represent $\mathbb{P}_{|\widetilde{L}_i|}(\ell)$.}
\begin{equation}  \label{equ::ordered:L}
    \mathbb{P}_{i}(\ell) = f_i(\ell)\cdot \mathbb{P}_{|L|}(\ell),
\end{equation}
where $f_i(\ell)$ is derived from the ordering coefficient in \eqref{equ::order:coefficient1}:
\begin{equation}
    f_i(\ell) = \frac{1}{\mathrm{B}(i,n-i+1)} [1 - F_{|L|}(\ell)]^{n-i} F_{|L|}(\ell)^{i}.
\end{equation}
%Here, $F_{|L|}(\ell)$ denotes the cumulative distribution function of $|L_i|$, which can be derived directly from the distribution of $L_i$.
}

   % The joint pdf of $|\widetilde{L}_{i}|$ and $|\widetilde{L}_{j}|$, $1\leq i < j \leq n$, is given by \cite{balakrishnan2014order}
    %\begin{align} \label{equ::joint:ordered:L}
    %    \mathbb{P}_{|\widetilde{L}_{i,j}|}(\ell,\hbar) = f_{i,j}(\ell,\hbar)\cdot \mathbb{P}_{L}(\ell) \cdot  \mathbb{P}_{L}(\hbar),
    %\end{align}

    The joint probability density function of $|\widetilde{L}_{i}|$ and $|\widetilde{L}_{j}|$, \black{for $1\leq i < j \leq n$}, is given by \cite{balakrishnan2014order}
    \begin{align} \label{equ::joint:ordered:L}
        \mathbb{P}_{i,j}(\ell_i,\ell_j) = f_{i,j}(\ell_i,\ell_j)\cdot \mathbb{P}_{L}(\ell_i) \cdot  \mathbb{P}_{L}(\ell_j),
    \end{align}
    where 
    \begin{align}
        f_{i,j}(\ell_i,\ell_j) &=  \frac{n!}{(i\!-\!1)!(j\!-\!i\!-\!1)!(n\!-\!j)!}(1\!-\! F_{|L|}(\ell_i)) ^{i-1} \notag \\
        &\cdot \left(F_{|L|}(\ell_i) \!-\! F_{|L|}(\ell_j)\right)^{j-i-1}  \cdot F_{|L|}(\ell_j)^{n-j} .
    \end{align}
    
    \black{
    Using the relationship $L_i = \frac{2 Y_i}{\sigma^2}$, we can express the sets $\mathcal{Y}_{a,b}^{-}$ and $\mathcal{Y}_{a,b}^{+}$ in terms of received signals:
    }
%Since $L_i = \frac{2 Y_i}{\sigma^2}$, sets $\mathcal{Y}_{a,b}^{-}$ and $\mathcal{Y}_{a,b}^{+}$ are equivalent to
    $$\mathcal{Y}_{a,b}^{(-)} =  \left\{y: -|\tilde{\ell}_{a-1}|\sigma^2/2 \leq y \leq   -|\tilde{\ell}_{b+1}|\sigma^2/2 \right\},$$
    and
    $$\mathcal{Y}_{a,b}^{(+)} =  \left\{y: |\tilde{\ell}_{b+1}|\sigma^2/2 \leq y \leq   |\tilde{\ell}_{a-1}|\sigma^2/2 \right\}.$$
    Thus, 
    \begin{align} \label{equ::setY:AWGN}
        \mathcal{Y}_{a,b} = \frac{\sigma^2}{2} \left( \left[ -|\tilde{\ell}_{a-1}|, -|\tilde{\ell}_{b+1}| \right] \cup \left[ |\tilde{\ell}_{b+1}|, |\tilde{\ell}_{a-1}| \right] \right).
    \end{align}
    Given $\mathcal{Y}_{a,b}$, distributions regarding $(\hat{X},\hat{Y})$, including $\mathbb{P}_{\hat{X},\hat{Y}}$ defined in \eqref{equ::auxiliary:iidvariable}, $\mathbb{P}_{\hat{Y}}$ and $\mathbb{P}_{\hat{X}|\hat{Y}}$, can be obtained based on the distribution of $(X,Y)$ over AWGN. For example, we define
    \begin{align} \label{equ::tau}
        \tau(x) :=& \int_{\mathcal{Y}_{a,b}} \mathbb{P}_{X,Y}(x,y') dy' \notag \\
        =& Q\left(\frac{-|\tilde{\ell}_{a-1}|\sigma}{2}-\frac{x}{\sigma}\right) - Q\left(\frac{-|\tilde{\ell}_{b+1}|\sigma}{2}-\frac{x}{\sigma}\right) \notag\\
        + &  Q\left(\frac{|\tilde{\ell}_{b+1}|\sigma}{2}-\frac{x}{\sigma}\right) - Q\left(\frac{|\tilde{\ell}_{a-1}|\sigma}{2}-\frac{x}{\sigma}\right),
    \end{align}
    \black{where $Q(\cdot)$ denotes the standard Q-function.}
    It can be seen $\tau(x)$ is symmetric, i.e., $\tau(x) = \tau(-x) $. Then, \black{the conditional probability $\mathbb{P}_{\hat{X}|\hat{Y}}$ can be derived as follows }
    \begin{align}
        \mathbb{P}_{\hat{X}|\hat{Y}}(x|y) &= \left(1 + \frac{\tau(x)}{\tau(-x)}\exp\left(-2xy/\sigma^2\right)\right)^{-1} \notag\\
        &= \left(1 + \exp\left(-2xy/\sigma^2\right)\right)^{-1} \notag\\
        & = \mathbb{P}_{X|Y}(x|y)
    \end{align}
    
    \black{We can denote  $\tau(x)$ in  \eqref{equ::tau} simple as $\tau$} because it does not depend on the value of $x\in\{-1,1\}$. Then, the expectation $\mathbb{E}[\mathbb{P}_{\hat{X}|\hat{Y}}(\bar{x}|Y)^q]$ in \eqref{equ::Newbound:XnYn:ordered} is simplified to
    \begin{align} \label{equ::expectation:Phatxy}
        \mathbb{E}\left[\mathbb{P}_{\hat{X}|\hat{Y}}(\bar{x}|Y)^q\right] &=  \int_{-\frac{\sigma^2}{2}|\tilde{\ell}_{a-1}|}^{-\frac{\sigma^2}{2}|\tilde{\ell}_{b+1}|} \mathbb{P}_{X|Y}(-1|y)^q \, \mathbb{P}_{\hat{Y}}(y) \, \mathrm{d}y  \notag \\
        & +  \int_{\frac{\sigma^2}{2}|\tilde{\ell}_{b+1}|}^{\frac{\sigma^2}{2}|\tilde{\ell}_{a-1}|} \mathbb{P}_{X|Y}(1|y)^q \, \mathbb{P}_{\hat{Y}}(y)  \, \mathrm{d}y \notag \\
         &\overset{(a)}{=}  \frac{2}{\tau}\int_{\frac{\sigma^2}{2}|\tilde{\ell}_{b+1}|}^{\frac{\sigma^2}{2}|\tilde{\ell}_{a-1}|} \mathbb{P}_{X|Y}(1|y)^q \, \mathbb{P}_{Y}(y)  \, \mathrm{d}y .   
    \end{align}
    Step (a) comes from the channel symmetry, i.e., $\mathbb{P}_{Y}(y) = \mathbb{P}_{Y}(-y)$ and $\mathbb{P}_{X|Y}(-1|y) = \mathbb{P}_{X|Y}(+1|-y)$.

    In a similar approach, $\mathbb{E}[\mathbb{P}_{\hat{X}|\hat{Y}}(1-\bar{x}|Y)^q]$ can be obtained by changing $\mathbb{P}_{X|Y}(1|y)$ to $\mathbb{P}_{X|Y}(-1|y)$  in \eqref{equ::expectation:Phatxy}, \black{i.e.,
    \begin{equation}  \label{equ::expectation:Phatxy-}
        \mathbb{E}\left[\mathbb{P}_{\hat{X}|\hat{Y}}(1-\bar{x}|Y)^q\right] =   \frac{2}{\tau}\int_{\frac{\sigma^2}{2}|\tilde{\ell}_{b+1}|}^{\frac{\sigma^2}{2}|\tilde{\ell}_{a-1}|} \mathbb{P}_{X|Y}(-1|y)^q \, \mathbb{P}_{Y}(y)  \, \mathrm{d}y .
    \end{equation}}
    %By substituting \eqref{equ::joint:ordered:L} and \eqref{equ::expectation:Phatxy} into \eqref{equ::Newbound:XnYn:ordered}, we can compute the upper bound for $\mathbb{E}[G^*(\widetilde{X}_a^b \mid \widetilde{Y}_a^b)^\omega]$ over a length-$n$ BI-OSC channel originates from the BI-AWGN channel.
    \black{
    By substituting \eqref{equ::joint:ordered:L},  \eqref{equ::expectation:Phatxy} and \eqref{equ::expectation:Phatxy-} into \eqref{equ::Newbound:XnYn:ordered}, we obtain the upper bound for $\mathbb{E}[G^*(\widetilde{X}_a^b \mid \widetilde{Y}_a^b)^\omega]$. This bound is applicable to a length-$n$ BI-OSC channel that originates from the BI-AWGN channel.
    }

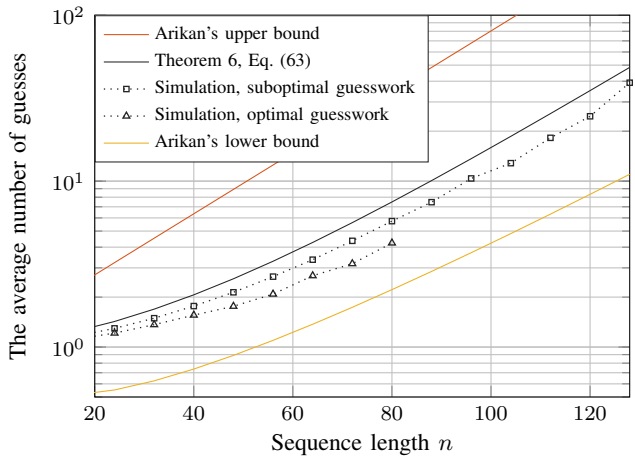
\begin{figure}  [t]
     \centering
\definecolor{mycolor1}{rgb}{0.85000,0.32500,0.09800}%
\definecolor{mycolor2}{rgb}{0.92900,0.69400,0.12500}%
\definecolor{mycolor3}{rgb}{0.00000,0.44706,0.74118}%
\definecolor{mycolor4}{rgb}{0.14902,0.14902,0.14902}%
\begin{tikzpicture}

\begin{axis}[%
width=2.8in,
height=2in,
at={(1.01in,0.685in)},
scale only axis,
xmin=20,
xmax=128,
xlabel style={at={(0.5,1ex)},font=\color{white!15!black}, font=\small},
xlabel={Sequence length $n$},
ymode=log,
ymin=0.5,
ymax=100,
yminorticks=true,
ylabel style={at={(1.5ex,0.5)},font=\color{white!15!black}, font=\small},
ylabel={The average number of guesses},
axis background/.style={fill=white},
tick label style={font=\footnotesize},
xmajorgrids,
ymajorgrids,
yminorgrids,
legend style={at={(0,1)}, anchor=north west , fill opacity=0.5, text opacity=1, font = \scriptsize	 , legend cell align=left, align=left, draw=white!15!black}
]

\addplot [color=mycolor1]
  table[row sep=crcr]{%
16	2.29664516375815\\
24	3.22690395986878\\
32	4.52949422038178\\
40	6.35543620120405\\
48	8.91575989043853\\
56	12.5061816463682\\
64	17.5413190470987\\
72	24.6025817383766\\
80	34.5053035044729\\
88	48.3928750547782\\
96	67.8687474983127\\
104	95.181546383253\\
112	133.484666637801\\
120	187.200316590497\\
128	262.530036731707\\
};
\addlegendentry{Arikan's upper bound}

\addplot [color=mycolor4]
  table[row sep=crcr]{%
16	1.24006533882665\\
24	1.42606198134773\\
32	1.69237176678038\\
40	2.0640266814058\\
48	2.57901711974827\\
56	3.28903780750321\\
64	4.26889185520611\\
72	5.62132700099438\\
80	7.48788605737229\\
88	10.0707330081555\\
96	13.6513231845674\\
104	18.6182985705944\\
112	25.5185790107975\\
120	35.1115842630846\\
128	48.4684409003205\\
};
\addlegendentry{Theorem \ref{The::Newbound:XnYn:ordered}, Eq. \eqref{equ::Newbound:XnYn:ordered}}

\addplot [color= mycolor4, mark=square, mark size = 1pt, dotted, line width=0.5pt, mark options={solid}]
  table[row sep=crcr]{%
16	1.160695\\
24	1.296015\\
32	1.4924325\\
40	1.76589\\
48	2.138045\\
56	2.6570475\\
64	3.3613325\\
72	4.36665\\
80	5.7391825\\
88	7.4675275\\
96	10.377345\\
104	12.830795\\
112	18.235605\\
120	24.588385\\
128	39.1822725\\
};
\addlegendentry{Simulation, suboptimal guesswork}

\addplot [color= mycolor4, mark=triangle, mark size = 1.5pt, dotted, line width=0.5pt, mark options={solid}]
  table[row sep=crcr]{%
16	1.11393333333333\\
24	1.21376666666667\\
32	1.3639\\
40	1.55716666666667\\
48	1.7607\\
56	2.08463333333333\\
64	2.69003333333333\\
72	3.1788\\
80	4.23693333333333\\
};
\addlegendentry{Simulation, optimal guesswork}

\addplot [color=mycolor2]
  table[row sep=crcr]{%
16	0.514281456348169\\
24	0.551416143515881\\
32	0.625765554188177\\
40	0.736894720961369\\
48	0.890604041924068\\
56	1.09730224861584\\
64	1.37218415984178\\
72	1.73626924414942\\
80	2.21812176120448\\
88	2.85632008823478\\
96	3.70287163603235\\
104	4.82788092878874\\
112	6.3259072935038\\
120	8.32461569723381\\
128	10.9965489526181\\
};
\addlegendentry{Arikan's lower bound}

\end{axis}
\end{tikzpicture}%
	\vspace{-0.5em}
    \caption{The average number of guesses at different blocklength $n$ in a BI-OSC derived from an AWGN channel at \black{SNR = 3 dB}. The guesswork focuses on the $[n/4, n/2]$ ordered channel outputs range. The suboptimal guesswork is performed in ascending order of the Hamming distance to the most likely sequence $\widetilde{X}_a^b$ when $\widetilde{Y}_a^b$ is given.}
    \vspace{-0.25em}
	\label{Fig::quater_and_half}
    \end{figure}

    We validate Theorem \ref{The::Newbound:XnYn:ordered} with the BI-OSC originating from the BI-AWGN channel. \black{For comparison,} the Arikan's lower and upper bounds introduced in Theorem \ref{The::Arikan:lower} are included. The Arikan's bounds are also extended for $(\widetilde{X}_a^b, \widetilde{Y}_a^b)$ leveraging Lemma \ref{lem::independent}. Specifically, they are first applied to the pair $(\hat{X}_a^b, \hat{Y}_a^b)$, and then the conditions $\{|\widetilde{L}_{a-1}| = |\tilde{\ell}_{a-1}|\}$ and $\{|\widetilde{L}_{b+1}| = |\tilde{\ell}_{b+1}|\}$ are relaxed. %We evaluate the average number of guesses over sequence lengths $n$ ranging from 8 to 150, with $a = n/4$ and $b = n/2$, i.e., the guesswork focuses on the $(n/4)_{\text{th}}$ to $(n/2)_{\text{th}}$ ordered channel outputs.
    \black{To demonstrate the effectiveness and generality of the proposed bound, we analyze sequences of length $n$ from 8 to 150, focusing on the middle section of ordered outputs with $a = n/4$ and $b = n/2$.}
    The results are illustrated in Fig. \ref{Fig::quater_and_half}. \black{Since the bound in Theorem \ref{The::Newbound:XnYn:ordered} is obtained using Hölder's inequality, we present the bound using the tightest results obtained across $q$ values in the range $(1,3]$, with $p$ determined correspondingly by $1/p + 1/q = 1$.} As shown in the figure, Theorem \ref{The::Newbound:XnYn:ordered} provides a much tighter evaluation than Arikan's bound for both optimal and suboptimal guesswork. Simulation results for optimal guesswork are provided only for $n \leq 80$, due to the prohibitive computation cost of $2^n$ posterior probabilities for large $n$.

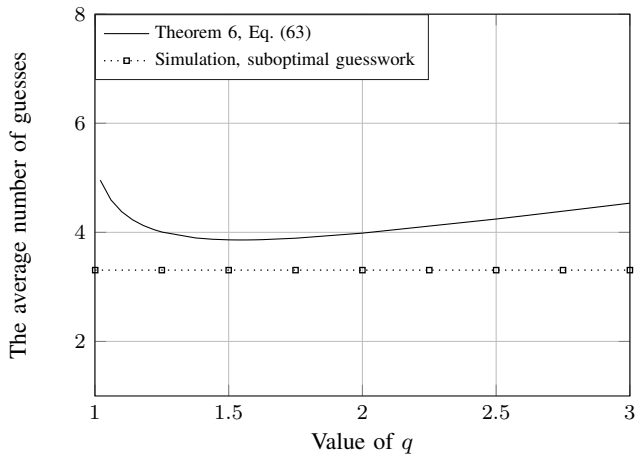
\begin{figure}  [t]
     \centering
\definecolor{mycolor1}{rgb}{0.85000,0.32500,0.09800}%
\definecolor{mycolor2}{rgb}{0.92900,0.69400,0.12500}%
\definecolor{mycolor3}{rgb}{0.00000,0.44706,0.74118}%
\definecolor{mycolor4}{rgb}{0.14902,0.14902,0.14902}%
\begin{tikzpicture}

\begin{axis}[%
width=2.8in,
height=2in,
at={(1.01in,0.685in)},
scale only axis,
xmin=1,
xmax=3,
xlabel style={at={(0.5,1ex)},font=\color{white!15!black}, font=\small},
xlabel={Value of $q$},
%ymode=log,
ymin=1,
ymax=8,
yminorticks=true,
ylabel style={at={(1.5ex,0.5)},font=\color{white!15!black}, font=\small},
ylabel={The average number of guesses},
axis background/.style={fill=white},
tick label style={font=\footnotesize},
xmajorgrids,
ymajorgrids,
yminorgrids,
legend style={at={(0,1)}, anchor=north west , fill opacity=0.5, text opacity=1, font = \scriptsize	 , legend cell align=left, align=left, draw=white!15!black}
]

\addplot [color=black]
  table[row sep=newline]{%
    1.0200    4.9562
    1.0600    4.5943
    1.1000    4.3785
    1.1400    4.2322
    1.1800    4.1276
    1.2200    4.0507
1.25100000000000	4.00483391722062
1.37600000000000	3.89737872219526
1.43850000000000	3.87354951323366
1.50100000000000	3.86231974150498
1.50100000000000	3.86231974150498
1.53225000000000	3.86039703463933
1.56350000000000	3.86052369916856
1.56350000000000	3.86052369916856
1.59475000000000	3.86245251552657
1.62600000000000	3.86597413687118
1.75100000000000	3.89277049203240
2.00100000000000	3.98607149277046
2.50100000000000	4.24478553584401
    2.6000    4.3012
    2.8000    4.4173
    3.0000    4.5351
};
\addlegendentry{Theorem \ref{The::Newbound:XnYn:ordered}, Eq. \eqref{equ::Newbound:XnYn:ordered}}

\addplot [color= black, mark=square, mark size = 1pt, dotted, line width=0.5pt, mark options={solid}]
 table[row sep=crcr]{%
1.001	 3.3072\\
1.25	 3.3072\\
1.5	 3.3072\\
1.75	 3.3072\\
2	 3.3072\\
2.25	 3.3072\\
2.5    3.3072\\
2.75    3.3072\\
3    3.3072\\
};
\addlegendentry{Simulation, suboptimal guesswork}

\end{axis}
\end{tikzpicture}%
	\vspace{-0.5em}
        \caption{\black{The performance of Theorem \ref{The::Newbound:XnYn:ordered} for the average number of guesses with various values of parameter $q$ for $n = 64$ in a BI-OSC derived from an AWGN channel at SNR = 3 dB.}}
    \vspace{-0.25em}
	\label{Fig::order:qValues}      
    \end{figure}

   \black{In Fig. \ref{Fig::order:qValues}, we further investigate the performance of the bound \eqref{The::Newbound:XnYn:ordered} with varying values of $q$. As shown in the figure, at SNR = 3 dB and $n = 64$, the bound achieves its tightest value of $ \mathbb{E}[G^*(\widetilde{X}_a^b | \widetilde{Y}_a^b)] \leq 3.862$ when $q = 1.5$. For comparison, our simulations result in an average number of guesses of 1.372 and 3.307 for the optimal and suboptimal guessing strategies, respectively.
   }

    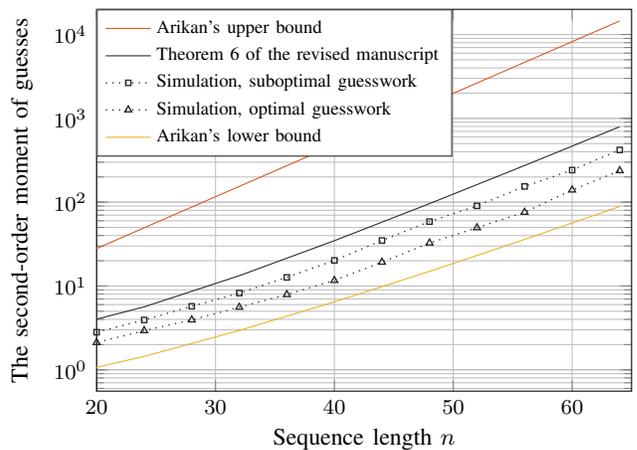
\begin{figure}  [t]
     \centering
\definecolor{mycolor1}{rgb}{0.85000,0.32500,0.09800}%
\definecolor{mycolor2}{rgb}{0.92900,0.69400,0.12500}%
\definecolor{mycolor3}{rgb}{0.00000,0.44706,0.74118}%
\definecolor{mycolor4}{rgb}{0.14902,0.14902,0.14902}%
\begin{tikzpicture}

\begin{axis}[%
width=2.8in,
height=2in,
at={(1.01in,0.685in)},
scale only axis,
xmin=20,
xmax=65,
xlabel style={at={(0.5,1ex)},font=\color{white!15!black}, font=\small},
xlabel={Sequence length $n$},
ymode=log,
ymin=0,
ymax=2e4,
yminorticks=true,
ylabel style={at={(1.5ex,0.5)},font=\color{white!15!black}, font=\small},
ylabel={The second-order moment of guesses},
axis background/.style={fill=white},
tick label style={font=\footnotesize},
xmajorgrids,
ymajorgrids,
yminorgrids,
legend style={at={(0,1)}, anchor=north west , fill opacity=0.5, text opacity=1, font = \scriptsize	 , legend cell align=left, align=left, draw=white!15!black}
]

\addplot [color=mycolor1]
  table[row sep=crcr]{%
8	4.99266424385759\\
16	15.8109655900111\\
24	49.4797197731517\\
32	154.372333162345\\
40	481.036282425077\\
48	1498.02436465957\\
56	4663.44813365984\\
64	14514.4298701004\\
};
\addlegendentry{Arikan's upper bound}

\addplot [color=mycolor4]
  table[row sep=crcr]{%
8	1.87256510244083\\
16	2.87139232676503\\
24	5.63700112520781\\
32	13.2722119421299\\
40	34.729314389795\\
48	96.0709597279241\\
56	273.821988824704\\
64	794.515254582067\\
};
\addlegendentry{Theorem 6 of the revised manuscript}

\addplot [color= mycolor4, mark=square, mark size = 1pt, dotted, line width=0.5pt, mark options={solid}]
  table[row sep=crcr]{%
8	1.32182724376388\\
12	1.59576120549195\\
16	2.03166563394782\\
20	2.81000437781918\\
24	3.93573085668221\\
28	5.7243960400376\\
32	8.24114678350165\\
36	12.6712502964685\\
40	20.2022636972256\\
44	34.9021039240132\\
48	58.6581367566323\\
52	90.4630796976018\\
56	154.142539810462\\
60	241.555533271831\\
64	420.153929204342\\
};
\addlegendentry{Simulation, suboptimal guesswork}

\addplot [color= mycolor4, mark=triangle, mark size = 1.5pt, dotted, line width=0.5pt, mark options={solid}]
  table[row sep=crcr]{%
8	1.21038203735713\\
12	1.379045677894\\
16	1.64068795817841\\
20	2.115849472774\\
24	2.93084885776075\\
28	3.943986641890\\
32	5.58361470966405\\
36	7.901763395675\\
40	11.6578057475194\\
44	19.244831433561\\
48	32.5880655366003\\
52	49.558914607427\\
56	75.8367652511463\\
60	139.763847147490\\
64	237.48850800057\\
};
\addlegendentry{Simulation, optimal guesswork}

\addplot [color=mycolor2]
  table[row sep=crcr]{%
8	0.526487946170799\\
16	0.792815845654481\\
24	1.44482119681465\\
32	2.94641118911257\\
40	6.46692759060838\\
48	14.9476001734951\\
56	35.901304118868\\
64	88.8178567667248\\
};
\addlegendentry{Arikan's lower bound}

\end{axis}
\end{tikzpicture}%
	\vspace{-0.5em}
    \caption{\black{The second-order moment of the number of guesses in a BI-OSC derived from an AWGN channel at SNR = 3 dB. The guesswork focuses on the $[n/4, n/2]$ ordered channel outputs range.}}
    \vspace{-0.25em}
	\label{Fig::secondOrder}
    \end{figure}

\begin{figure}  [t]
     \centering
\definecolor{mycolor1}{rgb}{0.85000,0.32500,0.09800}%
\definecolor{mycolor2}{rgb}{0.92900,0.69400,0.12500}%
\definecolor{mycolor3}{rgb}{0.00000,0.44706,0.74118}%
\definecolor{mycolor4}{rgb}{0.14902,0.14902,0.14902}%
\begin{tikzpicture}

\begin{axis}[%
width=2.4in,
height=2in,
at={(1.01in,0.685in)},
scale only axis,
xmin=1,
xmax=2.5,
xlabel style={at={(0.5,1ex)},font=\color{white!15!black}, font=\small},
xlabel={Value of $q$},
%ymode=log,
ymin=400,
ymax=1100,
yminorticks=true,
ylabel style={at={(1.5ex,0.5)},font=\color{white!15!black}, font=\small},
ylabel={The second-order moment of guesses},
axis background/.style={fill=white},
tick label style={font=\footnotesize},
xmajorgrids,
ymajorgrids,
yminorgrids,
legend style={at={(0,1)}, anchor=north west , fill opacity=0.5, text opacity=1, font = \scriptsize	 , legend cell align=left, align=left, draw=white!15!black}
]

\addplot [color=black]
  table[row sep=newline]{%
1.01	nan
1.05	945.68431618554
1.1	882.481098893695
1.15	844.933722310603
1.251	806.569176178955
1.32	796.216445789375
1.376	793.551106766837
1.501	799.825230579944
1.501	799.825230579944
1.501	799.825230579944
1.626	817.22275527141
1.751	841.865962604631
2.001	905.250139497245
2.501	1064.76866900211
};
\addlegendentry{Hamming subset bound, Eq. \eqref{equ::Newbound:XnYn:binary}}

\addplot [color= black, mark=square, mark size = 1pt, dotted, line width=0.5pt, mark options={solid}]
 table[row sep=crcr]{%
1.001	 420\\
2.5    420\\
};
\addlegendentry{Simulation, sub-optimal guesswork}

\end{axis}
\end{tikzpicture}%
	\vspace{-0.5em}
    \caption{\black{The performance of Theorem \ref{The::Newbound:XnYn:ordered} for the second-order moment with various values of parameter $q$ for $n = 64$ in a BI-OSC derived from an AWGN channel at SNR = 3 dB.}}
    \vspace{-0.25em}
	\label{Fig::secondOrder:q}
    \end{figure}
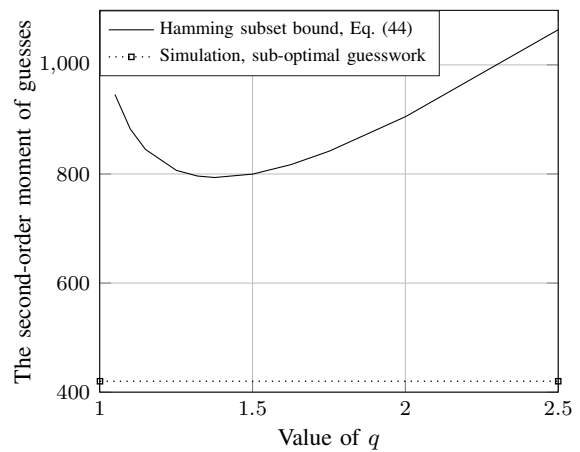

    \black{Furthermore, Fig. \ref{Fig::secondOrder} illustrates the performance of the proposed bound from Theorem \ref{The::Newbound:XnYn:ordered} for the second-order moment ($\omega = 2$) of guesswork within the $[n/4, n/2]$ range of ordered channel outputs. The results show that our bound provides tight estimates for the second-order moments, particularly for suboptimal guesswork based on Hamming distance ordering. Figure \ref{Fig::secondOrder:q} demonstrates how the accuracy of bound varies with different values of $q$ when evaluating second-order moments. As shown, for second-order moments, the bound achieves its best performance at $q \approx 1.4$.}

\section{\black{The Achievable Guesswork Complexity of Ordered Statistics Decoding}} \label{sec::OSD:bound}

\subsection{\black{The OSD Algorithm}}
\black{
We examine the OSD algorithm for a binary linear codebook $\mathcal{C}(n,k)$ with blocklength $n$ and information length $k$. Consider a codeword $\mathbf{c}$ of $\mathcal{C}$ transmitted over a BI-AWGN channel with BPSK modulation, resulting in received signal $\mathbf{y} = \mathbf{x} + \mathbf{w}$, where $\mathbf{x}$ is the BPSK symbol vector and $\mathbf{w}$ is Gaussian noise with variance $\sigma^2$.
}

\black{
OSD first orders the received symbols based on their reliability measured by LLR. For each received symbol $y_i$, its LLR magnitude $|\ell_i|$ indicates how reliable the bit decision of $y_i$ is, with larger magnitudes suggesting higher reliability. After ordering these reliability values in descending order, we obtain the ordered sequence $\widetilde{\mathbf{y}} = [\widetilde{y}_1, \widetilde{y}_2, ..., \widetilde{y}_n]$ with corresponding reliabilities $|\tilde{\ell}_i|$ satisfying \eqref{equ::ordering::BI-continuous}. The first $k$ ordered symbols $\tilde{y}_1^k = [\widetilde{y}_1, ..., \widetilde{y}_k]$ represent the most reliable bits (MRB) in the received sequence.
}

\black{
This reliability-based ordering defines a permutation $\pi$, which will be also applied to the generator matrix $\mathbf{G}$ to produce $\widetilde{\mathbf{G}} = \pi(\mathbf{G})$. Then, $\widetilde{\mathbf{G}}$ is transformed into systematic form $[\mathbf{I} \ \widetilde{\mathbf{P}}]$ through Gaussian elimination (GE)\footnote{\black{During GE,} an additional column permutations \black{beyond $\pi$} might be required to ensure that the first $k$ columns of $\widetilde{\mathbf{G}}$ are linearly independent, these permutations are typically minor and can usually be omitted \cite{Fossorier1995OSD}.}. Since $\tilde{\mathbf{c}} = \pi(\mathbf{c})$ is a codeword of the codebook $\widetilde{\mathcal{C}}$ defined by $\widetilde{\mathbf{G}}$, finding $\tilde{\mathbf{c}}$ suffices to recover the transmitted codeword $\mathbf{c}$ via the inverse permutation $\pi^{-1}$. We note that even if $\mathbf{G}$ is originally systematic, after permutation $\widetilde{\mathbf{G}}$ loses its systematic structure. Therefore, GE is always required in OSD.
}

\black{
OSD generates codeword candidates through re-encoding of test error patterns (TEPs). We assume that the hard decision of $[\widetilde{y}_1, ..., \widetilde{y}_k]$ is obtained as $\boldsymbol{b}_0$. We denote a TEP by  $\mathbf{e}_{\xi}$, where $\xi = 1,2,...,\xi_{\max}$, and $\xi_{\max}$ is the maximum allowed number of TEPs. Each TEP $\mathbf{e}_{\xi}$ has the length $k$ and represents a potential transmission error pattern in MRB. By applying TEPs to $\boldsymbol{b}_0$, we obtain different guesses $\boldsymbol{b}_{\xi} = \boldsymbol{b}_0 \oplus \mathbf{e}_{\xi}$ of MRB, which are then re-encoded to produce a codeword estimate $\boldsymbol{v}_{\xi} = \boldsymbol{b}_{\xi}\widetilde{\mathbf{G}}$. Note that MRB is also the set of information bits of the code $\widetilde{\mathcal{C}}$ defined by $\widetilde{\mathbf{G}}$. }

\black{
We summarize the OSD process in Algorithm \ref{Algo::OSD}.}

\begin{algorithm} 
   \caption{\black{Basic Operation of OSD (with Genie)}}
   \label{Algo::OSD}
   \begin{algorithmic}[1]
       \REQUIRE \black{Received signal $\mathbf{y} = \mathbf{x} + \mathbf{w}$, generator matrix $\mathbf{G}$, maximum guesses $\xi_{\max}$}
       \ENSURE \black{Decoded codeword $\mathbf{c}$}
           
       \STATE \black{Calculate LLR magnitude $|\ell_i|$ of each received bit $y_i$}
       \STATE \black{Order bits by reliability to obtain permutation $\pi$ and ordered sequence with $|\tilde{\ell}_i|$}
       \STATE \black{Apply permutation $\pi$ to $\mathbf{G}$ to obtain $\widetilde{\mathbf{G}}$. Transform $\widetilde{\mathbf{G}}$ to systematic form $[\mathbf{I} \ \widetilde{\mathbf{P}}]$ using GE}
       \STATE \black{Get hard decision $\mathbf{b}_0$ from first $k$ ordered bits}
       \FOR{\black{$\xi = 1$ to $\xi_{\max}$}}
           \STATE \black{Generate a TEP $\mathbf{e}_{\xi}$ of length $k$}
           \STATE \black{Obtain a guess of MRB $\boldsymbol{b}_{\xi} = \boldsymbol{b}_0 \oplus \mathbf{e}_{\xi}$}
           \STATE \black{Re-encoding to get a codeword estimate $\boldsymbol{v}_{\xi} = \boldsymbol{b}_{\xi}\widetilde{\mathbf{G}}$}
           \IF{\black{Genie identifies $\boldsymbol{v}_{\xi}$ as correct}}
               \RETURN \black{$\mathbf{c} = \pi^{-1}(\boldsymbol{v}_{\xi})$}
           \ENDIF
       \ENDFOR
       \RETURN \black{The codeword estimate with minimum Euclidean distance to $\mathbf{y}$}
   \end{algorithmic}
\end{algorithm}

\black{
For analysis purposes, Algorithm \ref{Algo::OSD} assumes an infallible genie that can identify the correct codeword. In practice, OSD must use error detection mechanisms like CRC or other verification methods to determine decoding success (e.g., \cite{yue2021probability, wu2007preprocessing_and_diversification, jin2006probabilisticConditions, LCOSD2022}). However, this idealized model allows us to focus on the \textit{achievable guesswork complexity}; that is, the minimum number of TEPs needed to find the correct error pattern in the MRB, which ensures optimal error performance.
}

%From a guesswork perspective, OSD essentially generates a sequence of codeword guesses $[\boldsymbol{v}1, \boldsymbol{v}2, \ldots, \boldsymbol{v}{\xi{\max}}]$ until either reaching the maximum allowed guesses $\xi_{\max}$ or finding the correct codeword $\mathbf{c}$ at the $\xi_{\rm{th}}$ guess. This number of guesses $\xi$ represents the achievable guesswork complexity - the minimum number of TEPs needed to ensure optimal error performance.
%Let $X_{G}$ denote the random variable representing the number of guesses until finding the correct codeword, with $\mathbb{P}{X{\xi}}(\xi)$ denoting the probability of success at the $\xi_{\rm{th}}$ guess. The $\omega_{\rm{th}}$ moment of $X_{G}$ is given by
%\begin{equation}
%\mathbb{E}[(X_{G})^\omega] = \sum_{\xi=1}^{\xi_{\max}} \xi^{\omega} \mathbb{P}{X{\xi}}(\xi),
%\end{equation}
%which follows the same form as the general guesswork expression in \eqref{equ::mydefine::guesswork}. Note that $\mathbb{P}{X{\xi}}(\xi) = \mathbb{P}(\boldsymbol{v}_{\xi} = \mathbf{c})$, as these events are equivalent.
%}

\black{
From a guesswork perspective, OSD essentially generates a sequence of TEPs $[\mathbf{e}_1, \ldots, \mathbf{e}_\xi, \ldots, \mathbf{e}_{\xi_{\max}}]$ to guess errors on MRB, until either reaching the maximum allowed guesses $\xi_{\max}$ or finding the correct error pattern. Equivalently, OSD generates $[\boldsymbol{b}_1, \ldots, \boldsymbol{b}_\xi, \ldots, \boldsymbol{b}_{\xi_{\max}}]$ as the guesses of the correct MRB.
}

\black{
Let $X_{G}$ denote the random variable representing the number of TEPs (guesses) until finding the correct error pattern,  This number of guesses $X_{G}$ represents the \textit{achievable guesswork complexity} of OSD. We denote $\mathbb{P}_{X_{G}}(\xi)$ as the probability of $\{X_{G} = \xi\}$, i.e., the decoding successes at the $\xi_{\rm{th}}$ guess. The $\omega_{\rm{th}}$ moment of $X_{G}$ is given by
\begin{equation} \label{equ::Xxi}
    \mathbb{E}[(X_{G})^\omega] = \sum_{\xi=1}^{\xi_{\max}} \xi^{\omega} \mathbb{P}_{X_{G}}(\xi),
\end{equation}
which follows the same form as the general guesswork expression in \eqref{equ::mydefine::guesswork}. We note the following equivalent probabilities, $$\mathbb{P}_{X_{G}}(\xi) = \mathbb{P}(\mathbf{e}_{\xi} = \mathbf{e}) =  \mathbb{P}(\boldsymbol{b}_{\xi} = \boldsymbol{b}),$$ where $\mathbf{e}$ is the actual error pattern in the MRB, and $\boldsymbol{b}$ is the correct transmitted MRB. Correctly guessing $\boldsymbol{b}$ will result in the correct codeword estimate $\mathbf{c} = \pi^{-1}(\boldsymbol{v}_{\xi})$, because $\boldsymbol{v}_{\xi} = \boldsymbol{b}_{\xi}\widetilde{\mathbf{G}}$.
}

\black{
The overall computational complexity of OSD consists of several components. Given a received sequence $\mathbf{y}$, OSD first performs reliability ordering requiring $C_{\rm{sort}} = O(n\log n)$ operations \cite{Fossorier1995OSD}. Then, GE to transform $\widetilde{\mathbf{G}}$ into systematic form requires $ C_{\rm{GE}} = O(n\cdot\min(k,n-k)^2)$ operations \cite{Fossorier1995OSD}. Finally, the decoder conducts guessing through re-encoding TEPs, where each re-encoding requires $C_{\rm{re-encoding}} = O(k(n-k))$ operations. Therefore, the total computational complexity can be expressed as
\begin{align} 
C_{\rm{OSD}} &= C_{\rm{sort}} + C_{\rm{GE}} + X_{G}  \cdot C_{\rm{re-encoding}} \notag \\
 &=  O(n\log n) + O(n\cdot\min(k,n-k)^2) \label{equ::OSD_complexity} \\
 &\hspace{0.2cm}+ X_{G}  \cdot O(k(n-k)) \notag
\end{align}
where $ X_{G}$ is the number of TEPs processed before finding the correct codeword, which is the \textit{achievable guesswork complexity} of OSD we defined earlier in Section I.
Among these components, $C_{\rm{sort}}$, $C_{\rm{GE}}$, and $C_{\rm{re-encoding}}$ are fixed for given code parameters ($n$ and $k$). However, $\mathbb{E}[X_{G}]$ varies with channel conditions and significantly impacts the overall complexity. Therefore, we focus on analyzing the the moments of guesswork complexity of OSD, represented by the random variable $X_{G}$.
}

\black{
\begin{remark} 
    As reported in \cite{yue2023efficient}, although the sorting and GE stages of OSD introduce fixed computational overhead, it can become significant at very high SNRs. This issue was discussed and partially addressed in our recent work \cite{yue2022ordered}. Moreover, from an implementation perspective, sorting and GE operations are not circuit-friendly and will incur substantial overhead in hardware implementations \cite{kim2021fpga}, which represents a significant open problem in practical OSD application. Nevertheless, this paper focuses specifically on the guesswork complexity of OSD, i.e., the iterative process of re-encoding $X_{G}$ TEPs to identify the correct codeword estimate.
\end{remark}
}

\subsection{Bounds for the Guesswork Complexity of OSD}

 \black{
 In analyzing the guesswork complexity of OSD, we first introduce the optimal guessing strategy for transmission errors in the most reliable bits (MRB). Given the ordered received sequence $[\widetilde{y}_1,\ldots,\widetilde{y}_k]$, the optimal strategy processes TEPs in descending order of their posterior probabilities $\mathbb{P}(\mathbf{e} = \mathbf{e}_{\xi} \mid \tilde{y}_1^k)$, where $\mathbf{e}$ represents the actual error pattern in the MRB. This ensures that the most likely error patterns are examined first, minimizing the expected number of guesses needed to identify the correct pattern. We refer to this strategy as ``optimal processing''.
  }

  \black{
However, computing and sorting these posterior probabilities introduces significant computational overhead in the real-time decoding. Therefore, OSD implementations typically adopt a suboptimal strategy based on Hamming weight ordering of TEPs. Starting from the hard decision $\boldsymbol{b}_0$ of the MRB $[\widetilde{y}_1,\ldots,\widetilde{y}_k]$, this approach processes TEPs $\mathbf{e}_{\xi}$ in ascending order of their Hamming weights. For instance, it first tests the all-zero error pattern ($\mathbf{e}_1 = \mathbf{0}$), then patterns with single errors, double errors, and so on, up to $m$ (the decoding order of OSD). This strategy is motivated by the observation that error patterns with lower Hamming weights are more likely to occur in the most reliable positions. We refer to this strategy as ``Hamming processing''.
}

\black{
The relationship between these strategies can be understood through the channel reliability ordering. Since the MRB positions have the highest reliability metrics, the probability of multiple simultaneous errors in these positions decreases rapidly with the Hamming weight of the error pattern. Consequently, the Hamming weight-based ordering of TEPs results in similar decoding performance (including both BLER and required number of TEPs) to the optimal probability-based ordering, particularly at moderate to high SNRs, as reported by \cite{yue2021probability}.
}

\subsubsection{Order-$k$ OSD}
\black{
We now analyze the guesswork complexity of OSD theoretically. Recall that $X_{G}$ denotes the random variable representing the number of guesses until finding the correct error pattern in OSD, with $\mathbb{E}[(X_{G})^\omega]$ being its $\omega_{\mathrm{th}}$ moment. We first consider the case where all possible error patterns in the MRB might be examined, corresponding to $\xi_{\max} = 2^k$. In this case, OSD potentially examines error patterns up to Hamming weight $k$ in the MRB; hence, we refer to it as order-$k$ OSD. 
}

\black{
For order-$k$ OSD, we can characterize $X_{G}$ through the guesswork over ordered statistics. Specifically, let $G^*(\widetilde{X}_1^k | \widetilde{Y}_1^k)$ represent the optimal guesswork for identifying the first $k$ positions in a BI-OSC, where $\widetilde{X}_1^k$ and $\widetilde{Y}_1^k$ denote the transmitted and received sequences in these positions, respectively. Then, $X_{G}$ is equivalent to $G^*(\widetilde{X}_1^k | \widetilde{Y}_1^k)$ since identifying the correct MRB $\boldsymbol{b} = \widetilde{X}_1^k$ is equivalent to finding the correct error pattern in OSD, as discussed right below \eqref{equ::Xxi}.
}

\black{
The moments of guesswork $G^*(\widetilde{X}_1^k | \widetilde{Y}_1^k)$ are bounded by Theorem \ref{The::Newbound:XnYn:ordered} with $a = 1$ and $b = k$. Setting $a = 1$ and $b = k$ indicates that we focus on guessing errors in the first $k$ most reliable positions within the $n$ received symbols. Utilizing Theorem \ref{The::Newbound:XnYn:ordered}, the $\omega_{\mathrm{th}}$ moment of $G^*(\widetilde{X}_1^k | \widetilde{Y}_1^k)$ is upper bounded by the following Corollary.
}

    \begin{corollary}[Guesswork of order-$k$ OSD] \label{Cor::MRB}
    \black{For optimal processing of TEPs in OSD with $\xi_{\max} = 2^k$, the $\omega_{\rm{th}}$ moment of guesswork is upper bounded by}
    \begin{align}   \label{equ::Newbound:XnYn:ordered:1k}
        \mathbb{E}[G^*(\widetilde{X}_1^k | \widetilde{Y}_1^k)^\omega]  &\leq \sum_{j=0}^{k} (\gamma_j)^{\frac{1}{p}} \binom{k}{j}^{1/q} \int_{\mathbb{R}^+} 
        \left[ \mathbb{E}\left(\mathbb{P}_{\hat{X}|\hat{Y}}(\bar{x}|Y)^q\right)^{\frac{k-j}{q}} \right. \notag \\
        & \cdot \left. \mathbb{E}\left(\mathbb{P}_{\hat{X}|\hat{Y}}(1-\bar{x}|Y)^q\right)^{\frac{j}{q}} \right] \cdot \mathbb{P}_{k+1}(\tilde{\ell}_{k+1}) \, d\tilde{\ell}_{k+1},
    \end{align}
    where 
    \begin{equation} \label{equ::gamma3}
    \gamma_j = \frac{(\beta_j)^{\omega p + 1} - (\beta_{j-1})^{\omega p +1}}{\omega p +1} ,
    \end{equation} 
    \begin{equation}
        \beta_j = \sum_{i=0}^{j}\binom{k}{i},
    \end{equation}
    \black{and $\mathbb{P}_{k+1}(\cdot)$ denotes the distribution of the $(k+1)$-th ordered LLR magnitude given by \eqref{equ::ordered:L}.}
    \end{corollary}

    \begin{IEEEproof}
    \black{
    This result follows directly from Theorem \ref{The::Newbound:XnYn:ordered} by setting $a = 1$ and $b = k$ to by focus on the first $k$ ordered positions, i.e., MRB. With these parameters, the set $\mathcal{Y}_{a,b}$ in \eqref{equ::setY:AWGN} simplifies to
    \begin{align} 
        \mathcal{Y}_{1,k} &= \frac{\sigma^2}{2} \left( \left[ -|\tilde{\ell}_{0}|, -|\tilde{\ell}_{k+1}| \right] \cup \left[ |\tilde{\ell}_{k+1}|, |\tilde{\ell}_{0}| \right] \right) \notag\\
        & = \frac{\sigma^2}{2} \left( \left[ -\infty, -|\tilde{\ell}_{k+1}| \right] \cup \left[ |\tilde{\ell}_{k+1}|, +\infty \right] \right).
    \end{align}
    }

    \black{
    As $\mathcal{Y}_{1,k}$ defines the support of $\hat{Y}$, the expectation terms in \eqref{equ::Newbound:XnYn:ordered:1k} can be evaluated as
    \begin{align} 
        \mathbb{E}\left[\mathbb{P}_{\hat{X}|\hat{Y}}(\bar{x}|Y)^q\right] = \frac{2}{\tau}\int_{\frac{\sigma^2}{2}|\tilde{\ell}_{k+1}|}^{\infty} \mathbb{P}_{X|Y}(1|y)^q \, \mathbb{P}_{Y}(y)  \, \mathrm{d}y,
    \end{align}
    where $\tau$ is given in $\eqref{equ::tau}$. By taking $|\tilde{\ell}_{0}| \to \infty$, $\tau\black{(x)}$ in \eqref{equ::tau} is also simplified to
    \begin{align} \label{equ::tau:1k}
        \tau(x) = Q\left(\frac{|\tilde{\ell}_{k+1}|\sigma}{2}+\frac{\black{x}}{\sigma}\right) 
        +  Q\left(\frac{|\tilde{\ell}_{k+1}|\sigma}{2}-\frac{\black{x}}{\sigma}\right).
    \end{align}
    Due to the symmetry of $\tau(x)$ with respect to $x \in \{-1,+1\}$, we denote it simply as $\tau$. The expectation $\mathbb{E}[\mathbb{P}_{\hat{X}|\hat{Y}}(1-\bar{x}|Y)^q]$ is obtained similarly with $\mathbb{P}_{X|Y}(-1|y)$ replacing $\mathbb{P}_{X|Y}(1|y)$.
    }
    \black{Thus, compared to Theorem \ref{The::Newbound:XnYn:ordered}, we only need to remove the condition $\{|\widetilde{L}_{k+1}| = \tilde{\ell}_{k+1}\}$, since $|\widetilde{L}_{0}| = \infty$ is deterministic. This is why \eqref{equ::Newbound:XnYn:ordered:1k} only requires integration over $|\widetilde{L}_{k+1}|$ with its distribution $\mathbb{P}_{k+1}$. The condition $|\widetilde{L}_{0}| = \infty$ reflects that there is no upper limit on the reliability of the most reliable bit (i.e., $|\widetilde{L}_{1}|$ ) in the ordered sequence \footnote{\black{Note that although $\mathbb{P}_{i,j}(\ell_i,\ell_j)$ in \eqref{equ::joint:ordered:L} is only defined for $1\leq i \leq j \leq n$, when $i = 0$, the joint distribution reduces to the marginal distribution $\mathbb{P}_{j}(\ell_j)$ due to the deterministic nature of $|\widetilde{L}_0| = \infty$.}}.}
    \end{IEEEproof}

    %The bound given in \eqref{equ::Newbound:XnYn:ordered:1k} suggests that guesswork is executed within the $k$-radius Hamming sphere, centered around the most probable estimate of $\tilde{c}_1^k$ conditioning on $\tilde{y}_1^k$, i.e., the hard-decision $\tilde{h}_1^k$.
    \black{The bound in \eqref{equ::Newbound:XnYn:ordered:1k} represents a complete analysis where all possible error patterns in the MRB are considered, examining TEPs that could flip any number of bits from the initial hard decision $\boldsymbol{b}_0$ of $\tilde{y}_1^k$. However, a practical OSD typically examines only a subset of these patterns by limiting the maximum Hamming weight of TEPs to a value $m < k$. This variant is known as an order-$m$ OSD, where only TEPs that flip at most $m$ bits in the MRB from $\boldsymbol{b}_0$ are considered. The order-$m$ OSD has a maximum of $\xi_{\max} = \beta_m = \sum_{i=0}^{m} \binom{k}{i}$ TEPs to examine. Equivalently, the order-$m$ OSD potentially checks $\xi_{\max} = \beta_m$ guesses of the transmitted MRB through the sequence $[\boldsymbol{b}_1,\ldots, \boldsymbol{b}_\xi,\ldots, \boldsymbol{b}_{\xi{\max}}]$, where each guess $\boldsymbol{b}_\xi$ is located within a radius-$m$ Hamming sphere centered at $\boldsymbol{b}_0$.}
    
    \subsubsection{Order-$m$ OSD}
    \black{Theorem \ref{The::Newbound:XnYn:ordered} and Corollary \ref{Cor::MRB} are built based on the proposed Hamming subset bound in Theorem \ref{The::Newbound:Xn}, all of which analyze guesswork by partitioning sequences according to their Hamming distance from the most likely sequence. Therefore, Corollary \ref{Cor::MRB} provides a natural framework for analyzing order-$m$ OSD where TEPs are restricted to maximum weight $m$. We formalize this connection in the following corollary.}
    
    %Since the Hamming subset bound is derived by dividing $\mathcal{X}^n$ according to the Hamming distance, it can be used to evaluate the complexity of an order-$m$ OSD with slight modification, which is detailed in the following corollary.
    
    \begin{corollary} [\black{Guesswork of order-$m$ OSD}] \label{Cor::Newbound:XnYn:ordered:1k:order}
        For an order-$m$ OSD $(m<k)$, the $\omega_{\rm{th}}$ moment of its complexity with optimal processing is upper bounded by 
    \begin{align}   \label{equ::Newbound:XnYn:ordered:1k:order}
        &\mathbb{E}[G^*(\widetilde{X}_1^k | \widetilde{Y}_1^k)^\omega]  \leq \notag\\
        &\hspace{1cm} \sum_{j=0}^{m} (\gamma_j)^{\frac{1}{p}} \binom{k}{j}^{1/q} \int_{\mathbb{R}^+} 
        \left[ \mathbb{E}\left(\mathbb{P}_{\hat{X}|\hat{Y}}(\bar{x}|Y)^q\right)^{\frac{n-j}{q}} \right. \notag \\
        &\hspace{1.5cm} \cdot \left. \mathbb{E}\left(\mathbb{P}_{\hat{X}|\hat{Y}}(1-\bar{x}|Y)^q\right)^{\frac{j}{q}} \mathbb{P}_{k+1}(|\tilde{\ell}_{k+1}|) \right]  \ d|\tilde{\ell}_{k+1}| \notag\\
        &\hspace{1cm} + (\beta_m)^{\omega}  \sum_{j=m+1}^{k} \binom{k}{j} \int_{\mathbb{R}^+} 
        \left[ \mathbb{E}\left(\mathbb{P}_{\hat{X}|\hat{Y}}(\bar{x}|Y)\right)^{n-j} \right. \notag \\
        &\hspace{1.5cm} \cdot \left. \mathbb{E}\left(\mathbb{P}_{\hat{X}|\hat{Y}}(1-\bar{x}|Y)\right)^{j}   \mathbb{P}_{k+1}(|\tilde{\ell}_{k+1}|) \right]\ d|\tilde{\ell}_{k+1}|
    \end{align}
    \black{where $\gamma_j$ is given in \eqref{equ::gamma3}, $\beta_m = \sum_{i=0}^m \binom{k}{i}$, and $\mathbb{P}_{k+1}(\cdot)$ denotes the distribution of the $(k+1)$-th ordered LLR magnitude given by \eqref{equ::ordered:L}.}
    \end{corollary}

    \begin{IEEEproof}
        Similar to Theorem \ref{The::Newbound:Xn}, we define the subsets of $\mathcal{X}^k$ according to Hamming shells. Let $\bar{x}^k$ denote the most likely sequence of $\hat{X}_1^k$ \black{of length-$k$ from pairs} $(\hat{X}_1^k, \hat{Y}_1^k)$, where $(\hat{X}_1^k, \hat{Y}_1^k)$ are the conditionally independent pairs \black{corresponding to} $(\widetilde{X}_1^k, \widetilde{Y}_1^k)$ \black{for a length-$n$ BI-OSC,} as introduced in Section \ref{sec::conditional::indep}. We construct subsets $\{\mathcal{X}_0,\ldots,\mathcal{X}_k\}$ defined as
    \begin{equation}
        \mathcal{X}_j := \left\{x^k \in \mathcal{X}^k \mid d_{\mathrm{H}} (x^k,\bar{x}^k) = j \right\}.
    \end{equation} 
    
    The order-$m$ OSD only conducts maximum $\beta_m$ guesses. Under the condition $|\widetilde{L}_{k+1}| = |\tilde{\ell}_{k+1}|$, $\mathbb{E}[G(\widetilde{X}_1^k | \widetilde{Y}_1^k)^\omega]$ is equivalent to $\mathbb{E}[G(\hat{X}_1^k | \hat{Y}_1^k)^\omega] $ given by
        \begin{align} \label{equ::Cor3::line1}
            &\mathbb{E}[G(\hat{X}_1^k | \hat{Y}_1^k)^\omega]  \notag\\
              &=  \sum_{j = 0}^{m} \mathbb{P}(\black{\hat{X}_1^k} \! \in\! \mathcal{X}_j| \hat{Y}_1^k) \cdot \mathbb{E}[G(\hat{X}_1^k|\hat{Y}_1^k)^\omega \mid \hat{X}_1^k \in \mathcal{X}_j] \notag \\
              &+ (\beta_m)^\omega  \sum_{j = m+1}^{k} \mathbb{P}(\hat{X}_1^k \! \in\! \mathcal{X}_j| \hat{Y}_1^k).
        \end{align}
        \black{The first term represents the guesswork for sequences within Hamming distance $m$ (i.e., $\widetilde{X}_1^k \in \black{\{\mathcal{X}_0\cup\ldots\cup\mathcal{X}_m\}}$), while the second term accounts for sequences beyond distance $m$ (i.e., $\widetilde{X}_1^k \notin \black{\{\mathcal{X}_0\cup\ldots\cup\mathcal{X}_m\}}$) where the decoder stops after $\beta_m$ guesses.} 

        According to Corollary \ref{Cor::Newbound:XnYn:ordered:condition}, \black{the first term of} \eqref{equ::Cor3::line1} is directly obtained from \eqref{equ::Newbound:XnYn:ordered:condition} by setting the upper summation limit to $m$. \black{For the second term, we apply} the derivation techniques used for \eqref{equ::subset:proof1} and \eqref{equ::cor1::Jensen}, and it follows that
        \begin{align} \label{equ::cor3::term1}
            & \sum_{j = m+1}^{k}\mathbb{P}(\hat{X}_1^k \! \in\! \mathcal{X}_j| \hat{Y}_1^k) =  \notag\\
            &\hspace{1cm}  \sum_{j = m+1}^{k}\binom{k}{j}  \mathbb{E}\left(\mathbb{P}_{\hat{X}|\hat{Y}}(\bar{x}|Y)\right)^{n-j}  \mathbb{E}\left(\mathbb{P}_{\hat{X}|\hat{Y}}(1-\bar{x}|Y)\right)^{j} 
        \end{align}

        %Finally, similar to Theorem \ref{The::Newbound:XnYn:ordered}, the proof completes by eliminating the condition ${|\widetilde{L}_{k+1}| = |\tilde{\ell}_{k+1}|}$ through the integration of $|\widetilde{L}_{k+1}|$ with its distribution.

        \black{Finally, following the approach of Theorem \ref{The::Newbound:XnYn:ordered}, we obtain the unconditional bound \eqref{equ::Newbound:XnYn:ordered:1k:order} by integrating \eqref{equ::Cor3::line1} over $|\widetilde{L}_{k+1}|$ using its distribution $\mathbb{P}_{k+1}$. This completes the proof.}
    \end{IEEEproof}

\subsection{A Simplified Bound for the \black{Guesswork} Complexity of OSD}

\black{By taking the limit as $q$ approaches 1 (and consequently $p \to \infty$ since $1/p + 1/q = 1$), we can derive a significantly simpler form of the bounds in \eqref{equ::Newbound:XnYn:ordered:1k} and \eqref{equ::Newbound:XnYn:ordered:1k:order}. We present the simplified bound in the following Theorem.} 
\begin{theorem} \label{The::Newbound:XnYn:ordered:qto1}
    The achievable \black{guesswork} complexity of an order-$k$ OSD, characterized by $\mathbb{E}[G^*(\widetilde{X}_1^k | \widetilde{Y}_1^k)^{\omega}]$, is upper bounded by
    \begin{align}   \label{equ::Newbound:XnYn:ordered:1k:?}
        \mathbb{E}[G^*(\widetilde{X}_1^k | \widetilde{Y}_1^k)^\omega] \leq \sum_{j=0}^{k} (\beta_j)^{\omega} \cdot \mathbb{P}_{E}(j),
    \end{align}
    where $E$ \black{denotes the} the random variable \black{representing} the number of \black{bit errors in the hard decision of} $\tilde{Y}_1^k$, and $\mathbb{P}_{E}(j)$ is its pmf, given by
    \begin{align} \label{equ::E1k}
        \mathbb{P}_{E}(j) &=  \binom{k}{j}  \int_{\mathbb{R}^+} \frac{1}{\tau^k}  Q\left(\frac{|\tilde{\ell}_{k+1}|\sigma}{2} - \frac{1}{\sigma} \right)^{k-j} \notag \\
        &\hspace{0.5cm} \cdot  Q\left(\frac{|\tilde{\ell}_{k+1}|\sigma}{2} + \frac{1}{\sigma} \right) ^j \mathbb{P}_{k+1}(|\tilde{\ell}_{k+1}|) \, d|\tilde{\ell}_{k+1}| . 
    \end{align}
\end{theorem}

\begin{IEEEproof}
\black{First, we evaluate the limit of the probability terms as $q \to 1$.} Since 
\begin{align}
     \lim_{q \downarrow 1} \mathbb{P}_{X|Y}(x|y)^q \, \mathbb{P}_{Y}(y) &=  \lim_{q \downarrow 1} \frac{1}{2}\frac{[\mathbb{P}(y|x)]^q}{[\mathbb{P}(y|0) +\mathbb{P}(y|1)]^{q-1}}\\
     &=  \frac{1}{2} \mathbb{P}(y|x),
\end{align}
\black{we have}
\begin{align}  \label{equ::ordered:correctP}
    \lim_{q \downarrow 1} \mathbb{E}\left[\mathbb{P}_{\hat{X}|\hat{Y}}(\bar{x}|Y)^q\right] =  \frac{1}{\tau}Q\left(\frac{|\tilde{\ell}_{k+1}|\sigma}{2} - \frac{1}{\sigma} \right) 
\end{align}
and
\begin{align}   \label{equ::ordered:errorP}
   \lim_{q \downarrow 1} \mathbb{E}\left[\mathbb{P}_{\hat{X}|\hat{Y}}(1-\bar{x}|Y)^q\right] = \frac{1}{\tau}Q\left(\frac{|\tilde{\ell}_{k+1}|\sigma}{2} + \frac{1}{\sigma} \right)\black{.}
\end{align}
It appears that these two are the average error probability and average correct probability of estimating $\hat{X}$ based on $\hat{Y}$. \black{Specifically, \eqref{equ::ordered:correctP} and \eqref{equ::ordered:errorP} represent the probabilities of correct and incorrect bit decisions of $\hat{X}$, respectively.}

\black{Given that} $(\hat{X}_1^k,\hat{Y}_1^k)$ are i.i.d. pairs under the condition $\{\widetilde{L}_{k+1} = \Tilde{\ell}_{k+1}\}$, \black{the probability of exactly $j$ errors in the hard decision of $\hat{Y}_1^k$ follows a binomial distribution:}%the probability of there are $j$ errors of estimating $\hat{X}_1^k$ based on $\hat{Y}_1^k$ is given by $\hat{Y}_1^k$ is exactly
    \begin{align} \label{equ::E1k:conditionalIID}
         \frac{1}{\tau^k} \binom{k}{j} Q\left(\frac{|\tilde{\ell}_{k+1}|\sigma}{2} - \frac{1}{\sigma} \right)^{k-j}  Q\left(\frac{|\tilde{\ell}_{k+1}|\sigma}{2} + \frac{1}{\sigma} \right) ^j. 
    \end{align}
By relaxing the condition $\{\widetilde{L}_{k+1} = \Tilde{\ell}_{k+1}\}$ in \eqref{equ::E1k:conditionalIID}, we can derive \eqref{equ::E1k}.  \black{Finally, since $\beta_{j-1} < \beta_j$, as $p \to \infty$, the term $(\beta_{j-1})^{\omega p +1}$ becomes negligible compared to $(\beta_j)^{\omega p +1}$ in the expression of $\gamma_j$. Therefore,
\begin{equation}
    \lim_{p \to \infty} \gamma_j = \lim_{p \to \infty} \frac{(\beta_j)^{\omega p + 1} - (\beta_{j-1})^{\omega p + 1}}{\omega p +1} = (\beta_j)^\omega.
\end{equation}
This completes the proof.}
    
\end{IEEEproof}

\black{
\begin{remark}
    Eq. \eqref{equ::Newbound:XnYn:ordered:1k:?} can be obtained directly by applying $l_1$ and $l_\infty$ norms in Lemma \ref{lem::holder}, corresponding to taking $q \downarrow 1$ and $p\to \infty$ in Hölder's inequality. Specifically, we can apply $l_1$ norm to $a_g = \mathbb{P}(X = \mathbf{x}_{(g)})$ and $l_\infty$ norm to $b_g = g^\omega$. Since $a_g$ and $b_g$ are finite sequences, their $l_1$ and $l_\infty$ norms are existing and finite. This approach naturally extends to $X^n$ and pairs $(X^n,Y^n)$, yielding results equivalent to Theorem \ref{The::Newbound:Xn} and Corollary \ref{Cor::HSB}. Furthermore, incorporating the conditional independence from Lemma \ref{lem::independent} allows us to extend this $l_1$-$l_\infty$ bound to BI-OSC, leading to an alternative derivation of Theorem \ref{The::Newbound:XnYn:ordered:qto1}. However, while this direct approach is simpler, it bypasses the more general bounds with arbitrary $(p,q)$ parameters established in Theorem \ref{The::Newbound:XnYn:ordered}. As shown in Fig. \ref{Fig::order:qValues} and Fig. \ref{Fig::secondOrder:q}, these general bounds can provide tighter results for BI-OSC through appropriate choice of $(p,q)$ parameters.
\end{remark}
}

%Similarly, the bound in Corollary \ref{Cor::Newbound:XnYn:ordered:1k:order} is simplified to
\black{Following the same limit analysis with $q \downarrow 1$, we can similarly simplify the bound for order-$m$ OSD given in Corollary \ref{Cor::Newbound:XnYn:ordered:1k:order} to}
\begin{align}   \label{equ::Newbound:XnYn:ordered:1k:order:?}
    \mathbb{E}[G^*(\widetilde{X}_1^k | \widetilde{Y}_1^k)^\omega]  &\leq \sum_{j=0}^{m} (\beta_j)^{\omega} \mathbb{P}_{E}(j) + (\beta_m)^{\omega}  \sum_{j=m+1}^{k}  \mathbb{P}_{E}(j),
\end{align}
%for an order-$m$ OSD decoder, which is obtained by taking $q\downarrow 1$ in \eqref{equ::Newbound:XnYn:ordered:1k:order} 
\black{where the first term accounts for OSD examining error patterns within Hamming weight $m$, and the second term represents the complexity contribution from the case that the true error pattern has a Hamming weight larger than $m$.}

    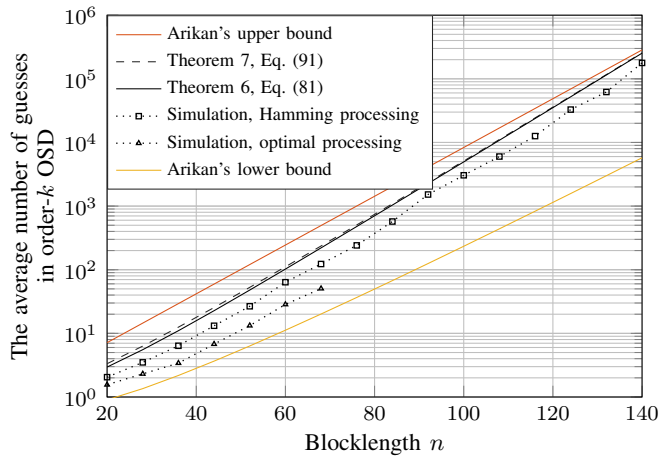
\begin{figure}  [t]
         \centering
\definecolor{mycolor1}{rgb}{0.85000,0.32500,0.09800}%
\definecolor{mycolor2}{rgb}{0.92900,0.69400,0.12500}%
\definecolor{mycolor3}{rgb}{0.00000,0.44706,0.74118}%
\definecolor{mycolor4}{rgb}{0.14902,0.14902,0.14902}%
\begin{tikzpicture}

\begin{axis}[%
width=2.8in,
height=2in,
at={(1.01in,0.685in)},
scale only axis,
xmin=20,
xmax=140,
xlabel style={at={(0.5,1ex)},font=\color{white!15!black}, font=\small},
xlabel={Blocklength $n$},
ymode=log,
ymin=1,
ymax=1000000,
yminorticks=true,
ylabel style={at={(1.5ex,0.5)},font=\color{white!15!black},align=center, font=\small},
ylabel={The average number of guesses \\ in order-$k$ OSD},
axis background/.style={fill=white},
tick label style={font=\footnotesize},
xmajorgrids,
ymajorgrids,
yminorgrids,
legend style={at={(0,1)}, anchor=north west , fill opacity=0.5, text opacity=1, font = \scriptsize	 , legend cell align=left, align=left, draw=white!15!black}
]

\addplot [color=mycolor1]
  table[row sep=crcr]{%
20	7.15198357162594\\
28	14.4966304182157\\
36	29.3750785535895\\
44	59.517454893027\\
52	120.582975502214\\
60	244.294653601037\\
68	494.918086286158\\
76	1002.64439502603\\
84	2031.21781589468\\
92	4114.93635893319\\
100	8336.18878482383\\
108	16887.6905012188\\
116	34211.4629686119\\
124	69306.1801713207\\
132	140401.391749361\\
140	284426.559534383\\
};
\addlegendentry{Arikan's upper bound}

\addplot [color=mycolor4,dashed] %checked, no problem. 
  table[row sep=crcr]{%
20	3.3364634141491\\
28	6.31291514148637\\
36	12.5186925883123\\
44	25.4732468985985\\
52	52.8575138952759\\
60	111.195011724896\\
68	236.228982175296\\
76	505.609081445453\\
84	1088.45297405809\\
92	2354.04800756965\\
100	5110.61334647009\\
108	11130.5291434665\\
116	24307.4483042412\\
124	53208.7810346127\\
132	116712.895685918\\
140	256472.434651789\\
};
\addlegendentry{Theorem \ref{The::Newbound:XnYn:ordered:qto1}, Eq. \eqref{equ::Newbound:XnYn:ordered:1k:?}}

\addplot [color=black]
  table[row sep=crcr]{%
20	2.9550741344605\\
28	5.5412030052554\\
36	11.0102114246781\\
44	22.7183946735823\\
52	47.9189768014309\\
60	102.38026769624\\
68	220.513137274427\\
76	477.52579169706\\
84	1038.08567182489\\
92	2263.38306157398\\
100	4946.80055790533\\
108	10833.5799657527\\
116	23767.7063585259\\
124	52225.8208227396\\
132	114918.983160736\\
140	253195.028452267\\
};
\addlegendentry{Theorem \ref{The::Newbound:XnYn:ordered}, Eq. \eqref{equ::Newbound:XnYn:ordered:1k}}

%\addplot [color= black, mark=square, mark size = 1pt,  only marks]
%  table[row sep=crcr]{%
%20	1.1992388\\
%28	1.365058\\
%36	1.59771355\\
%44	1.9192687\\
%52	2.34806405\\
%60	2.93604765\\
%68	3.728925\\
%76	4.79305075\\
%84	6.1802416\\
%92	8.1722657\\
%100	10.70880165\\
%108	13.91339955\\
%116	18.60785045\\
%124	24.31012235\\
%132	32.92295875\\
%140	41.92501815\\
%};
%\addlegendentry{Simulation, Hamming processing}

\addplot [color= black, mark=square, mark size = 1pt, dotted, line width=0.5pt, mark options={solid}]
  table[row sep=crcr]{%
20	2.050092\\
28	3.501392\\
36	6.38393\\
44	13.170404\\
52	26.649404\\
60	63.660376\\
68	123.030812\\
76	241.260294\\
84	572.554008\\
92	1520.418184\\
100	3049.247448\\
108	6007.63104\\
116	12588.366784\\
124	32863.177498\\
132	62061.935392\\
140	177802.982406\\
};
\addlegendentry{Simulation, Hamming processing}

\addplot [color= black, mark=triangle, mark size = 1pt,  dotted, line width=0.5pt, mark options={solid}]
  table[row sep=crcr]{%
20	1.56\\
28	2.31\\
36	3.408\\
44	6.78\\
52	13.24\\
60	28.49\\
68	50.54\\
};
\addlegendentry{Simulation, optimal processing}

\addplot [color=mycolor2]
  table[row sep=crcr]{%
20	0.901722120045461\\
28	1.35431132704386\\
36	2.17970194285625\\
44	3.66278437145282\\
52	6.33919010190059\\
60	11.2090482322707\\
68	20.1456426467446\\
76	36.6737134716485\\
84	67.4550201638827\\
92	125.131978136784\\
100	233.785928404806\\
108	439.440891540495\\
116	830.32419597432\\
124	1576.0314465153\\
132	3003.38519224219\\
140	5743.6353248224\\
};
\addlegendentry{Arikan's lower bound}

\end{axis}
\end{tikzpicture}%
    	\vspace{-0.5em}
        \caption{The average number of guesses versus blocklength $n$ with AWGN channel at SNR = 2 dB, where $k = n/2$.}
        \vspace{-0.25em}
    	\label{Fig::step1App:exmp}
        
    \end{figure}

Figure \ref{Fig::step1App:exmp} compares the simplified bound \eqref{equ::Newbound:XnYn:ordered:1k:?} to the original bound \eqref{equ::Newbound:XnYn:ordered:1k} obtained from Theorem \ref{The::Newbound:XnYn:ordered}. \black{ For the original bound \eqref{equ::Newbound:XnYn:ordered:1k}, we present the tightest results obtained across $q$ values in the range $(1, 3]$.} The information length is set to $k = n/2$. \black{As shown, bound from \black{\eqref{equ::Newbound:XnYn:ordered:1k:?}} is slightly looser than the bound \eqref{equ::Newbound:XnYn:ordered:1k} when the blocklength is smaller than 100.} For $n>100$, \black{both bounds show similar} tightness for in characterizing the achievable guesswork complexity of order-$k$ OSD.

Bounds given in \eqref{equ::Newbound:XnYn:ordered:1k:?} and \eqref{equ::Newbound:XnYn:ordered:1k:order:?} are efficiently computed with the approximation introduced in the next subsection.

\subsection{\black{The Approximation} of \black{OSD Guesswork} Bounds}

 For large $n$ and $k$, calculating \eqref{equ::Newbound:XnYn:ordered:1k:?} becomes \black{computationally intensive} due to the summation. We introduce a further approximation that allows \black{efficient computation} of \black{bounds of} the \black{achievable guesswork} complexity.

\subsubsection{\black{Binomial Approximation}}
First, according to \cite[Eq. (62-63)]{yue2021revisit}, $\mathbb{P}_{E}(j)$ can be approximated by the binomial distribution
\begin{equation} \label{equ:PE:approx}
    \mathbb{P}_{E}(j) \approx \binom{k}{j} p_e^j (1-p_e)^{k-j}
\end{equation}
for \black{$j\geq 0$}, where 
\begin{equation} \label{equ::pe}
    p_e  = \int_{0}^{\infty} \frac{1}{\tau}Q\Big(\frac{|\tilde{\ell}_{k+1}|}{2} + \frac{1}{\sigma^2} \Big) \cdot \mathbb{P}_{k+1}( |\tilde{\ell}_{k+1}|) d |\tilde{\ell}_{k+1}|,
\end{equation}
\black{where $\tau$ is given in \eqref{equ::tau:1k}, and $\mathbb{P}_{k+1}(\cdot)$ denotes the distribution of the $(k+1)$-th ordered LLR magnitude given by \eqref{equ::ordered:L}. Note that $\tau$ is a function of $|\tilde{\ell}_{k+1}|$ as shown in \eqref{equ::tau:1k}.} Then, as reported by \cite{yue2021revisit}, \black{the distribution $\mathbb{P}_{k+1}(\cdot)$} can be tightly approximated by a normal distribution with mean \cite[Eq. (20)]{yue2021revisit}
\begin{equation} \label{equ::OrderStat::Amean}
      \mu_{r} = F_{|L|}^{-1}\left(1-\frac{k+1}{n}\right) = F_{|L|}^{-1}\left(1-r-\frac{1}{n}\right)
\end{equation}
and variance 
\begin{align} 
\label{equ::OrderStat::Avar}
         \sigma_{r}^2 &= \frac{\pi N_0}{n}\left(1\!-\!r\!-\!\frac{1}{n}\right)\left(r\!+\!\frac{1}{n}\right)\left(e^{\!-\frac{(\mu_r+1)^2}{N_0}\!} + e^{\!-\frac{(\mu_r-1)^2}{N_0}\!}\right)^{\!\!-2}\!\!, 
\end{align}
where $r= k/n$ is the coding rate\black{, and $F_{|L|}^{-1}(\cdot)$ is the inverse function of $F_{|L|}(\cdot)$ given in \eqref{equ::cdf::L}.}
Thus, $ p_e$ is approximately given by
\begin{equation} \label{equ:pe_approx}
    p_e  \approx \int_{0}^{\infty} \frac{1}{\tau}Q\Big(\frac{|\tilde{\ell}_{k+1}| \sigma}{2} + \frac{1}{\sigma} \Big) \cdot \phi\left(\frac{|\tilde{\ell}_{k+1}|-\mu_{r}}{\sigma_{r}^2}\right) d |\tilde{\ell}_{k+1}|,
\end{equation}
\black{where $\phi(x)$ is the $\mathrm{pdf}$ of the standard normal distribution.}

\black{Using \eqref{equ:PE:approx} and \eqref{equ:pe_approx}, we can efficiently approximate the bounds in \eqref{equ::Newbound:XnYn:ordered:1k:?} and \eqref{equ::Newbound:XnYn:ordered:1k:order:?} introduced in Theorem \ref{The::Newbound:XnYn:ordered:qto1}. That is,
\begin{align}
        \mathbb{E}[G^*(\widetilde{X}_1^k | \widetilde{Y}_1^k)^\omega] &\leq \sum_{j=0}^{k} (\beta_j)^{\omega} \cdot \mathbb{P}_{E}(j) \notag\\
        &\approx \sum_{j=0}^{k} (\beta_j)^{\omega} \binom{k}{j} p_e^j (1-p_e)^{k-j}  
\end{align}
Since this approximation requires only a single integral for computing $p_e$ in \eqref{equ:pe_approx}, the computational cost is significantly reduced compared to the original bounds.}

%Therefore, bounds \eqref{equ::Newbound:XnYn:ordered:1k:?} and \eqref{equ::Newbound:XnYn:ordered:1k:order:?} are approximately computed by exploiting \eqref{equ:PE:approx} and \eqref{equ:pe_approx}. We note that a single evaluation of the integral in \eqref{equ:pe_approx}  suffices to deduce \eqref{equ::Newbound:XnYn:ordered:1k:?} and \eqref{equ::Newbound:XnYn:ordered:1k:order:?}; thus these two bounds are computationally efficient.

\black{Next, we introduce an asymptotic approximation for the OSD guesswork bounds, which helps us to characterize the asymptotic behavior of the \textit{achievable guesswork complexity} of OSD.}

\subsubsection{\black{Bessel Approximation for the Order-$k$ OSD}}

If the block length $n$ is not small, this approximation derives the following result for the bound of achievable complexity.
\begin{theorem} [\black{Asymptotic} Bessel Approximation]
 \label{The::Newbound:XnYn:ordered:1k:expapp}
    For a fixed rate $r$ and $\omega = 1$, the \black{guesswork} complexity bound \black{of OSD} given by \eqref{equ::Newbound:XnYn:ordered:1k:?} \black{exhibits exponential growth} with the blocklength $n$ (or information length $k = rn$). \black{This growth can be characterized through modified Bessel functions as}
    \begin{equation}  \label{equ::Newbound:XnYn:ordered:1k:expapp}
        e^{-k p_e}I_0(2k\sqrt{p_e}) \approx \frac{1}{\sqrt{4\pi k\sqrt{p_e}}} \ 2^{k(2\sqrt{p_e}-p_e)\log_2(e)}
    \end{equation}
    with 
    \begin{equation}   \label{equ:pe_approx2}
        p_e = \frac{1}{r} Q\left(\frac{1}{2} F_{|L|}^{-1}\left(1-r\right) + \frac{1}{\sigma^2} \right)
    \end{equation}
    where $I_0$ is the modified Bessel function, \black{and $F_{|L|}^{-1}(\cdot)$ is the inverse cumulative distribution function of the bit LLR magnitude in an AWGN channel with noise power $\sigma^2$.}
\end{theorem}

\begin{IEEEproof}
    For sufficiently large values of $n$, the expressions given in \eqref{equ::OrderStat::Amean} and \eqref{equ::OrderStat::Avar} simplify to
    \begin{equation} \label{equ::OrderStat::Amean:limit}
          \lim_{n\to\infty}\mu_{r} = F_{|L|}^{-1}\left(1-r\right)
    \end{equation}
    and 
    \begin{equation} \label{equ::OrderStat::Avar:limit}
            \lim_{n\to\infty} \sigma_{r}^2 = 0,
    \end{equation}
    respectively. This indicates that as $ n $ approaches infinity, the random variable $ |\widetilde{L}|_{k+1} $ tends towards $ \mu_{r} $. Thus, \black{following \eqref{equ:pe_approx}, we have}
    \begin{align}
        \lim_{n\to\infty} p_e  &\approx \int_{0}^{\infty} \frac{1}{\tau}Q\Big(\frac{|\tilde{\ell}_{k+1}| \sigma}{2} + \frac{1}{\sigma} \Big) \phi\left(\frac{|\tilde{\ell}_{k+1}|-\mu_{r}}{\sigma_{r}^2}\right) d |\tilde{\ell}_{k+1}| \notag\\
       & = \frac{Q\big(\frac{\mu_r \sigma}{2} + \frac{1}{\sigma} \big)}{Q\big(\frac{\mu_r \sigma}{2} - \frac{1}{\sigma} \big) + Q\big(\frac{\mu_r \sigma}{2} + \frac{1}{\sigma} \big)}  \notag\\
    & \overset{(a)}{=} \black{\frac{1}{r} Q\left(\frac{1}{2} F_{|L|}^{-1}\left(1-r\right) + \frac{1}{\sigma^2} \right)}.
    \end{align}
    \black{where step (a) follows from 
    \begin{align}
    Q\big(\frac{\mu_r \sigma}{2} - \frac{1}{\sigma} \big) + Q\big(\frac{\mu_r \sigma}{2} + \frac{1}{\sigma} \big) &= 1 - F_{|L|}(\mu_r) \notag\\
    &= 1 - F_{|L|}\left(F_{|L|}^{-1}\left(1-r\right)\right) \notag \\
    &= r
    \end{align}}
    \black{As shown, for large $n$, $p_e$} is only dependent on the \black{code} rate $r$ \black{and the channel noise power $\sigma^2$}. On the other hand, when $k = nr$ is large, the binomial distribution in \eqref{equ:PE:approx} is approximated by the Poisson distribution with parameter $kp_e$ \cite[Chapter VII]{feller1968}, i.e.,
    \begin{equation} \label{equ::BioToPois}
        \mathbb{P}_{E}(j) \approx \frac{(k p_e)^j e^{-kp_e}}{j!}
    \end{equation}
    As a result, the bound given in \eqref{equ::Newbound:XnYn:ordered:1k:?} is approximated by 
    \begin{equation} \label{equ::Newbound:XnYn:ordered:1k:app1}
        \sum_{j=0}^{k} \beta_j\cdot \mathbb{P}_{E}(j) \approx e^{-k p_e}\sum_{j=0}^{k} \sum_{i=0}^{j}\binom{k}{i}\frac{(k p_e)^j }{j!} .
    \end{equation}
    Focusing on \eqref{equ::Newbound:XnYn:ordered:1k:app1}, we observe that
    \begin{itemize}
        \item The term $\frac{(k p_e)^j }{j!}$ decreases super-exponentially as $j$ increases due to the factorial in the denominator. 
        \item The \black{combinatorial} sum $\sum_{i=0}^{j}\binom{k}{i}$ \black{has its rate of increase with respect to $j$ peaking at $j = k/2$, where it grows exponentially.}
    \end{itemize}
    Consequently, the combined term $\sum_{i=0}^{j}\binom{k}{i}\frac{(k p_e)^j }{j!}$ diminishes rapidly with increasing $j$, and the series is predominantly supported by small $j$ values compared to $k$.
    
    Using Stirling's approximation, i.e.,
    $$ n! \approx \sqrt{2\pi n} \left(\frac{n}{e}\right)^n, $$
    we can approximate $\binom{k}{j}$ as
    \begin{align} \label{equ::Stirling}
        \binom{k}{j} &\approx \frac{e^{-j} k^j}{j!(1-\frac{j}{k})^{k}}\left(1 - \frac{j}{k}\right)^{j-\frac{1}{2}} \approx \frac{k^j}{j!},
    \end{align}
    which is valid for $j \ll k$.

    Therefore, \eqref{equ::Newbound:XnYn:ordered:1k:app1} is further approximated by
    \begin{align}  \label{equ::Newbound:XnYn:ordered:1k:app2}
        \sum_{j=0}^{k} \sum_{i=0}^{j}\binom{k}{i}\frac{(k p_e)^j }{j!} &\approx  \sum_{j=0}^{k} \sum_{i=0}^{j} \frac{k^i}{i!}\frac{(k p_e)^j }{j!} \notag \\
        &\overset{(a)}{\approx}  \sum_{j=0}^{k} \frac{k^j}{j!} \frac{k^j}{j!} (p_e)^j \notag\\
        &= \sum_{j=0}^{k} \left(\frac{1}{j!}\right)^2 (k \sqrt{p_e})^{2j},
    \end{align}
    where step (a) takes $\sum_{i=0}^{j} \frac{k^i}{i!} \approx \frac{k^j}{j!}$, because $\frac{k^{j-1}}{(j-1)!} \ll \frac{k^j}{j!}$ for $j \ll k$.

    Eq. \eqref{equ::Newbound:XnYn:ordered:1k:app2} resembles the series expansion of the modified Bessel function of the first kind of order zero; that is 
    \begin{equation}
        I_0(x) = \sum_{j=0}^{\infty} \frac{1}{j! \Gamma(j+1)} \left(\frac{x}{2}\right)^{2j},
    \end{equation}
    where $\Gamma(j+1) = j!$ is the Gamma function. Consequently,
    \begin{align}  \label{equ::Newbound:XnYn:ordered:1k:app3}
        \sum_{j=0}^{k} \sum_{i=0}^{j}\binom{k}{i}\frac{(k p_e)^j }{j!} &\approx  I_0(2k\sqrt{p_e}) \approx \frac{1}{\sqrt{4\pi k\sqrt{p_e}}}e^{2k\sqrt{p_e}} .
    \end{align}
    The last step considers the asymptotic expansion of modified Bessel function, i.e., $$I_0(2k\sqrt{p_e}) = \frac{1}{4\pi k\sqrt{p_e}}e^{2k\sqrt{p_e}} \left(1 + O\left(\frac{1}{2k\sqrt{p_e}}\right)\right).$$

    %Substituting \eqref{equ::Newbound:XnYn:ordered:1k:app3} into \eqref{equ::Newbound:XnYn:ordered:1k:app1} and changing the exponential base from $e$ to 2 complete the proof.
    
    \black{Substituting \eqref{equ::Newbound:XnYn:ordered:1k:app3} into \eqref{equ::Newbound:XnYn:ordered:1k:app1} results in
    \begin{align}
        \sum_{j=0}^{k} \beta_j\cdot \mathbb{P}_{E}(j) &\approx 
        \frac{1}{\sqrt{4\pi k\sqrt{p_e}}} \ e^{k(2\sqrt{p_e}-p_e)} \notag \\
        &= \frac{1}{\sqrt{4\pi k\sqrt{p_e}}} \ 2^{k(2\sqrt{p_e}-p_e)\log_2(e)}
    \end{align}
    This completes the proof.}
\end{IEEEproof}

       \begin{figure}  [t]
         \centering
            \input{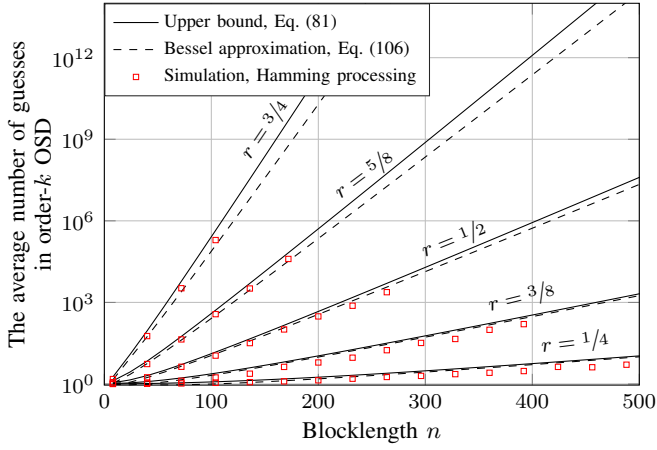}
    	\vspace{-0.5em}
        \caption{The average number of guesses for various blocklength $n$ with AWGN channel at SNR = 4 dB. }
        \vspace{-0.25em}
    	\label{Fig::OSD:cmp:exp}
        
    \end{figure}

    \begin{figure}  [t]
         \centering
            \definecolor{mycolor2}{rgb}{0.00000, 0.00000, 0.00000}% Black
%\definecolor{mycolor1}{rgb}{0.00000,0.44700,0.74100}%
\definecolor{mycolor1}{rgb}{0.00000, 0.00000, 0.00000}%
\begin{tikzpicture}

\begin{axis}[%
width=2.8in,
height=2in,
at={(0.758in,0.481in)},
scale only axis,
xlabel style={at={(0.5,1ex)},font=\color{white!15!black}, font=\small},
xlabel={Blocklength $n$},
ylabel style={at={(1.5ex,0.5)},font=\color{white!15!black},align=center, font=\small},
ylabel={The average number of guesses \\ in  order-$k$ OSD},
xmin=5,
xmax=80,
ymode=log,
ymin=0.909791807764386,
ymax=1e+4,
yminorticks=true,
axis background/.style={fill=white},
tick label style={font=\footnotesize},
xmajorgrids,
ymajorgrids,
legend style={at={(0.0,1)}, anchor=north west ,fill opacity=0.5, text opacity=1,  font = \scriptsize, legend cell align=left, align=left, draw=white!15!black}
]

\addplot [color=mycolor1]
  table[row sep=crcr]{%
8	1\\
16	1.03913411287711\\
24	1.06569495170677\\
32	1.10018567303294\\
40	1.14302621641203\\
48	1.19453438125119\\
56	1.25520748111964\\
64	1.32564334664742\\
72	1.40655254238196\\
80	1.49876420151232\\
};
\addlegendentry{Upper bound, Eq. \eqref{equ::Newbound:XnYn:ordered:1k}}

\addplot [color=mycolor1, forget plot]
  table[row sep=crcr]{%
8	1.07086261031027\\
16	1.14479938105738\\
24	1.25790948023397\\
32	1.41676479330429\\
40	1.63127730809393\\
48	1.91534029644989\\
56	2.28763546076997\\
64	2.77288566484458\\
72	3.40363755218166\\
80	4.22243087396133\\
};

\addplot [color=mycolor1, forget plot]
  table[row sep=crcr]{%
8	1.20808123223354\\
16	1.54064749782113\\
24	2.0894339681574\\
32	3.01229037747691\\
40	4.54161734605612\\
48	7.06868869883416\\
56	11.2486731901482\\
64	18.1758183449082\\
72	29.6791966015694\\
80	48.8191227256144\\
};

\addplot [color=mycolor1, forget plot]
  table[row sep=crcr]{%
8	1.59606292662482\\
16	3.01591759894483\\
24	6.38892727148034\\
32	14.4955689685163\\
40	34.2732314207359\\
48	82.9152342172698\\
56	203.219082847328\\
64	502.188812411432\\
72	1248.20831520688\\
80	3116.1269205924\\
};

\addplot [color=mycolor1, forget plot]
  table[row sep=crcr]{%
8	2.30452492815105\\
16	9.53671355248831\\
24	42.5520003268969\\
32	193.346312967662\\
40	887.545965886938\\
48	4110.54518992334\\
56	19193.7700901314\\
64	90281.402820638\\
72	427351.947271272\\
80	2033863.6975719\\
};

\addplot [color=mycolor2, dashed]
  table[row sep=newline]{%
8	1.02435636081899
16	0.947467875917874
24	0.922208644112697
32	0.927239257571851
40	0.952154249591703
48	0.992141930768635
56	1.04493462670884
64	1.10955658691345
72	1.18576999521360
80	1.27380964166421
};
\addlegendentry{Bessel approximation \eqref{equ::Newbound:XnYn:ordered:1k:expapp}}

\iffalse
\addplot [color=mycolor1, forget plot]
  table[row sep=crcr]{%
8	1.08441003443907\\
16	1.17560950100797\\
24	1.31538259290016\\
32	1.50938395600133\\
40	1.7673642819444\\
48	2.10310532214668\\
56	2.53504130892903\\
64	3.08733255839355\\
72	3.79133950042642\\
80	4.6875653793137\\
};
\fi

\addplot [color=mycolor2, dashed, forget plot]
  table[row sep=newline]{%
8	0.910532621871185
16	0.958373233375016
24	1.07758267096389
32	1.25774113118462
40	1.50260217681552
48	1.82373742236972
56	2.23891434753164
64	2.77240522269335
72	3.45623011712301
80	4.33210295996264
};

\iffalse
\addplot [color=mycolor1, forget plot]
  table[row sep=crcr]{%
8	1.25128160507242\\
16	1.62071802740311\\
24	2.23130281842985\\
32	3.20879131708466\\
40	4.7545734054228\\
48	7.1926086212703\\
56	11.0456057081956\\
64	17.1552873658238\\
72	26.877829455758\\
80	42.4011947112598\\
};
\fi

\addplot [color=mycolor2, dashed, forget plot]
  table[row sep=newline]{%
8	0.988082557805553
16	1.31931069023384
24	1.89280556653761
32	2.82867039862349
40	4.33343435792273
48	6.74985117501498
56	10.6394009961225
64	16.9205537295467
72	27.0973604026913
80	43.6367275156068
};
\addplot [color=mycolor2, dashed, forget plot]
  table[row sep=newline]{%
8	1.29275280846046
16	2.64547567791754
24	5.61624122037353
32	12.4039281753124
40	28.1079397579406
48	64.7969560652728
56	151.212530416102
64	356.114296741457
72	844.636632358920
80	2014.69129478432
};

\iffalse
\addplot [color=mycolor1, forget plot]
  table[row sep=crcr]{%
8	1.30993913049209\\
16	2.44901721405893\\
24	5.13745503252334\\
32	11.3028827577688\\
40	25.5570853757773\\
48	58.8310773005044\\
56	137.147907894904\\
64	322.739534870662\\
72	765.013158385033\\
80	1823.87987106018\\
};

\addplot [color=mycolor1, forget plot]
  table[row sep=newline]{%
8	1.92811486358174
16	8.28112056766757
24	34.4194600911714
32	142.508405009693
40	598.171285093811
48	2547.26624667679
56	10974.3889429738
64	47709.7907721171
};
\fi

\addplot [color=mycolor2, dashed, forget plot]
  table[row sep=crcr]{%
8	2.64212707913791\\
16	8.3897727837881\\
24	31.6885256241457\\
32	127.660074651015\\
40	531.016909089645\\
48	2252.14237338977\\
56	9679.52233948664\\
64	42009.4623096999\\
72	183692.566091288\\
80	808003.818416708\\
};

\addplot [red, mark=square, mark size = 1pt, only marks]
  table[row sep=newline]{%
8	1.00794000000000
16	1.01764000000000
24	1.02938000000000
32	1.04728000000000
40	1.06700000000000
48	1.08920000000000
56	1.12426000000000
64	1.15706000000000
72	1.18050000000000
};
\addlegendentry{Simulation, Hamming processing}

\addplot [red, mark=square, mark size = 1pt, only marks,forget plot]
  table[row sep=newline]{%
8	1.03068000000000
16	1.07632000000000
24	1.14374000000000
32	1.24810000000000
40	1.36854000000000
48	1.54584000000000
56	1.78510000000000
64	2.04464000000000
72	2.44786000000000
};

\addplot [red, mark=square, mark size = 1pt, only marks,forget plot]
  table[row sep=newline]{%
    8.0000    1.0943
   16.0000    1.2706
   24.0000    1.5741
   32.0000    2.0887
   40.0000    2.9542
   48.0000    4.2454
   56.0000    6.5358
   64.0000    11.4860
   72.0000   18.5729
};

\addplot [red, mark=square, mark size = 1pt, only marks,forget plot]
  table[row sep=newline]{%
8	1.25946000000000
16	1.96365500000000
24	3.63519500000000
32	7.35268000000000
40	18.0936650000000
48	41.6851450000000
56	96.2699350000000
64	209.682090000000
72	520.296330000000
};

\addplot [red, mark=square, mark size = 1pt, only marks,forget plot]
  table[row sep=newline]{%
8	1.69218000000000
16	4.95722000000000
24	18.7846500000000
32	82.4603650000000
40	333.601940000000
48	1285.43837500000
56	5152.72271000000
64	25040.6023550000
72	149538.525165000
};

\node[rotate=6] at (axis cs: 72,1.75) {\footnotesize $r = \sfrac{1}{4}$};

\node[rotate=12] at (axis cs: 68,4.75) {\footnotesize $r = \sfrac{3}{8}$};

\node[rotate=22] at (axis cs: 60,20) {\footnotesize $r = \sfrac{1}{2}$};

\node[rotate=36] at (axis cs: 48,120) {\footnotesize $r = \sfrac{5}{8}$};

\node[rotate=48] at (axis cs: 32,400) {\footnotesize $r = \sfrac{3}{4}$};

\end{axis}

\end{tikzpicture}
    	\vspace{-0.5em}
        \caption{The average number of guesses for short blocklength $n$ with AWGN channel at SNR = 3 dB. }
        \vspace{-0.25em}
    	\label{Fig::OSD:cmp:exp:short}
        
    \end{figure}
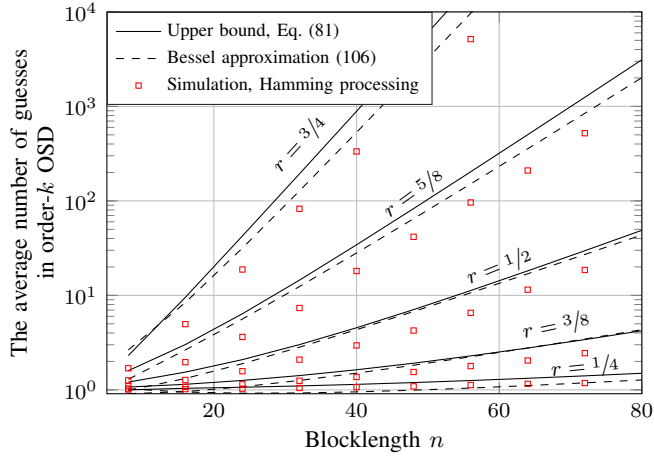

\begin{figure}  [t]
         \centering
\definecolor{mycolor1}{rgb}{0.85000,0.32500,0.09800}%
\definecolor{mycolor2}{rgb}{0.92900,0.69400,0.12500}%
\definecolor{mycolor3}{rgb}{0.00000,0.44706,0.74118}%
\definecolor{mycolor4}{rgb}{0.14902,0.14902,0.14902}%
\begin{tikzpicture}

\begin{axis}[%
width=2.8in,
height=2in,
at={(1.01in,0.685in)},
scale only axis,
xmin=-2,
xmax=4,
xlabel style={at={(0.5,1ex)},font=\color{white!15!black}, font=\small},
xlabel={SNR (dB)},
ymode=log,
ymin=1,
ymax=1e7,
yminorticks=true,
ylabel style={at={(1.5ex,0.5)},font=\color{white!15!black},align=center, font=\small},
ylabel={The average number of guesses \\ in  order-$k$ OSD},
axis background/.style={fill=white},
tick label style={font=\footnotesize},
xmajorgrids,
ymajorgrids,
%yminorgrids,
legend style={at={(1,1)}, anchor=north east , fill opacity=0.5, text opacity=1, font = \scriptsize	 , legend cell align=left, align=left, draw=white!15!black}
]

\addplot [color=mycolor1]
  table[row sep=crcr]{%
-2	1104340.91880537\\
-1.8	790233.724255444\\
-1.6	559684.194404092\\
-1.4	392369.18343464\\
-1.2	272311.100510823\\
-1	187127.161083604\\
-0.8	127357.166473743\\
-0.6	85876.4993042583\\
-0.4	57395.5242969433\\
-0.2	38042.2129752708\\
0	25021.6342649834\\
0.2	16343.9252209801\\
0.4	10611.3456783474\\
0.6	6854.84177538365\\
0.8	4410.98671715733\\
1	2831.02623018085\\
1.2	1814.84672688387\\
1.4	1163.85636771199\\
1.6	747.912190600105\\
1.8	482.467513180057\\
2	313.013161956008\\
2.2	204.629814238384\\
2.4	135.061164161475\\
2.6	90.1741673409424\\
2.8	61.0142014973764\\
3	41.911809447854\\
3.2	29.2746157425236\\
3.4	20.8210748425786\\
3.6	15.0966318265842\\
3.8	11.1690984172899\\
4	8.43712186887616\\
};
\addlegendentry{Arikan's upper bound}

\addplot [color=mycolor4] %checked, no problem. 
  table[row sep=crcr]{%
-2	391154.634806617\\
-1.8	287059.302112385\\
-1.6	208172.015778512\\
-1.4	149191.129523861\\
-1.2	105680.671811605\\
-1	74006.9186768206\\
-0.8	51248.440206792\\
-0.6	35104.6744869662\\
-0.4	23795.221620452\\
-0.2	15968.2808972422\\
0	10614.859441557\\
0.2	6994.13505753736\\
0.4	4571.47211185385\\
0.6	2966.64248387066\\
0.8	1913.4055952475\\
1	1228.01256155372\\
1.2	785.301041163053\\
1.4	501.171291296977\\
1.6	319.745653355149\\
1.8	204.331266653578\\
2	131.069840655824\\
2.2	84.5917087488478\\
2.4	55.068710961839\\
2.6	36.2579347641255\\
2.8	24.2133536219186\\
3	16.4485976291793\\
3.2	11.4000073765839\\
3.4	8.0839457306034\\
3.6	5.88079085940647\\
3.8	4.39877418624936\\
4	3.38886963009376\\
};
\addlegendentry{Upper bound, Eq. \eqref{equ::Newbound:XnYn:ordered:1k}}

\addplot [color=black,dashed]
  table[row sep=crcr]{%
-2	1194227.33065068\\
-1.8	799206.709947083\\
-1.6	529395.553508787\\
-1.4	347185.676972292\\
-1.2	225501.913499524\\
-1	145121.169872332\\
-0.8	92582.834474662\\
-0.6	58589.7106798385\\
-0.4	36806.3887791306\\
-0.2	22972.3823033014\\
0	14259.0928691359\\
0.2	8811.62705527428\\
0.4	5427.84624099896\\
0.6	3337.24361705788\\
0.8	2051.0202210414\\
1	1261.98488360773\\
1.2	778.689482876026\\
1.4	482.682322606396\\
1.6	301.117588598883\\
1.8	189.408763136824\\
2	120.358148577368\\
2.2	77.4077021525816\\
2.4	50.4818590113816\\
2.6	33.4436169356184\\
2.8	22.5457447408791\\
3	15.4912447365068\\
3.2	10.8646093579064\\
3.4	7.78775134644916\\
3.6	5.7116796808678\\
3.8	4.29008180115677\\
4	3.30233337387585\\
};
\addlegendentry{Bessel approximation \eqref{equ::Newbound:XnYn:ordered:1k:expapp}}

\addplot [color= black, mark=square, mark size = 1pt, dotted, only marks, line width=0.5pt, mark options={solid}]
  table[row sep=crcr]{%
-2	277780.65076\\
-1.5	120162.71526\\
-1	47594.63008\\
-0.5	19781.2173\\
0	6216.39284\\
0.5	2032.05442\\
1	697.11458\\
1.5	242.77854\\
2	88.99142\\
2.5	27.15516\\
3	9.76654\\
3.5	4.44244\\
4	2.34554\\
};
\addlegendentry{Simulation, Hamming processing}

%\addplot [color= black, mark=triangle, mark size = 1pt,  dotted, only marks, line width=0.5pt, mark options={solid}]
%  table[row sep=crcr]{%
%-2	277780.65076\\
%-1.5	120162.71526\\
%-1	47594.63008\\
%-0.5	19781.2173\\
%0	6216.39284\\
%0.5	2032.05442\\
%1	697.11458\\
%1.5	242.77854\\
%2	88.99142\\
%2.5	27.15516\\
%3	9.76654\\
%3.5	4.44244\\
%4	2.34554\\
%};
%\addlegendentry{Simulation, optimal processing}

\addplot [color=mycolor2]
  table[row sep=crcr]{%
-2	47640.5135729433\\
-1.8	34090.1435644656\\
-1.6	24144.3941866371\\
-1.4	16926.5387985836\\
-1.2	11747.3150356347\\
-1	8072.53802305312\\
-0.8	5494.10124598785\\
-0.6	3704.65357303526\\
-0.4	2476.00374823802\\
-0.2	1641.11510560861\\
0	1079.41622602164\\
0.2	705.065780019782\\
0.4	457.766210785128\\
0.6	295.713196060697\\
0.8	190.28695667288\\
1	122.128539518564\\
1.2	78.2912492443471\\
1.4	50.2079694220867\\
1.6	32.2644214851678\\
1.8	20.8133192556361\\
2	13.5031741890053\\
2.2	8.82759053535594\\
2.4	5.82644644859552\\
2.6	3.89005203916756\\
2.8	2.63211101307597\\
3	1.80804685660569\\
3.2	1.26288694448909\\
3.4	0.898206959236967\\
3.6	0.65125839420867\\
3.8	0.481827283301285\\
4	0.363971679457084\\
};
\addlegendentry{Arikan's lower bound}

\end{axis}
\end{tikzpicture}%
    	\vspace{-0.5em}
        \caption{The average number of guesses for $n=64$ and $k=32$  with AWGN channel at various SNRs.}
        \vspace{-0.25em}
    	\label{Fig::OSD:cmp:exp:SNR}
        
    \end{figure}
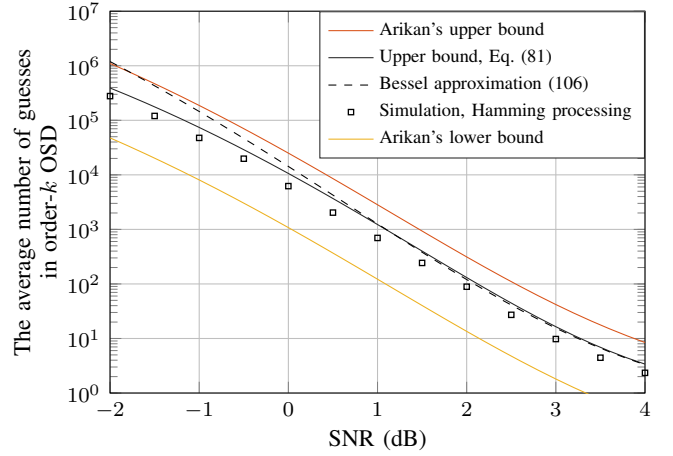

    The accuracy of \eqref{equ::Newbound:XnYn:ordered:1k:expapp} is verified in Figs. \ref{Fig::OSD:cmp:exp}-\ref{Fig::OSD:cmp:exp:SNR}. Only the simulation results of Hamming processing are included for comparison, and the optimal processing is omitted due to its prohibitively high complexity for simulation. As shown by Fig. \ref{Fig::OSD:cmp:exp}, \eqref{equ::Newbound:XnYn:ordered:1k:expapp} can well approximate the \black{guesswork} complexity bound \eqref{equ::Newbound:XnYn:ordered:1k} \black{for $\omega = 1$ (i.e,, the first order moment)}, particularly when the code rate is not high. The simulated results of Hamming processing also align closely with the value provided by \eqref{equ::Newbound:XnYn:ordered:1k:expapp}, despite the fact that \eqref{equ::Newbound:XnYn:ordered:1k:expapp} is derived from the upper bound \eqref{equ::Newbound:XnYn:ordered:1k:?}. Furthermore, Fig. \ref{Fig::OSD:cmp:exp:short} indicates that \eqref{equ::Newbound:XnYn:ordered:1k:expapp} also provides a reliable prediction of the average \black{guesswork} complexity even for very short block lengths. The tightness of \eqref{equ::Newbound:XnYn:ordered:1k} and \eqref{equ::Newbound:XnYn:ordered:1k:expapp} at various SNRs are demonstrated in Fig. \ref{Fig::OSD:cmp:exp:SNR}, showing that the Bessel approximation slightly loses accuracy at very low SNRs. Nevertheless, these results show that \eqref{equ::Newbound:XnYn:ordered:1k:expapp} can be directly used to estimate the average achievable \black{guesswork} complexity of OSD in typical scenarios.

     For a general $\omega > 1$, the proof of Theorem \ref{The::Newbound:XnYn:ordered:1k:expapp}, or more specifically the combination of \eqref{equ::Newbound:XnYn:ordered:1k:app1} and \eqref{equ::Newbound:XnYn:ordered:1k:app2}, indicates that the bound of \eqref{equ::Newbound:XnYn:ordered:1k:?} can be approximated by a simple series, i.e.,
    \begin{equation} \label{equ::Newbound:XnYn:ordered:1k:omega:app}
        \sum_{j=0}^{k} (\beta_j)^{\omega} \cdot \mathbb{P}_{E}(j) \approx e^{-kp_e} \sum_{j=0}^{k} \left(\frac{k^j}{j!}\right)^{\omega+1}p_e^{j},
    \end{equation}
    for large $n$, which is also computationally efficient.

    \subsubsection{\black{Bessel Approximation for the Order-$m$ OSD}}

    \black{We have introduced the asymptotic Bessel approximation of order-$k$ OSD guesswork. Now we consider the case of order-$m$ OSD where $m < k$. This analysis is particularly important as most OSD algorithms restrict the maximum Hamming weight of TEPs within the decoding order $m$ to reduce complexity. Using similar techniques based on Bessel functions, we can derive a tight approximation for the guesswork complexity of order-$m$ OSD, which is summarized in the following Theorem.}

    \black{\begin{theorem}[Bessel Approximation for Order-$m$ OSD]
    For an order-$m$ OSD with $kp_e < m < k$, the guesswork complexity bound in \eqref{equ::Newbound:XnYn:ordered:1k:order:?} can be approximated as
    \begin{align}   \label{equ::Newbound:XnYn:ordered:1k:order:app}
        \mathbb{E}[G^*(\widetilde{X}_1^k | \widetilde{Y}_1^k)^\omega] \leq e^{-kp_e} &\left(\sum_{j=0}^{m} \left(\frac{k^j}{j!}\right)^{\omega+1}p_e^{j} \right.\notag\\
        &\left.+ \left(\frac{k^m}{m!}\right)^{\omega}\frac{(kp_e) ^{m+1}}{(m+1)!}\right),
    \end{align}
    where $p_e$ is given by \eqref{equ:pe_approx2}. For $\omega = 1$, this simplifies to
    \begin{equation}  \label{equ::Newbound:XnYn:ordered:1k:order:app:w1}
        e^{-kp_e} \left( \sum_{j=0}^{m} \left(\frac{k^j}{j!}\right)^{2}p_e^{j} + \frac{k^m}{m!}\frac{(kp_e) ^{m+1}}{(m+1)!}\right).
    \end{equation}
    \end{theorem}}
    
    \begin{IEEEproof}
    \black{Following the approach in Theorem \ref{The::Newbound:XnYn:ordered:1k:expapp}, we apply Poisson approximation and Stirling's formula to both terms in \eqref{equ::Newbound:XnYn:ordered:1k:order:?}. For the first term ($j \leq m$), we have
    \begin{align}  \label{equ::Newbound:XnYn:ordered:1k:order:app:last3}
        \sum_{j=0}^{m} (\beta_j)^{\omega} \mathbb{P}_{E}(j) &\approx e^{-kp_e} \sum_{j=0}^{m} \left(\frac{k^j}{j!}\right)^{\omega+1}p_e^{j},
    \end{align}
    where we used \eqref{equ::BioToPois} and \eqref{equ::Stirling} as in Theorem \ref{The::Newbound:XnYn:ordered:1k:expapp}.}
    
    \black{For the second term ($j > m$), we have
    \begin{align}
        (\beta_m)^{\omega} \sum_{j=m+1}^{k} \mathbb{P}_{E}(j) &\approx \left(\frac{k^m}{m!}\right)^{\omega} \sum_{j=m+1}^{k} \frac{(k p_e)^j}{j!} e^{-kp_e} \label{equ::Newbound:XnYn:ordered:1k:order:app:last2}\\
        &\approx e^{-kp_e} \left(\frac{k^m}{m!}\right)^{\omega}\frac{(kp_e)^{m+1}}{(m+1)!}. \label{equ::Newbound:XnYn:ordered:1k:order:app:last}
    \end{align}
    The last approximation follows because for $j > kp_e$, each term in the sum is dominated by its predecessor by a factor of at least $\frac{kp_e}{j}$, making the first term ($j = m+1$) dominant.}
    
    \black{Combining these approximations yields \eqref{equ::Newbound:XnYn:ordered:1k:order:app}. The case $\omega = 1$ follows directly by substitution.}
    \end{IEEEproof}

\begin{figure}  [t]
         \centering
            \input{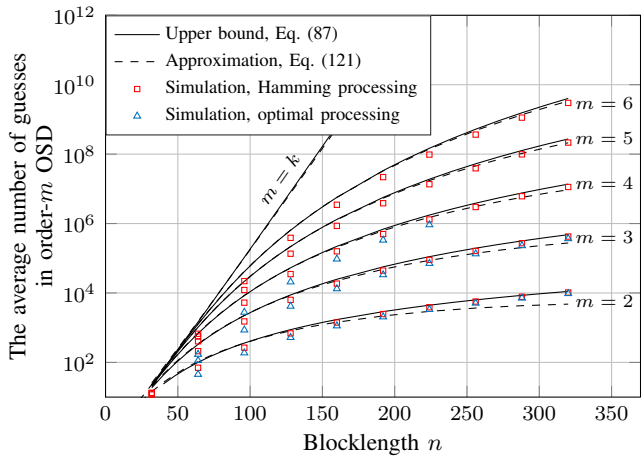}
    	\vspace{-0.5em}
        \caption{The average number of guesses for different orders $m$ with AWGN channel at SNR = 1 dB, where $k = n/2$. }
        \vspace{-0.25em}
    	\label{Fig::OSD:cmp:orders:exp}
        
    \end{figure}

    We validate the performance of \eqref{equ::Newbound:XnYn:ordered:1k:order:app:w1} in Fig. \ref{Fig::OSD:cmp:orders:exp} with simulation. As shown, \eqref{equ::Newbound:XnYn:ordered:1k:order:app:w1} is an accurate approximation of \eqref{equ::Newbound:XnYn:ordered:1k}. However, the accuracy of \eqref{equ::Newbound:XnYn:ordered:1k:order:app:w1} slightly diminishes for large values of $n$ and low orders of $m$. This loss of accuracy is attributed to the approximation in \eqref{equ::Newbound:XnYn:ordered:1k:order:app:last}.

\section{\black{Properties of Guesswork Complexity in OSD}} \label{sec::property}

    \subsection{\black{The Asymptotic Behavior of the OSD Guesswork}} \label{sec::asymtotic}

    \subsubsection{Guesswork Complexity of Order-$k$ OSD}
    \black{With Theorem \ref{The::Newbound:XnYn:ordered:1k:expapp}, we can characterize the asymptotic behavior of the guesswork complexity of order-$k$ OSD. For a fixed code rate $r$, the quantity $p_e$ given by \eqref{equ:pe_approx2} remains constant, and according to \eqref{equ::Newbound:XnYn:ordered:1k:expapp}, the average achievable guesswork complexity grows exponentially, i.e., $O(2^{k(2\sqrt{p_e}-p_e)})$, with information length $k$, governed by the factor $2\sqrt{p_e}-p_e$.}

    \black{Furthermore, the behavior of $p_e$ with respect to code rate $r$ is particularly insightful. As $r$ increases (higher rate codes), $F_{|L|}^{-1}(1-r)$ decreases, leading to larger $p_e$. Moreover, $2\sqrt{p_e}-p_e$ is monotonically increasing with $p_e$ for $p_e<1$. Thus, the a larger $p_e$ results in a higher guesswork complexity of the order-$k$ OSD. }

    \begin{figure}  [t]
         \centering
            \input{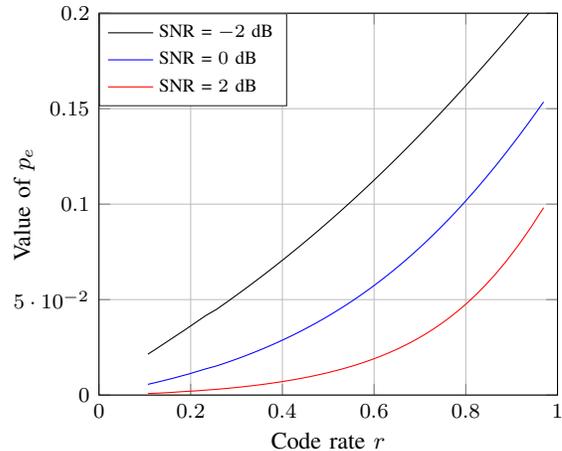}
    	\vspace{-0.5em}
        \caption{\black{The values of $p_e$ from \eqref{equ:pe_approx2} at different code rates and SNRs.}}
        \vspace{-0.25em}
    	\label{Fig::OSD:pe_rate}
        
    \end{figure}

    \black{
    Figure \ref{Fig::OSD:pe_rate} illustrates the values of $p_e$ at different code rates and SNRs. As shown, $p_e$ increases with code rate, and it decreases with SNR. These relationships indicate that for a fixed information length $k$, higher rate codes require more guesses in order-$k$ OSD at a specific SNR. Similarly, worse channel conditions (lower SNR) lead to increased guesswork complexity.
    }

    \begin{remark}[\black{The exponential growth of the guesswork complexity of the order-$k$ OSD}]
    \black{The analysis presented in Theorem \ref{The::Newbound:XnYn:ordered:1k:expapp} reveals that the guesswork complexity of an order-$k$ OSD has an exponential growth with block length $n = \frac{k}{r}$ for a fixed code rate $r$, which indicates that this complexity becomes prohibitive for long codes. This explains why the designs of OSD typically focus on shorter codes, or restrict the decoding order $m \ll k$. Nevertheless, our asymptotic analysis of the order-$k$ OSD helps understand this fundamental growth behavior.}
    
    \end{remark}

    \subsubsection{Guesswork Complexity of Order-$m$ OSD}
    \black{For order-$m$ OSD ($m<k$), the asymptotic behavior of guesswork complexity is fundamentally different from that of order-$k$ OSD, due to the restricted Hamming weight of examined error patterns. According to \eqref{equ::Newbound:XnYn:ordered:1k:order:?}, the guesswork complexity consists of two components: one from examining error patterns within Hamming weight $m$, and another from the probability mass of error patterns beyond Hamming weight $m$.}

    \black{Interestingly, for order-$m$ OSD with fixed $m$, as the information length $k$ approaches infinity, the guesswork complexity exhibits convergent behavior. This convergence can be proven by analyzing the expression in \eqref{equ::Newbound:XnYn:ordered:1k:order:?}, i.e.,
    \begin{align}  
        \mathbb{E}[G^*(\widetilde{X}_1^k | \widetilde{Y}_1^k)^\omega]  \leq \sum_{j=0}^{m} (\beta_j)^{\omega} \mathbb{P}_{E}(j) + (\beta_m)^{\omega}  \sum_{j=m+1}^{k}  \mathbb{P}_{E}(j) \label{equ::orderm::Asym1}
    \end{align}
    As $k \to \infty$, the term $\sum_{j=m+1}^{k} \mathbb{P}_{E}(j)$ represents the probability of having more than $m$ errors, which approaches 1 for large $k$ since the expected number of errors $kp_e$ grows with $k$. Meanwhile, the first term in \eqref{equ::orderm::Asym1} becomes negligible compared to the second term as $k$ grows large. Therefore, we have
    \begin{align}
    \lim_{k \to \infty} \mathbb{E}[G^*(\widetilde{X}_1^k | \widetilde{Y}_1^k)^\omega] \leq (\beta_m)^{\omega} .
    \end{align}
    This result indicates that for sufficiently large $k$, the average guesswork complexity $\omega = 1$) of order-$m$ OSD converges to $\beta_m = \sum_{i=0}^m\binom{k}{i}$, which grows polynomial at rate $ O(k^m)$ with $k$. This is fundamentally different from order-$k$ OSD where the complexity grows exponentially with $k$.
    }

\black{
    \begin{remark}
    While our analysis shows that order-$m$ OSD has polynomial complexity growth $O(k^m)$ for any fixed $m$, this does not imply that larger $m$ always leads to better decoding performance. In fact, there exists an optimal order $m_e \!=\! \lceil d_{\min}/4\!-\!1\rceil$ that achieves near-ML decoding performance, beyond which increasing $m$ provides negligible error-correction gain. Therefore, in practice, it is unnecessary to use very large $m$ or approach order-$k$ OSD. The relationship between guesswork complexity and optimal selection of $m$ will be discussed in detail in Section \ref{sec::cmp_to_MLD}.
    \end{remark}
}

    \subsubsection{Overall Complexity of Order-$m$ OSD}

\black{As shown in \eqref{equ::OSD_complexity}, the computational complexity of OSD consists of several components that must be considered together for a comprehensive view for the overall complexity behavior. }

\black{For each received block, OSD first performs reliability ordering to sort received bits by their reliability metrics, requiring $C_{\rm{sort}} = O(n\log n)$ operations. This sorting complexity exhibits quasi-linear growth with blocklength $n$.}

\black{Following the reliability ordering, OSD performs GE to transform the permuted generator matrix into systematic form. This operation requires $C_{\rm{GE}} = O(n\cdot\min(k,n-k)^2)$ operations, showing cubic growth with respect to dimension $k$.}

\black{The third component is TEP processing, where each TEP must be re-encoded with complexity $C_{\rm{re-encoding}} = O(k(n-k))$. Since order-$m$ OSD processes $O(k^m)$ TEPs, the total Reprocessing complexity is $O(k^m)\cdot O(k(n-k))$. Therefore, The overall computational complexity behavior of OSD can thus be expressed as:
\begin{align}
C_{\rm{OSD}} &= O(n\log n) + O(n\cdot\min(k,n-k)^2) \notag \\
&\hspace{0.5cm}+ O(k^m)\cdot O(k(n-k)),
\end{align}}

\black{When all terms are considered together, the overall complexity grows faster than $O(k^{m+2})$. This compounding polynomial growth effectively limits the application of OSD to only short block codes.}

\black{Particularly, we note that if considering the order-$k$ OSD, the overall complexity can be as high as $O(k^2\cdot2^{k(2\sqrt{p_e}-p_e)})$.}

\subsection{\black{Guesswork} Complexity Saturation Threshold for OSD}

    \begin{figure}  [t]
         \centering
\definecolor{mycolor1}{rgb}{0.00000,0.44700,0.74100}%
\definecolor{mycolor2}{rgb}{0.85000,0.32500,0.09800}%
\definecolor{mycolor3}{rgb}{0.92900,0.69400,0.12500}%
\definecolor{mycolor4}{rgb}{0.49400,0.18400,0.55600}%
\definecolor{mycolor5}{rgb}{0.46600,0.67400,0.18800}%
\begin{tikzpicture}

\begin{axis}[%
width=2.8in,
height=2in,
at={(1.01in,0.685in)},
scale only axis,
xmode=log,
xmin=1,
xmax=100000000,
xlabel style={at={(0.5,1ex)},font=\color{white!15!black}, font=\small},
xlabel={Upper bound of \black{achievable guesswork} complexity},
ymode=log,
ymin=1e-12,
ymax=1,
ylabel style={at={(1.5ex,0.5)},font=\color{white!15!black}, font=\small},
ylabel={\black{Absolute performance gap to MLD} $\epsilon_m - \epsilon_{\mathrm{ML}}$},
axis background/.style={fill=white},
tick label style={font=\footnotesize},
xmajorgrids,
ymajorgrids,
legend style={at={(0.8,0.2)}, anchor=south east ,  font = \scriptsize, legend cell align=left, align=left, draw=white!15!black}
]

\addplot [color=mycolor1, mark=square, mark size = 1.5pt, mark options={solid, mycolor1}, forget plot]
  table[row sep=newline]{%
1.00000000000000	0.0176459989150386
1.28233598264062	0.000146880559569368
1.29996164978894	7.61926952976303e-07
1.30038832888261	2.75431328943833e-09
1.30039334173280	7.35492824760637e-12
1.30039337385912	1.50055577818970e-14
1.30039337397929	2.38580734127695e-17
1.30039337397956	2.98745808860479e-20
1.30039337397956	2.95587515061902e-23
1.30039337397956	2.30323121175076e-26
1.30039337397956	1.39849841559177e-29
1.30039337397956	6.48673567100971e-33
1.30039337397956	2.22190679414597e-36
1.30039337397956	5.30039307939970e-40
1.30039337397956	7.86757211593337e-44
};
\addplot [color=mycolor2, mark=square, mark size = 1.5pt, mark options={solid, mycolor2}, forget plot]
  table[row sep=newline]{%
1.00000000000000	0.0931146576127747
3.97966904360879	0.00434085933763064
6.13273527507359	0.000131665351642144
6.78579541921862	2.90522882528170e-06
6.89026744777575	4.95995218624553e-08
6.90025560109033	6.81115775938945e-10
6.90087282275756	7.72486107437198e-12
6.90089882352755	7.37412809285811e-14
6.90089959916047	6.00858471092600e-16
6.90089961601383	4.22359294395774e-18
6.90089961628630	2.58202820459850e-20
6.90089961628963	1.38133620217417e-22
6.90089961628966	6.49720832087113e-25
6.90089961628966	2.69603427824960e-27
6.90089961628966	9.89249727776616e-30
};
\addplot [color=mycolor3, mark=square, mark size = 1.5pt, mark options={solid, mycolor3}, forget plot]
  table[row sep=newline]{%
1	0.539051285104712
35.4992822467015	0.179892308875231
398.162176939166	0.0421550546336359
2154.51037319497	0.00749924906513722
6919.35324720560	0.00106496864439309
15039.2194560044	0.000124952368987256
24407.4443509268	1.24166793457862e-05
32120.8866114011	1.06462732213856e-06
36833.1031944774	7.99059046389353e-08
39033.7585141938	5.31033343310696e-09
39838.1317910514	3.15398033476303e-10
40072.6604383400	1.68700186938620e-11
40128.0652017752	8.17844686124103e-13
40138.8091119471	3.61305012572586e-14
40140.5381623747	1.46126491837115e-15
};
\addplot [color=mycolor4, mark=square, mark size = 1.5pt, mark options={solid, mycolor4}, forget plot]
  table[row sep=newline]{%
1	0.264864270852973
13.7134850009427	0.0379363359483657
56.5056719506992	0.00364100359338618
119.480470101907	0.000259181277543366
169.911963086295	1.45102754154542e-05
194.757965721279	6.63636546197245e-07
202.901789561577	2.54601113073419e-08
204.776393930153	8.35396657302990e-10
205.091630018377	2.37924886311490e-11
205.131532559043	5.95001437264920e-13
205.135424294401	1.31866655398907e-14
205.135722249751	2.60942455263035e-16
205.135740429230	4.63896620967553e-18
205.135741324218	7.44699090653748e-20
205.135741360136	1.08410436031565e-21
};
\addplot [color=mycolor5, mark=square, mark size = 1.5pt, mark options={solid, mycolor5}, forget plot]
  table[row sep=newline]{%
0.999999999999997	0.825151473470881
67.0121178776705	0.516896999959192
1700.40663774872	0.248570288699076
22122.9415572648	0.0948275263554548
172100.260690525	0.0296072194802526
883858.290711310	0.00776061983275542
3215926.10257828	0.00174263225119569
8751774.55412054	0.000340650401861064
18626390.7332306	5.87259865583299e-05
32244964.4694705	9.02327305289813e-06
47101712.3585257	1.24656361875455e-06
60162803.3960657	1.55980799921142e-07
69560122.9825336	1.77884759669320e-08
75165917.2517565	1.85880134950819e-09
77969272.6541470	1.78794330861422e-10
};

\addplot [dashed, color=gray!100]
  table[row sep=newline]{%
1.11480099937355	1.84206123785273e-10
1.30038832888261	2.75431328943833e-09
1.67426527735596	2.38913390112786e-08
2.40386077014343	1.47046504704828e-07
3.84481887745743	7.13540209316213e-07
6.78579541921862	2.90522882528170e-06
13.0016599945741	1.03285533809464e-05
26.4602806660752	3.29359355414943e-05
55.8368657088871	9.59935618437146e-05
119.480470101907	0.000259181277543366
254.657031632267	0.000654602381106904
533.638620805226	0.00155737054416066
1089.60901866223	0.00350713345396937
2154.51037319497	0.00749924906513722
4107.34885516096	0.0152516414252060
7523.27300941905	0.0295136924378166
13200.0969267112	0.0543022098166196
22122.9415572648	0.0948275263554548
53586.5773983472	0.244516402956687
105205.425502093	0.493103714683269
};

\addplot [dashed, color=gray!100]
  table[row sep=newline]{%
%1.30039337385912	1.50055577818970e-14
%1.67438280649776	4.29027383013935e-13
%2.40547189685698	7.04517068584474e-12
3.86025622627013	7.94216578362100e-11
6.90025560109033	6.81115775938945e-10
13.7028471543083	4.74079282429352e-09
30.1712988834774	2.79687749846462e-08
73.3448126176106	1.44168298224060e-07
194.757965721279	6.63636546197245e-07
554.190381903776	2.77207250261200e-06
1647.94138731021	1.06317469592667e-05
4989.61209429543	3.77614300007262e-05
15039.2194560044	0.000124952368987256
126145.665907023	0.00112152108615392
883858.290711310	0.00776061983275542
4861385.15369945	0.0407249176697770
19918790.7059706	0.155923195730730
};

\node[rotate=-90] at (axis cs: 2.5,4e-11) {\footnotesize $r = \sfrac{1}{8}$};
\node[rotate=-90] at (axis cs: 13,4e-11) {\footnotesize $r = \sfrac{1}{4}$};
\node[rotate=-90] at (axis cs: 400,4e-11) {\footnotesize $r = \sfrac{3}{8}$};
\node[rotate=-90] at (axis cs: 8e4,4e-11) {\footnotesize $r = \sfrac{1}{2}$};
\node[rotate=-90] at (axis cs: 4e7,4e-11) {\footnotesize $r = \sfrac{5}{8}$};
\node[rotate=28, text = gray!100] at (axis cs:7e2,5e-4) {\footnotesize $m = 3$};
\node[rotate=26, text = gray!100] at (axis cs:3e3,4.5e-6) {\footnotesize $m = 5$};
\end{axis}
\end{tikzpicture}%
    	\vspace{-0.5em}
        \caption{The \black{absolute} performance gap to MLD, $\epsilon_m - \epsilon_{\mathrm{ML}}$, versus the average \black{achievable guesswork} complexity \black{of OSD} evaluated by \eqref{equ::Newbound:XnYn:ordered:1k:order:?} for fixed block length $n=128$ and various coding rate $r$. The SNR is 2 dB.}
        \vspace{-0.25em}
    	\label{Fig::comp_error_OSD_n}
        
    \end{figure}
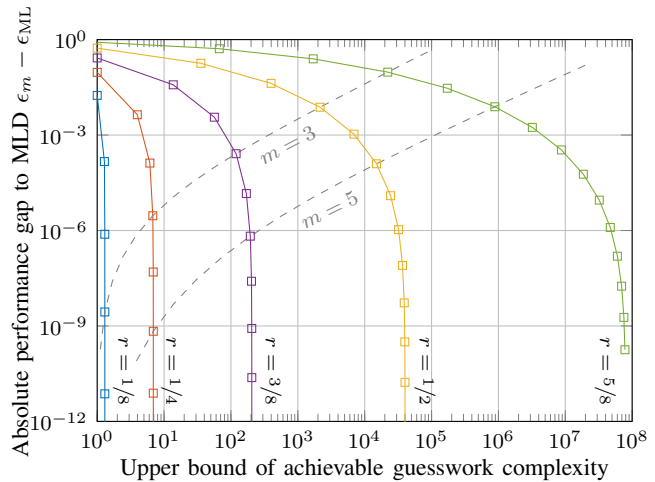

        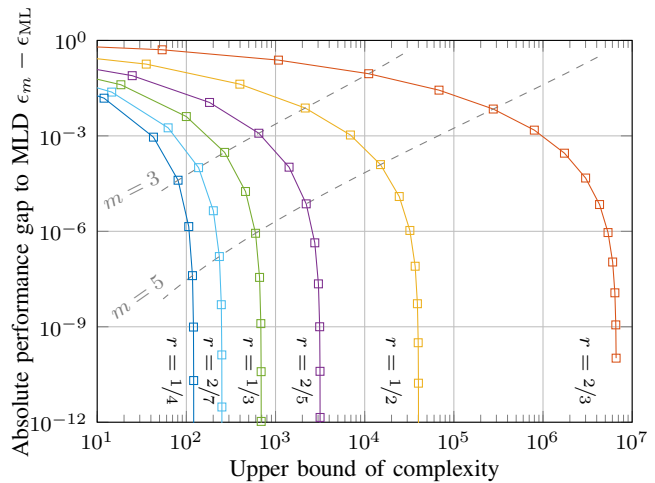
\begin{figure}  [t]
         \centering
\definecolor{mycolor1}{rgb}{0.85000,0.32500,0.09800}%
\definecolor{mycolor2}{rgb}{0.92900,0.69400,0.12500}%
\definecolor{mycolor3}{rgb}{0.49400,0.18400,0.55600}%
\definecolor{mycolor4}{rgb}{0.46600,0.67400,0.18800}%
\definecolor{mycolor5}{rgb}{0.30100,0.74500,0.93300}%
\definecolor{mycolor6}{rgb}{0.00000,0.44700,0.74100}%
\begin{tikzpicture}

\begin{axis}[%
width=2.8in,
height=2in,
at={(1.01in,0.685in)},
scale only axis,
xmode=log,
xmin=10,
xmax=10000000,
xlabel style={at={(0.5,1ex)},font=\color{white!15!black}, font=\small},
xlabel={Upper bound of complexity},
ymode=log,
ymin=1e-12,
ymax=1,
ylabel style={at={(1.5ex,0.5)},font=\color{white!15!black}, font=\small},
ylabel={\black{Absolute performance gap to MLD} $\epsilon_m - \epsilon_{\mathrm{ML}}$},
axis background/.style={fill=white},
tick label style={font=\footnotesize},
xmajorgrids,
ymajorgrids,
legend style={at={(0.8,0.2)}, anchor=south east ,  font = \scriptsize, legend cell align=left, align=left, draw=white!15!black}
]

\addplot [color=mycolor1, mark=square, mark size = 1.5pt, mark options={solid, mycolor1}, forget plot]
  table[row sep=newline]{%
1.00000000000000	0.820550932049008
53.5152596511365	0.508107377025071
1077.85973173368	0.240355506152503
11092.0315400716	0.0898150115707281
68158.3343318345	0.0273590970641890
276758.098206908	0.00696966130724553
799304.049891696	0.00151514356805756
1740535.76757371	0.000285620442176931
3004739.07713009	4.72941479856003e-05
4307247.55665055	6.95117798414903e-06
5360164.83266780	9.14737175281620e-07
6040359.53768482	1.08556449529299e-07
6396882.22534050	1.16890983919984e-08
6550440.26055837	1.14795296398352e-09
6605376.35316960	1.03280048775157e-10
};

\addplot [color=mycolor2, mark=square, mark size = 1.5pt, mark options={solid, mycolor2}, forget plot]
  table[row sep=crcr]{%
0.999999999999999	0.539051285104712\\
35.4992822467015	0.179892308875231\\
398.162176939166	0.0421550546336359\\
2154.51037319497	0.00749924906513722\\
6919.35324720560	0.00106496864439309\\
15039.2194560044	0.000124952368987256\\
24407.4443509268	1.24166793457862e-05\\
32120.8866114011	1.06462732213856e-06\\
36833.1031944774	7.99059046389353e-08\\
39033.7585141938	5.31033343310696e-09\\
39838.1317910514	3.15398033476303e-10\\
40072.6604383400	1.68700186938620e-11\\
40128.0652017752	8.17844686124103e-13\\
40138.8091119471	3.61305012572586e-14\\
40140.5381623747	1.46126491837115e-15\\
};
\addplot [color=mycolor3, mark=square, mark size = 1.5pt, mark options={solid, mycolor3}, forget plot]
  table[row sep=newline]{%
1.00000000000000	0.370299614928351
24.6991753554145	0.0780012549051522
181.949705244201	0.0112207991995764
649.453083095351	0.00121084111474610
1418.79246721827	0.000103670505757791
2209.22948241461	7.30796544933116e-06
2757.13957334405	4.35359441282793e-07
3027.59190760900	2.23463223158135e-08
3126.50042554542	1.00304234205918e-09
3154.12479793601	3.98364628955024e-11
3160.15895503769	1.41319842636199e-12
3161.20980342641	4.51301239790940e-14
3161.35802044314	1.30583789690200e-15
3161.37517505122	3.44230249749817e-17
3161.37682238918	8.30559971754688e-19
};
\addplot [color=mycolor4, mark=square, mark size = 1.5pt, mark options={solid, mycolor4}, forget plot]
  table[row sep=newline]{%
1	0.272376312624863
18.4320840079912	0.0404371846742575
99.9534483112943	0.00404804475360607
268.611184925538	0.000302375355716755
460.733228939427	1.78725827824090e-05
597.002950834898	8.68431237750258e-07
662.113034036681	3.56269639250201e-08
684.245080898704	1.25850048598267e-09
689.815412165371	3.88589625570518e-11
690.885610707722	1.06130061067643e-12
691.046369323108	2.58846789101934e-14
691.065617061159	5.68209774610311e-16
691.067483184055	1.12998267102931e-17
691.067631628284	2.04702102602883e-19
691.067641424446	3.39386735310025e-21
};
\addplot [color=mycolor5, mark=square, mark size = 1.5pt, mark options={solid, mycolor5}, forget plot]
  table[row sep=newline]{%
1.00000000000000	0.211439612824735
14.5321352207831	0.0237719540132754
62.4563945115464	0.00178958493520511
137.017661251932	0.000100235715989977
200.705029534780	4.43609641724041e-06
234.528099901186	1.61259651572991e-07
246.618440361771	4.94675802415949e-09
249.691446544285	1.30616266412699e-10
250.269575739179	3.01392735083990e-12
250.352581060097	6.15038173905793e-14
250.361897241041	1.12065513051197e-15
250.362730555469	1.83763850570160e-17
250.362790907463	2.72967766341901e-19
250.362794493402	3.69337129542972e-21
250.362794670151	4.57335201229962e-23
};

\addplot [color=mycolor6, mark=square, mark size = 1.5pt, mark options={solid, mycolor6}, forget plot]
  table[row sep=newline]{%
0.999999999999998	0.170874568739086
11.9359723993015	0.0152822029094245
42.7448934647013	0.000911205454628772
80.7093575263545	4.03513132301893e-05
106.347613521299	1.41065041660257e-06
117.103134550491	4.04860785918517e-08
120.138552705713	9.80220878118857e-10
120.747481786937	2.04235800329029e-11
120.837879929569	3.71821507706910e-13
120.848120111225	5.98582959281449e-15
120.849026804077	8.60359510561550e-17
120.849090780048	1.11282537841868e-18
120.849094434805	1.30381596086281e-20
120.849094606085	1.39139084100270e-22
120.849094612744	1.35884043830959e-24
};

\addplot [dashed, color=gray!100]
  table[row sep=newline]{%
26834.3014727344	0.373687142741448
17529.5176126765	0.184594844575427
11092.0315400716	0.0898150115707281
7070.84540529985	0.0450270596241496
4617.65757943681	0.0236235087468717
3106.67217413712	0.0130071085346874
2154.51037319497	0.00749924906513722
1537.87871487236	0.00450962892617996
1127.25064081759	0.00281633209426688
846.395106607989	0.00181914939794311
649.453083095351	0.00121084111474610
508.154219080547	0.000827794549380008
404.635607965903	0.000579612138418819
327.338087945526	0.000414626199394754
268.611184925538	0.000302375355716755
223.283618708832	0.000224384995179240
187.790825422397	0.000169157038512912
159.630802946444	0.000129364365754254
137.017661251932	0.000100235715989977
118.656713825758	7.86018844519544e-05
103.595779313587	6.23190312546024e-05
91.1251727254544	4.99123332145348e-05
80.7093575263545	4.03513132301893e-05
71.9395310570174	3.29055233273817e-05
64.5002673087039	2.70502227682152e-05
58.1457397011293	2.24037305030267e-05
52.6825627696492	1.86851855922418e-05
};

\addplot [dashed, color=gray!100]
  table[row sep=newline]{%
4085210.60201905	0.297182179020317
1761212.74229236	0.0883176842490906
689188.948344794	0.0241505348125528
276758.098206908	0.00696966130724553
119079.326633125	0.00221111007518334
55458.7387567979	0.000775397524737711
27887.6983160511	0.000298559334448658
15039.2194560044	0.000124952368987256
8631.02069738118	5.62641212413228e-05
5232.55382576893	2.70069599134447e-05
3328.79291005673	1.37084969311235e-05
2209.22948241461	7.30796544933116e-06
1521.84189218430	4.06787197210655e-06
1083.33936026989	2.35263880811056e-06
793.918797770426	1.40777201901420e-06
597.002950834898	8.68431237750258e-07
459.332942435760	5.50581627194048e-07
360.705872919242	3.57792828297427e-07
288.479833725034	2.37770989867362e-07
234.528099901186	1.61259651572991e-07
193.496746740803	1.11419955896633e-07
161.778717177754	7.83057123177959e-08
136.893549025468	5.59008311732297e-08
117.103134550491	4.04860785918517e-08
101.168307601943	2.97154146306793e-08
88.1914233682170	2.20811866237609e-08
77.5125868098945	1.65976563379578e-08
68.6400228444603	1.26099057554977e-08
61.2025795467455	9.67626232859683e-09
54.9168284448790	7.49468148880727e-09
};

\node[rotate=-90] at (axis cs: 3e6,4e-11) {\footnotesize $r = \sfrac{2}{3}$};
\node[rotate=-90] at (axis cs: 2e4,4e-11) {\footnotesize $r = \sfrac{1}{2}$};
\node[rotate=-90] at (axis cs: 2e3,4e-11) {\footnotesize $r = \sfrac{2}{5}$};
\node[rotate=-90] at (axis cs: 5e2,4e-11) {\footnotesize $r = \sfrac{1}{3}$};
\node[rotate=-90] at (axis cs: 1.8e2,4e-11) {\footnotesize $r = \sfrac{2}{7}$};
\node[rotate=-90] at (axis cs: 70,4e-11) {\footnotesize $r = \sfrac{1}{4}$};

\node[rotate=27, text = gray!100] at (axis cs:24,2e-5) {\footnotesize $m = 3$};
\node[rotate=35, text = gray!100] at (axis cs:28,9e-9) {\footnotesize $m = 5$};
\end{axis}
\end{tikzpicture}%
    	\vspace{-0.5em}
        \caption{The performance gap to MLD, $\epsilon_m - \epsilon_{\mathrm{ML}}$, versus the average OSD decoding complexity evaluated by \eqref{equ::Newbound:XnYn:ordered:1k:order:?} for fixed information length $k = 64$ and various coding rate $r$. The SNR is 2 dB.}
        \vspace{-0.25em}
    	\label{Fig::comp_error_OSD_k}
        
    \end{figure}

\black{The} bound described in Theorem \ref{The::Newbound:XnYn:ordered:qto1} \black{reveals that} $\mathbb{P}_{E}(j)$ given in \eqref{equ::E1k} describes the distribution of the number of errors in the most reliable bits in OSD, representing the probability that there are $j$ transmission errors in the $k$ most reliable bits. On the other hand, the BLER performance of an order-$m$ OSD algorithm is approximately given by \cite{yue2021revisit}
\begin{equation} \label{equ::OSD:error:rate}
    \epsilon_m \approx \epsilon_{\mathrm{ML}} + 1 - \sum_{j=0}^{m} \mathbb{P}_{E}(j) = \epsilon_{\mathrm{ML}} + \sum_{j=m+1}^{k} \mathbb{P}_{E}(j) , 
\end{equation}
where $\epsilon_{\mathrm{ML}}$ is the error rate of the MLD of code $\mathcal{C}$. The term $\epsilon_m - \epsilon_{\mathrm{ML}}$ represents the \black{absolute} performance gap between an order-$m$ OSD and the MLD. From \eqref{equ::Newbound:XnYn:ordered:1k:order:?} and \eqref{equ::OSD:error:rate}, one can infer a direct correlation between the complexity and error rate of an OSD decoder. \black{However, since $\epsilon_{\mathrm{ML}}$ can be very small (especially at high SNRs or low coderates), it is also insightful to consider the relative performance gap $(\epsilon_m - \epsilon_{\mathrm{ML}})/\epsilon_{\mathrm{ML}}$, which indicates how significant the gap is compared to MLD performance. }

\begin{figure}  [t]
         \centering
\definecolor{mycolor1}{rgb}{0.00000,0.44700,0.74100}%
\definecolor{mycolor2}{rgb}{0.85000,0.32500,0.09800}%
\definecolor{mycolor3}{rgb}{0.92900,0.69400,0.12500}%
\definecolor{mycolor4}{rgb}{0.49400,0.18400,0.55600}%
\definecolor{mycolor5}{rgb}{0.46600,0.67400,0.18800}%
\begin{tikzpicture}

\begin{axis}[%
width=2.8in,
height=2in,
at={(1.01in,0.685in)},
scale only axis,
xmode=log,
xmin=1,
xmax=100000000,
xlabel style={at={(0.5,1ex)},font=\color{white!15!black}, font=\small},
xlabel={\black{Upper bound of achievable guesswork complexity}},
ymode=log,
ymin=1e-12,
ymax=1e12,
ylabel style={at={(1.5ex,0.5)},font=\color{white!15!black}, font=\small},
ylabel={\black{Relative performance gap $(\epsilon_m - \epsilon_{\mathrm{ML}})/\epsilon_{\mathrm{ML}}$}},
axis background/.style={fill=white},
tick label style={font=\footnotesize},
xmajorgrids,
ymajorgrids,
grid style={dashed, gray!50},
legend style={at={(0.8,0.2)}, anchor=south east ,  font = \scriptsize, legend cell align=left, align=left, draw=white!15!black}
]

\addplot [color=mycolor1, mark=square, mark size = 1.5pt, mark options={solid, mycolor1}, forget plot]
table[
  row sep=newline,
  x index=0,              % The first column is x-values
  y expr=\thisrowno{1}/1.2736e-15  % The second column divided by A
]
{%
1.00000000000000	0.0176459989150386
1.28233598264062	0.000146880559569368
1.29996164978894	7.61926952976303e-07
1.30038832888261	2.75431328943833e-09
1.30039334173280	7.35492824760637e-12
1.30039337385912	1.50055577818970e-14
1.30039337397929	2.38580734127695e-17
1.30039337397956	2.98745808860479e-20
1.30039337397956	2.95587515061902e-23
1.30039337397956	2.30323121175076e-26
1.30039337397956	1.39849841559177e-29
1.30039337397956	6.48673567100971e-33
1.30039337397956	2.22190679414597e-36
1.30039337397956	5.30039307939970e-40
1.30039337397956	7.86757211593337e-44
};
\addplot [color=mycolor2, mark=square, mark size = 1.5pt, mark options={solid, mycolor2}, forget plot]
table[
  row sep=newline,
  x index=0,              % The first column is x-values
  y expr=\thisrowno{1}/5.4689e-10  % The second column divided by A
]
{%
1.00000000000000	0.0931146576127747
3.97966904360879	0.00434085933763064
6.13273527507359	0.000131665351642144
6.78579541921862	2.90522882528170e-06
6.89026744777575	4.95995218624553e-08
6.90025560109033	6.81115775938945e-10
6.90087282275756	7.72486107437198e-12
6.90089882352755	7.37412809285811e-14
6.90089959916047	6.00858471092600e-16
6.90089961601383	4.22359294395774e-18
6.90089961628630	2.58202820459850e-20
6.90089961628963	1.38133620217417e-22
6.90089961628966	6.49720832087113e-25
6.90089961628966	2.69603427824960e-27
6.90089961628966	9.89249727776616e-30
};

\addplot [color=mycolor4, mark=square, mark size = 1.5pt, mark options={solid, mycolor4}, forget plot]
table[
  row sep=newline,
  x index=0,              % The first column is x-values
  y expr=\thisrowno{1}/9.3919e-06  % The second column divided by A
]
{%
1	0.264864270852973
13.7134850009427	0.0379363359483657
56.5056719506992	0.00364100359338618
119.480470101907	0.000259181277543366
169.911963086295	1.45102754154542e-05
194.757965721279	6.63636546197245e-07
202.901789561577	2.54601113073419e-08
204.776393930153	8.35396657302990e-10
205.091630018377	2.37924886311490e-11
205.131532559043	5.95001437264920e-13
205.135424294401	1.31866655398907e-14
205.135722249751	2.60942455263035e-16
205.135740429230	4.63896620967553e-18
205.135741324218	7.44699090653748e-20
205.135741360136	1.08410436031565e-21
};

\addplot [color=mycolor3, mark=square, mark size = 1.5pt, mark options={solid, mycolor3}, forget plot]
table[
  row sep=newline,
  x index=0,              % The first column is x-values
  y expr=\thisrowno{1}/0.0069  % The second column divided by A
]
{%
1	0.539051285104712
35.4992822467015	0.179892308875231
398.162176939166	0.0421550546336359
2154.51037319497	0.00749924906513722
6919.35324720560	0.00106496864439309
15039.2194560044	0.000124952368987256
24407.4443509268	1.24166793457862e-05
32120.8866114011	1.06462732213856e-06
36833.1031944774	7.99059046389353e-08
39033.7585141938	5.31033343310696e-09
39838.1317910514	3.15398033476303e-10
40072.6604383400	1.68700186938620e-11
40128.0652017752	8.17844686124103e-13
40138.8091119471	3.61305012572586e-14
40140.5381623747	1.46126491837115e-15
};

\addplot [color=mycolor5, mark=square, mark size = 1.5pt, mark options={solid, mycolor5}, forget plot]
table[
  row sep=newline,
  x index=0,              % The first column is x-values
  y expr=\thisrowno{1}/0.2590  % The second column divided by A
]
{%
0.999999999999997	0.825151473470881
67.0121178776705	0.516896999959192
1700.40663774872	0.248570288699076
22122.9415572648	0.0948275263554548
172100.260690525	0.0296072194802526
883858.290711310	0.00776061983275542
3215926.10257828	0.00174263225119569
8751774.55412054	0.000340650401861064
18626390.7332306	5.87259865583299e-05
32244964.4694705	9.02327305289813e-06
47101712.3585257	1.24656361875455e-06
60162803.3960657	1.55980799921142e-07
69560122.9825336	1.77884759669320e-08
75165917.2517565	1.85880134950819e-09
77969272.6541470	1.78794330861422e-10
};

\node[rotate=-90] at (axis cs: 2.5,4e-10) {\footnotesize $r = \sfrac{1}{8}$};
\node[rotate=-90] at (axis cs: 13,4e-10) {\footnotesize $r = \sfrac{1}{4}$};
\node[rotate=-90] at (axis cs: 400,4e-10) {\footnotesize $r = \sfrac{3}{8}$};
\node[rotate=-90] at (axis cs: 8e4,4e-10) {\footnotesize $r = \sfrac{1}{2}$};
\node[rotate=-90] at (axis cs: 3e7,4e-10) {\footnotesize $r = \sfrac{5}{8}$};

\node[draw=mycolor1, circle, minimum size=7pt, inner sep=0pt] at (axis cs:1.30039337397956, 1.1533e+11) {};

\node[draw=mycolor2, circle, minimum size=7pt, inner sep=0pt] at (axis cs:6.13273527507359, 2.4075e+05) {};

\node[draw=mycolor4, circle, minimum size=7pt, inner sep=0pt] at (axis cs:169.911963086295, 1.545) {};

\node[draw=mycolor3, circle, minimum size=7pt, inner sep=0pt] at (axis cs:36833.1031944774, 1.1581e-05) {};

\node[draw=mycolor5, circle, minimum size=7pt, inner sep=0pt] at (axis cs:69560122.9825336, 6.8681e-08) {};

\end{axis}
\end{tikzpicture}%
    	\vspace{-0.5em}
        \caption{\black{The relative performance gap to MLD, $(\epsilon_m - \epsilon_{\mathrm{ML}})/\epsilon_{\mathrm{ML}}$, versus the average achievable guesswork complexity of OSD evaluated by \eqref{equ::Newbound:XnYn:ordered:1k:order:?} for fixed block length $n=128$ and various coding rate $r$. The SNR is 2 dB.}}
        \vspace{-0.25em}
    	\label{Fig::relative_error_OSD_n}
        
    \end{figure}
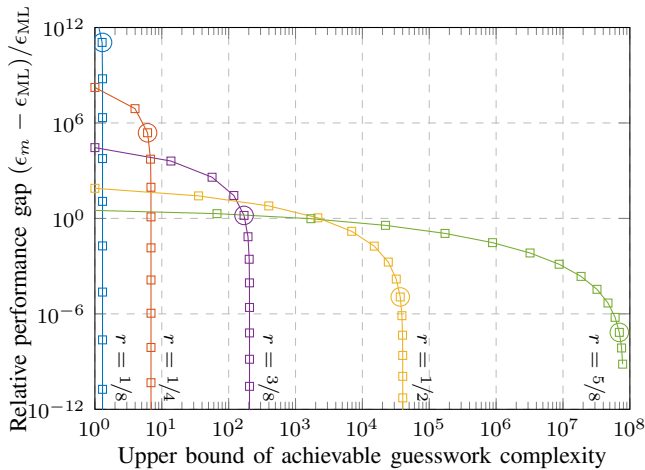

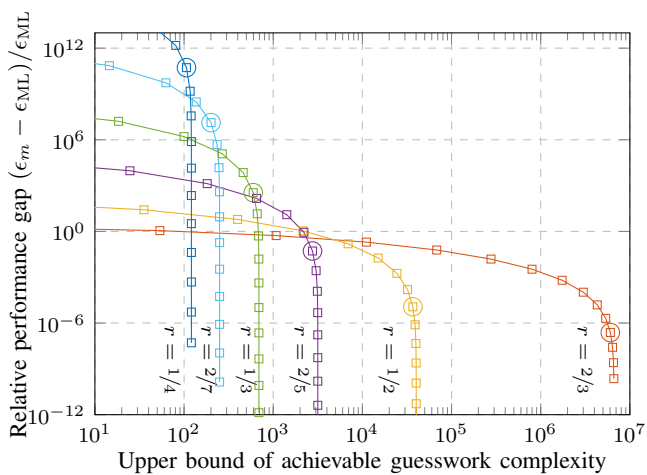
\begin{figure}  [t]
         \centering
\definecolor{mycolor1}{rgb}{0.85000,0.32500,0.09800}%
\definecolor{mycolor2}{rgb}{0.92900,0.69400,0.12500}%
\definecolor{mycolor3}{rgb}{0.49400,0.18400,0.55600}%
\definecolor{mycolor4}{rgb}{0.46600,0.67400,0.18800}%
\definecolor{mycolor5}{rgb}{0.30100,0.74500,0.93300}%
\definecolor{mycolor6}{rgb}{0.00000,0.44700,0.74100}%
\begin{tikzpicture}

\begin{axis}[%
width=2.8in,
height=2in,
at={(1.01in,0.685in)},
scale only axis,
xmode=log,
xmin=10,
xmax=10000000,
xlabel style={at={(0.5,1ex)},font=\color{white!15!black}, font=\small},
xlabel={\black{Upper bound of achievable guesswork complexity}},
ymode=log,
ymin=1e-12,
ymax=1e13,
ylabel style={at={(1.5ex,0.5)},font=\color{white!15!black}, font=\small},
ylabel={\black{Relative performance gap $(\epsilon_m - \epsilon_{\mathrm{ML}})/\epsilon_{\mathrm{ML}}$}},
axis background/.style={fill=white},
tick label style={font=\footnotesize},
xmajorgrids,
ymajorgrids,
grid style={dashed, gray!50},
legend style={at={(0.8,0.2)}, anchor=south east ,  font = \scriptsize, legend cell align=left, align=left, draw=white!15!black}
]

\addplot [color=mycolor1, mark=square, mark size = 1.5pt, mark options={solid, mycolor1}, forget plot]
  table[row sep=newline]{%
1.00000000000000 1.8193812461617738
53.5152596511365 1.1266017006320424
1077.85973173368 0.5329389714246407
11092.0315400716 0.1991685400237209
68158.3343318345 0.0606632307188226
276758.098206908 0.0154537279318726
799304.049891696 0.0033594978426553
1740535.76757371 0.0006332825991506
3004739.07713009 0.0001048872681499
4307247.55665055 0.0000154127561363
5360164.83266780 0.0000020282199230
6040359.53768482 0.0000002407459635
6396882.22534050 0.0000000259182669
6550440.26055837 0.0000000025453724
6605376.35316960 0.0000000002290246
};

\addplot [color=mycolor2, mark=square, mark size = 1.5pt, mark options={solid, mycolor2}, forget plot]
  table[row sep=newline]{%
0.999999999999999 78.1233746527120
35.4992822467015 26.0713490544537
398.162176939166 6.1094282078893
2154.51037319497 1.0868476906577
6919.35324720560 0.1543433687526
15039.2194560044 0.0181090390561
24407.4443509268 0.0017996636732
32120.8866114011 0.0001542939596
36833.1031944774 0.0000115806674
39033.7585141938 0.0000007696136
39838.1317910514 0.0000000457098
40072.6604383400 0.0000000024449
40128.0652017752 0.0000000001185
40138.8091119471 0.0000000000052
40140.5381623747 0.0000000000002
};
\addplot [color=mycolor3, mark=square, mark size = 1.5pt, mark options={solid, mycolor3}, forget plot]
  table[row sep=newline]{%
1	44414.8123407279
24.6991753554145	9355.69727671455
181.949705244201	1345.85527683739
649.453083095351	145.231803431099
1418.79246721827	12.4345418490148
2209.22948241461	0.876538621535888
2757.13957334405	0.0522182770540574
3027.59190760900	0.00268028286325471
3126.50042554542	0.000120307814527387
3154.12479793601	4.77810117130275e-06
3160.15895503769	1.69503127674666e-07
3161.20980342641	5.41303827127415e-09
3161.35802044314	1.56625993655260e-10
3161.37517505122	4.12879768929770e-12
3161.37682238918	9.96197775964267e-14
};
\addplot [color=mycolor4, mark=square, mark size = 1.5pt, mark options={solid, mycolor4}, forget plot]
  table[row sep=newline]{%
1 108780826.959888
18.4320840079912 16149680.3683284
99.9534483112943 1616695.85590721
268.611184925538 120761.753950539
460.733228939427 7137.89799209593
597.002950834898 346.831438056735
662.113034036681 14.2285889712129
684.245080898704 0.502616113256388
689.815412165371 0.0155193747981356
690.885610707722 0.000423859024192831
691.046369323108 1.03377446823729e-05
691.065617061159 2.26929899201370e-07
691.067483184055 4.51289057482052e-09
691.067631628284 8.17533058839742e-11
691.067641424446 1.35543246659222e-12
};
\addplot [color=mycolor5, mark=square, mark size = 1.5pt, mark options={solid, mycolor5}, forget plot]
  table[row sep=newline]{%
1	633716807507.073
14.5321352207831	71248176272.3675
62.4563945115464	5363659329.25254
137.017661251932	300421747.309987
200.705029534780	13295658.3762638
234.528099901186	483319.800908110
246.618440361771	14826.1891927454
249.691446544285	391.476896186720
250.269575739179	9.03320051203327
250.352581060097	0.184336332655715
250.361897241041	0.00335877455570799
250.362730555469	5.50768321804765e-05
250.362790907463	8.18126079250415e-07
250.362794493402	1.10695977684092e-08
250.362794670151	1.37070343542623e-10
};

\addplot [color=mycolor6, mark=square, mark size = 1.5pt, mark options={solid, mycolor6}, forget plot]
  table[row sep=newline]{%
0.999999999999998	6.39979658198824e+15
11.9359723993015	572367150165712
42.7448934647013	34127545117182.5
80.7093575263545	1511285139707.46
106.347613521299	52833348936.4258
117.103134550491	1516332531.53003
120.138552705713	36712392.4389085
120.747481786937	764928.091119959
120.837879929569	13925.8991650528
120.848120111225	224.188374262715
120.849026804077	3.22232026427547
120.849090780048	0.0416788531242951
120.849094434805	0.000488320584592813
120.849094606085	5.21120165169551e-06
120.849094612744	5.08929003112206e-08
};

\node[rotate=-90] at (axis cs: 3e6,4e-9) {\footnotesize $r = \sfrac{2}{3}$};
\node[rotate=-90] at (axis cs: 2e4,4e-9) {\footnotesize $r = \sfrac{1}{2}$};
\node[rotate=-90] at (axis cs: 2e3,4e-9) {\footnotesize $r = \sfrac{2}{5}$};
\node[rotate=-90] at (axis cs: 5e2,4e-9) {\footnotesize $r = \sfrac{1}{3}$};
\node[rotate=-90] at (axis cs: 1.8e2,4e-9) {\footnotesize $r = \sfrac{2}{7}$};
\node[rotate=-90] at (axis cs: 70,4e-9) {\footnotesize $r = \sfrac{1}{4}$};

\node[draw=mycolor1, circle, minimum size=7pt, inner sep=0pt] at (axis cs:6040359.53768482, 0.0000002407459635) {};

\node[draw=mycolor2, circle, minimum size=7pt, inner sep=0pt] at (axis cs:36833.1031944774, 0.0000115806674) {};

\node[draw=mycolor3, circle, minimum size=7pt, inner sep=0pt] at (axis cs:2757.13957334405,	0.0522182770540574) {};

\node[draw=mycolor4, circle, minimum size=7pt, inner sep=0pt] at (axis cs:597.002950834898, 346.831438056735) {};

\node[draw=mycolor5, circle, minimum size=7pt, inner sep=0pt] at (axis cs:200.705029534780,	13295658.3762638) {};

\node[draw=mycolor6, circle, minimum size=7pt, inner sep=0pt] at (axis cs:106.347613521299,	52833348936.4258) {};

\end{axis}
\end{tikzpicture}%
    	\vspace{-0.5em}
        \caption{\black{The relative performance gap to MLD, $(\epsilon_m - \epsilon_{\mathrm{ML}})/\epsilon_{\mathrm{ML}}$, versus the average achievable guesswork complexity of OSD evaluated by \eqref{equ::Newbound:XnYn:ordered:1k:order:?} for fixed information length $k=64$ and various coding rate $r$. The SNR is 2 dB.}}
        \vspace{-0.25em}
    	\label{Fig::relative_error_OSD_k}
        
    \end{figure}

We investigate the relationship between the average \black{guesswork} complexity and the \black{absolute } gap $\epsilon_m - \epsilon_{\mathrm{ML}}$  at various code rates. The results of fixing $n=128$ and $k = 64$ are depicted in Figs. \ref{Fig::comp_error_OSD_n} and \ref{Fig::comp_error_OSD_k}, respectively, at the SNR of  2dB. For each specified code rate, the OSD order $m$ \black{is increased from} $0$ to a maximum of 15. The points of $m = 3$ and $m=5$ are highlighted with dashed grey curves, which are evaluated incrementally from low to high rates. \black{Figures \ref{Fig::relative_error_OSD_n} and \ref{Fig::relative_error_OSD_k} also show the relative performance gaps versus guesswork complexity for various code rates, with fixing $n=128$ and $k = 64$, respectively}.

It was proved that the OSD of order $m_e = \lceil{d_{\min}/4} -1\rceil$ approximates MLD at high SNR (taking $\sigma^2 \to 0$) \cite{Fossorier1995OSD}, where $d_{\min}$ is the minimum Hamming distance of the code $\mathcal{C}$. This approximation is obtained by assuming the performance gap to MLD is less than the performance of MLD itself, i.e., $\epsilon_m - \epsilon_{\mathrm{ML}} < \epsilon_{\mathrm{ML}}$. A prevalent understanding based on $m_e$, therefore, is that OSD is not preferred for low-rate codes with large $d_{\min}$, which necessitates a high decoding order resulting in a large \black{list of TEPs}, i.e., $\xi_{\max} = \sum_{i=0}^{m} \binom{k}{i}$. However, what we observed in Figs. \ref{Fig::comp_error_OSD_n}-\ref{Fig::relative_error_OSD_k} contradicts this common belief to some extent. That is, there exists a \black{threshold} decoder order $m_s$, such that if the decoder order $m$ exceeds $m_s$, the increase in guesswork complexity will be negligible. For low-rate codes, it is possible that $m_s$ is much smaller than $\lceil{d_{\min}/4} -1\rceil$ for a certain range of SNRs.

This can be evidenced by comparing the coefficient of $e^{-kp_e}$ in \eqref{equ::Newbound:XnYn:ordered:1k:order:app:w1}, i.e.,
\begin{equation} \label{equ::Newbound:XnYn:ordered:1k:order:app:w1:coeff}
     \sum_{j=0}^{m} \left(\frac{k^j}{j!}\right)^{2}p_e^{j} + \frac{k^m}{m!}\frac{(kp_e) ^{m+1}}{(m+1)!},
\end{equation}
to \eqref{equ::Newbound:XnYn:ordered:1k:app2}. Specifically, \eqref{equ::Newbound:XnYn:ordered:1k:order:app:w1:coeff} can be regarded a truncated series of $I_0(2k\sqrt{p_e})$, \black{as the second term} $\frac{k^m}{m!}\frac{(kp_e) ^{m+1}}{(m+1)!}$ quickly vanishes to 0 \black{for increasing $m$}. Then, the difference between \eqref{equ::Newbound:XnYn:ordered:1k:order:app:w1:coeff} and \eqref{equ::Newbound:XnYn:ordered:1k:app2} is characterized the summation
\begin{equation} \label{equ::gaptoI0}
    \sum_{j = m+1}^{k}\left( \frac{(k\sqrt{p_e})^{j}}{j!}\right)^2 = \sum_{j = m+1}^{k}s(j)
\end{equation}
\black{This difference \eqref{equ::gaptoI0} is substantial for small values of $m$.} However, as $m$ increases, \black{each term} $s(j)$ tends towards zero, \black{causing the summation \eqref{equ::gaptoI0} to approach zero.} Furthermore, \black{for any $m > k\sqrt{p_e}$, the ratio between consecutive terms in the summation is}
\begin{equation}
    \frac{s(m)}{s(m-1)} = \left(\frac{k\sqrt{p_e}}{m}\right)^2 < 1,
\end{equation}
\black{indicating that} the gap \eqref{equ::gaptoI0}  shrinks significantly as $m$ increases. Therefore, we contend that when the order $m \geq \lceil k\sqrt{p_e} \rceil$, the average \black{guesswork} complexity \black{of OSD} approximately approaches the saturation point, and we refer to $m_s = \lceil k\sqrt{p_e} \rceil$ as the \black{guesswork} complexity saturation threshold. If $k$ and $d_{\min}$ are given, $m_s \leq \black{m_e} = \lceil{d_{\min}/4} -1\rceil$ will be satisfied with small $p_e$, which occurs at low code rates or high SNRs according to \eqref{equ:pe_approx2}.

\begin{table*}[t] 
    \centering
    \footnotesize	
    \tabcolsep=0.11cm
    \caption{Values of $k\sqrt{p_e}$ to determine the complexity saturation threshold at SNR = 2 dB.}
    \label{tab:sample_table}
    \begin{tabular}{|c|c|c|c|c|c|c|c|c|c|}
    \hline
    Rate $r$ & $\sfrac{1}{8}$ & $\sfrac{1}{4}$ & $\sfrac{1}{3}$ & $\sfrac{3}{8}$ & $\sfrac{1}{2}$ & $\sfrac{5}{8}$ & $\sfrac{2}{3}$ & $\sfrac{3}{4}$ & $\sfrac{7}{8}$ \\
    \hline
    \hline
    $k\sqrt{p_e}$ with fixed $n=128$  & 0.5336 & 1.7672 & - & 3.8370 & 7.0191 & 11.7472 & - & 18.6604 & 28.5450 \\
    \hline
    $k\sqrt{p_e}$ with fixed $k=64$  & 2.0321 & 3.4605 & 4.5055 & - & 7.0191 & - & 10.4154 & - & - \\
    \hline
    \end{tabular}
    \label{tab::CST}
\end{table*}

For instance, as \black{shown} in Fig. \ref{Fig::comp_error_OSD_k}, the average \black{guesswork} complexity for the $(n=256, k=64)$ code ($r = \frac{1}{4}$) is generally lower than those of the higher rate codes with the same $k$. \black{The guesswork complexity saturates when the order $m\geq 4$ (where $k\sqrt{p_e} = 3.4605$).} However, \black{when using the best known linear code \cite{Grasslcodetables},} a $(n=256, k=64)$ code can have the minimum distance of $d_{\min} = 65$, \black{requiring a} decoding order of $m_e = \lceil{d_{\min}/4} -1\rceil = 16$ to achieve MLD performance. 

Table \ref{tab::CST} \black{presents} the values of $k\sqrt{p_e}$ for various code parameters. \black{In Figs. \ref{Fig::relative_error_OSD_n}-\ref{Fig::relative_error_OSD_k}, we mark the guesswork complexity saturation threshold $m_s = \lceil k\sqrt{p_e} \rceil$ with circles for each code rate. The results demonstrate that for low rate codes, $m_s$ is typically reached before the decoding order $m_e$ needed for near-MLD performance where $(\epsilon_m - \epsilon_{\mathrm{ML}})/\epsilon_{\mathrm{ML}} < 1$. This indicates that for low-rate codes with $m_s < m_e = \lceil{d_{\min}/4} -1\rceil$, one can freely increase the decoding order beyond $m_s$ to approach MLD performance without introducing additional average guesswork complexity.}

\black{
Additionally, Figs. \ref{Fig::relative_error_OSD_n}-\ref{Fig::relative_error_OSD_k} show that for high-rate codes, the optimal order $m_e$ for near-MLD performance is reached before the saturation threshold $m_s$. Thus, although Section \ref{sec::asymtotic} showed that the guesswork complexity of order-$k$ OSD increases with code rate, high-rate codes need only a low decoding order ($m_e \ll k$) to approach MLD performance, keeping the guesswork complexity manageable. The relationship between guesswork complexity and MLD performance will be further discussed in the next subsection.
}

\subsection{\black{Guesswork} Complexity to achieve MLD} \label{sec::cmp_to_MLD}

    \black{In practical applications of channel coding, low-rate codes} typically operates at very low SNRs, in contrast to the scenarios depicted in Figs. \ref{Fig::comp_error_OSD_n}-\ref{Fig::relative_error_OSD_k} with SNR = 2 dB. On the other hand, the \black{guesswork} complexity of high-rate codes \black{is often lower than suggested by these figures}. \black{This is because} high-rate codes usually have a small \black{minimum distance} $d_{\min}$, \black{thus requiring only a low order} to approximate MLD. For example, the $(128,78)$ extended BCH code ($r\approx 5/8$) has $d_{\min} = 16$, and decoding order $m_e = \lceil{d_{\min}/4} -1\rceil = 3$ suffices to approach MLD.  As shown \black{in} Fig. \ref{Fig::comp_error_OSD_n}, the \black{guesswork} complexity at \black{order} $m=3$ for this code \black{is around $10^4$, which} remains reasonable. 

    \black{In this subsection}, we investigate the \black{\textit{achiable guesswork complexity}} of OSD approaching the MLD \black{performance} of a $(n,k)$ code. For given $n$ and $k$, the SNR is set to where the MLD performance of the $(n,k)$ code matches a target BLER $\epsilon_b$, i.e., $\epsilon_{\mathrm{ML}} = \epsilon_b$. This SNR is obtained by the normal approximation bound of the best error probability achieved by $(n, k)$ codes at short blocklengths \cite{erseghe2016coding}. Once the SNR is determined, the required OSD order is chosen such that $\epsilon_m - \epsilon_{\mathrm{ML}} < \epsilon_{\mathrm{ML}} $, i.e.,
    \begin{equation} \label{equ::ML:condition}
         \sum_{j=m+1}^{k} \mathbb{P}_{E}(j) \leq \epsilon_{\mathrm{ML}} = \epsilon_b,
    \end{equation}
     according to \eqref{equ::OSD:error:rate}. Then, \eqref{equ::Newbound:XnYn:ordered:1k:order:app:w1} is used to evaluate the complexity of this decoder, which represents the practical decoding complexity of OSD to achieve MLD with a target BLER. 
     
\begin{figure}  [t]
         \centering
            \input{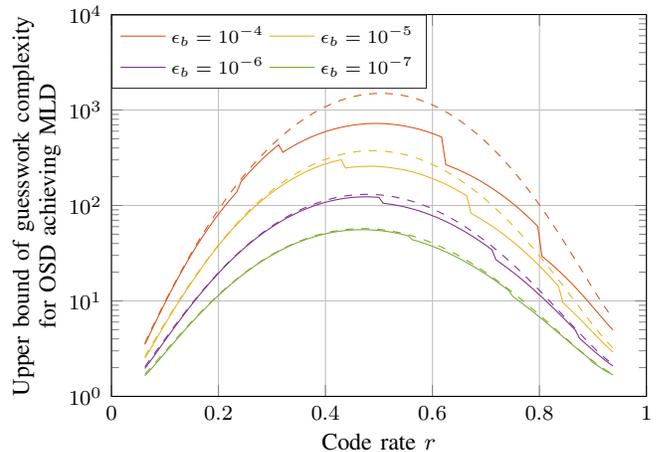}
    	\vspace{-0.5em}
        \caption{The average guesswork complexity \black{versus} the code rate for fixed block length $n = 128$, when the \black{relative} gap to MLD \black{$\epsilon_\mathrm{ML}  = \epsilon_b$ is less than 1}, i.e.,  $\epsilon_m - \epsilon_\mathrm{ML} < \epsilon_\mathrm{ML}$. The dashed curves are obtained with \eqref{equ::Newbound:XnYn:ordered:1k:?}, representing the guesswork complexity of \black{the order-$k$ OSD}.}
        \vspace{-0.25em}
    	\label{Fig::comp_r_reach_MLD_fixn}
        
    \end{figure}

     Figure \ref{Fig::comp_r_reach_MLD_fixn} depicts the \black{required guesswork} complexity of OSD approaching the MLD across a rate range from 0 to 1 with the fixed blocklength $n = 128$. As shown in the figure, \black{the guesswork complexity of OSD remains low} for both low-rate and high-rate codes, but \black{it increases significantly} for codes near the half rate. The \black{guesswork} complexity required to achieve $\epsilon_b\black{=\epsilon_{\mathrm{ML}}}$ is still effectively described by the bound given in \eqref{equ::Newbound:XnYn:ordered:1k:?} \black{for the order-$k$ OSD}, \black{especially at} extremely low values of $\epsilon_b$. We note that for $\epsilon_b = 10^{-4}$ and $10^{-5}$, there are significant gaps between the \black{guesswork} complexity and \eqref{equ::Newbound:XnYn:ordered:1k:?} for high-rate codes. \black{This is because for these codes,} the required decoding order $m_e$ to satisfy \eqref{equ::ML:condition} is lower than the complexity saturation threshold $m_s$. \black{Note that these results differ from Figs. \ref{Fig::comp_error_OSD_n} and \ref{Fig::relative_error_OSD_n} due to the dynamic SNR required for $\epsilon_b = \epsilon_{ML}$, which leads to lower guesswork complexity for high-rate codes compared to those for half-rate codes}

\begin{figure}  [t]
         \centering
            \input{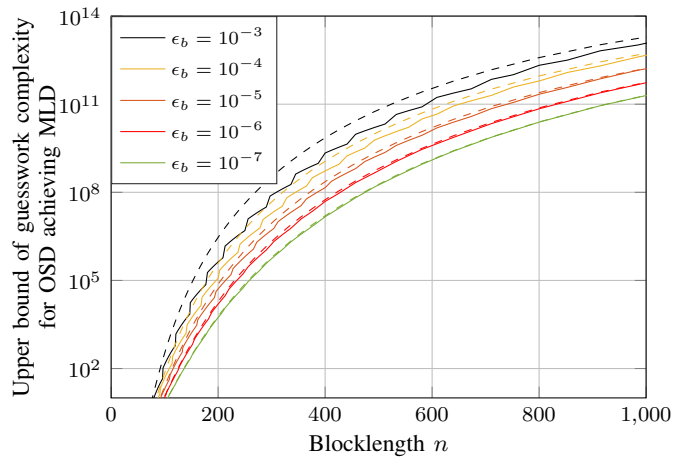}
    	\vspace{-0.5em}
        \caption{The average OSD decoding complexity \black{versus} blocklength for fixed information length $k = 64$, when the \black{relative} gap to MLD \black{$\epsilon_\mathrm{ML}  = \epsilon_b$ is less than 1}, i.e.,  $\epsilon_m - \epsilon_\mathrm{ML} < \epsilon_\mathrm{ML}$. The dashed curves are obtained with \eqref{equ::Newbound:XnYn:ordered:1k:?}, representing the guesswork complexity of \black{the order-$k$ OSD}.}
        \vspace{-0.25em}
    	\label{Fig::comp_r_reach_MLD_fixk}
        
    \end{figure}

    In Fig. \ref{Fig::comp_r_reach_MLD_fixk}, we present the \black{guesswork} complexity of OSD estimated by \eqref{equ::Newbound:XnYn:ordered:1k:order:app:w1} across a blocklength range from 100 to 1000 with the fixed information length $k = 64$, while the order $m$ is chosen as $m_e$ to satisfy \eqref{equ::ML:condition} for a given target BLER $\epsilon_b$. As depicted, the \black{guesswork} complexity increases with blocklength\black{, and are generally} close to the bound for \black{the order-$k$ OSD} given in \eqref{equ::Newbound:XnYn:ordered:1k:?}, especially for small $\epsilon_b$. This is because a smaller $\epsilon_b$ necessitates a higher order $m_e$ to achieve MLD performance as indicated by \eqref{equ::ML:condition}. When this required order $m_e$ \black{is larger than the} complexity saturation threshold $m_s = \lceil k\sqrt{p_e} \rceil$\black{, both order-$k$ and order-$m_e$ OSD exhibit similar guesswork complexity since increasing the decoding order beyond $m_s$ does not increase complexity further.}
    
\section{Discussions on the \black{Design of OSD Algorithms}} \label{sec::discussion}

\black{This section discusses the application of bounds and approximations derived in Section \ref{sec::OSD:bound} in the design of OSD algorithms.}

\subsection{OSD with Correct Codeword Identification} \label{sec::PB-OSD}

Section \ref{sec::OSD:bound} assumes \black{an infallible genie that identifies when the OSD decoder encounters the transmitted codeword $\mathbf{c}$ at the $\xi_{\rm{th}}$ guess $\boldsymbol{v}_{\xi} = \boldsymbol{b}_{\xi}\widetilde{\mathbf{G}}$.}  However, in a practical scenario, this becomes challenging since \black{the decoder} cannot know the transmitted \black{codeword} $\mathbf{c}$. \black{A common approach uses} CRC as an \black{additional} parity check for the information bits. The decoder can run an efficient CRC-check during decoding to determine if it has encountered the transmitted codeword. However, short CRCs, typically \black{used with} short block codes, have a non-negligible false positive rate. \black{This can be a critical concern} in URLLC scenarios \black{with} stringent BLER requirements. Moreover, executing CRC checks for each guess introduces extra complexity.

Apart from CRC, \black{several alternative approaches for correct codeword identification have been proposed for OSD}. For example, \cite{jin2006probabilisticConditions} computes a syndrome \black{for each guess $\boldsymbol{v}_{\xi} = \boldsymbol{b}_{\xi}\widetilde{\mathbf{G}}$} and compares the weight of syndrome with a threshold. If the weight is higher than the threshold, then the guess is regarded as correct. Similarly, in \cite{yue2021probability}, the posterior correct probability of a guess is estimated \black{and compared with a} predetermined threshold. These techniques leverages metrics such as \black{received signal distance and syndrome properties, and uses adjustable thresholds to balance complexity and BLER performance. However, stricter thresholds, while maintaining BLER performance, might fail to identify some correct codewords.}

\black{The identification approaches from \cite{jin2006probabilisticConditions,yue2021probability} can be effectively combined with CRC checking to enhance identification accuracy} while maintaining the near-optimal BLER performance. We next show that this \black{combined} approach \black{achieves} the \textit{achievable \black{guesswork} complexity} \black{predicted} by bound \eqref{equ::Newbound:XnYn:ordered:1k:order:?} and approximation \eqref{equ::Newbound:XnYn:ordered:1k:order:app:w1} for order-$m$ OSD. \black{Algorithm \ref{Algo} outlines this decoder design}, where \black{TEPs are examined in the order of increasing Hamming weight, and} $\mathbb{P}(\boldsymbol{v}_\xi)$ is computed \black{according to} \cite[Eq. (3)]{yue2021probability}. In the algorithm, \black{the preprocessing stage includes reliability ordering and Gaussian elimination, while each TEP $\mathbf{e}_{\xi}$ is used to flip bits in the hard decision $\boldsymbol{b}_0$ of the most reliable bits (MRB) to obtain information bit guesses.} Then, a codeword estimate $\boldsymbol{v}_\xi$ is obtained by re-encoding, i.e., $\boldsymbol{v}_{\xi} = \boldsymbol{b}_{\xi}\widetilde{\mathbf{G}}$. If \black{valid codeword is identified through this process}, the decoder selects the codeword candidate with the minimum Euclidean distance to the received signal $\mathbf{y}$. \black{For brevity, this final selection procedure is omitted from Algorithm \ref{Algo}.}

\begin{algorithm} 
    \caption{OSD with Correct Codeword Identification}
    \label{Algo}
    \begin{algorithmic} [1]
        \REQUIRE Received signal $\mathbf{y}$, Threshold parameter $\lambda$
        \ENSURE Decoded result $\hat{\mathbf{c}}$
            
        \STATE Preprocessing for OSD: \black{reliability ordering and GE}
        \STATE \black{Get hard decision $\boldsymbol{b}_0$ from MRB}
        \FOR{$\xi = 1$ to $\xi_{\max}$}
            \STATE \black{Generate TEP $\mathbf{e}_{\xi}$ following Hamming weight order}
            \STATE \black{Obtain information bit guess $\boldsymbol{b}_{\xi} = \boldsymbol{b}_0 \oplus \mathbf{e}_{\xi}$}
            \STATE \black{ Generate codeword guess by re-encoding: $\boldsymbol{v}_{\xi} = \boldsymbol{b}_{\xi}\widetilde{\mathbf{G}}$}
            \STATE Calculate posterior probability $\mathbb{P}(\boldsymbol{v}_i)$ using \cite{yue2021probability}
            \IF{$\mathbb{P}(\boldsymbol{v}_\xi) \geq \lambda$}
                \IF{CRC check for $\boldsymbol{v}_\xi$ is valid}
                    \RETURN $\boldsymbol{v}_\xi$
                \ENDIF
            \ENDIF
        \ENDFOR
        \RETURN \black{The codeword estimate with minimum Euclidean distance to $\mathbf{y}$}
    \end{algorithmic}
\end{algorithm}

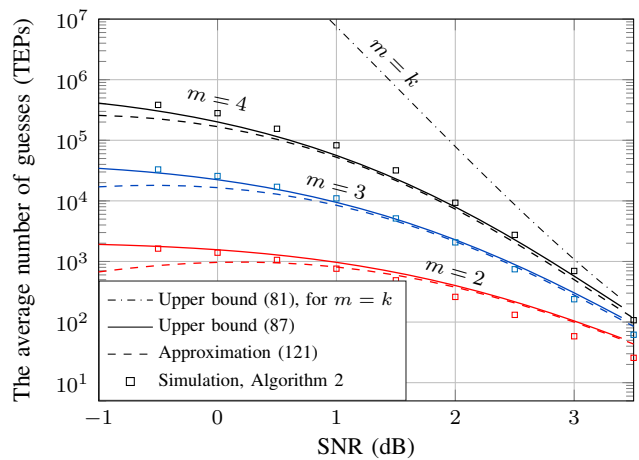
\begin{figure}  [t]
         \centering
\definecolor{mycolor1}{rgb}{0, 0 , 0}%
\definecolor{mycolor2}{rgb}{0.00000,0.2706,0.74118}%
\definecolor{mycolor3}{rgb}{0.00000,0.44700,0.74100}%
\begin{tikzpicture}

\begin{axis}[%
width=2.8in,
height=2in,
at={(1.01in,0.685in)},
scale only axis,
xmin=-1,
xmax=3.5,
xlabel style={at={(0.5,1ex)},font=\color{white!15!black}, font=\small},
xlabel={SNR (dB)},
ymode=log,
ymin=5,
ymax=10000000,
ylabel style={at={(1.5ex,0.5)},font=\color{white!15!black}, font=\small},
ylabel={The average number of guesses (TEPs)},
axis background/.style={fill=white},
tick label style={font=\footnotesize},
xmajorgrids,
ymajorgrids,
legend style={at={(0,0)}, anchor=south west , fill opacity=0.5, text opacity=1,  font = \scriptsize, legend cell align=left, align=left, draw=white!15!black}
]

\addplot [color=black, dashdotted]
  table[row sep=crcr]{%
-1	29094985116.5829\\
-0.8	13907037831.0979\\
-0.6	6503740543.22157\\
-0.4	2977387156.18522\\
-0.2	1335197632.29693\\
0	587038927.998087\\
0.2	253318514.615388\\
0.4	107429717.796584\\
0.6	44848261.9394964\\
0.8	18466269.7429772\\
1	7516737.27454266\\
1.2	3032956.13132306\\
1.4	1216833.03466801\\
1.6	487121.899980735\\
1.8	195328.558880087\\
2	78785.5703521136\\
2.2	32110.0362703049\\
2.4	13286.3438340331\\
2.6	5608.6072173981\\
2.8	2427.24380476748\\
3	1082.0876497342\\
3.2	499.215314131182\\
3.4	239.346107352474\\
};
\addlegendentry{Upper bound \eqref{equ::Newbound:XnYn:ordered:1k}, for $m=k$}

\addplot [color=black]
  table[row sep=crcr]{%
  1 10\\
};
\addlegendentry{Upper bound \eqref{equ::Newbound:XnYn:ordered:1k:order}}

\addplot [color=black, dashed,]
  table[row sep=crcr]{%
  1 10\\
};
\addlegendentry{Approximation \eqref{equ::Newbound:XnYn:ordered:1k:order:app:w1}}

\addplot [mark=square, mark size = 1.5pt, only marks,  mark options={solid, black}]
  table[row sep=crcr]{%
  1 0.1\\
};
\addlegendentry{Simulation, Algorithm \ref{Algo}}

\addplot [color=mycolor1, line width=0.5pt, forget plot]
  table[row sep=crcr]{%
-1	410226.351714721\\
-0.8	368812.090826834\\
-0.6	326086.54003782\\
-0.4	283110.421500437\\
-0.2	241008.762735779\\
0	200881.229465544\\
0.2	163709.954254343\\
0.4	130278.072953341\\
0.6	101111.167874246\\
0.8	76450.1467759079\\
1	56258.4778259577\\
1.2	40260.469630534\\
1.4	28001.969667748\\
1.6	18921.7292865785\\
1.8	12421.359118349\\
2	7924.05481156584\\
2.2	4916.19604905986\\
2.4	2970.26984670113\\
2.6	1751.20961455816\\
2.8	1010.47899859963\\
3	572.921638401332\\
3.2	320.856175953369\\
3.4	178.660697462562\\
};

\addplot [color=mycolor1, dashed, line width=0.5pt, forget plot]
  table[row sep=newline]{%
-1	257354.880407973
-0.750000000000000	245930.313247828
-0.500000000000000	225931.280507390
-0.250000000000000	199058.274642922
0	167816.541698325
0.250000000000000	135092.189940464
0.500000000000000	103648.258610191
0.750000000000000	75676.2820667433
1	52519.6153984617
1.25000000000000	34622.1454989160
1.50000000000000	21678.1921340714
1.75000000000000	12900.8846538881
2	7308.80353485205
2.25000000000000	3953.09250174433
2.50000000000000	2050.27663065459
2.75000000000000	1026.30900705102
3	500.315013469245
3.25000000000000	240.384177483352
3.50000000000000	115.551791098863
};

\addplot [color=mycolor2, line width=0.5pt, forget plot]
  table[row sep=crcr]{%
-1	34427.3450333459\\
-0.8	32410.774445107\\
-0.6	30158.3896605386\\
-0.4	27698.4646524633\\
-0.2	25074.0299318909\\
0	22341.4183489973\\
0.2	19567.1140636123\\
0.4	16823.1029575776\\
0.6	14181.2158117311\\
0.8	11707.1908220556\\
1	9455.29332412174\\
1.2	7464.27623957919\\
1.4	5755.24430672613\\
1.6	4331.64418911702\\
1.8	3181.22193511645\\
2	2279.46122960503\\
2.2	1593.81567702993\\
2.4	1088.01156945545\\
2.6	725.812293426429\\
2.8	473.8509155368\\
3	303.384662194833\\
3.2	191.040554615211\\
3.4	118.764839622033\\
};

\addplot [color=mycolor2, dashed, line width=0.5pt, forget plot]
  table[row sep=newline]{%
-1	17060.1187989859
-0.750000000000000	17858.3494189502
-0.500000000000000	18064.6875551664
-0.250000000000000	17621.0821055785
0	16541.9280812140
0.250000000000000	14917.7612366510
0.500000000000000	12903.0183841929
0.750000000000000	10689.7871238653
1	8474.05631615103
1.25000000000000	6423.46217944087
1.50000000000000	4654.73114835738
1.75000000000000	3225.30244220034
2	2138.69563537476
2.25000000000000	1359.20651194465
2.50000000000000	829.876632955586
2.75000000000000	488.484749469057
3	278.574458571643
3.25000000000000	154.963177454440
3.50000000000000	84.8480832156517

};

\addplot [color=red, line width=0.5pt, forget plot]
  table[row sep=crcr]{%
-1	1913.69375472683\\
-0.8	1864.90080909337\\
-0.6	1805.78245696577\\
-0.4	1735.52949929954\\
-0.2	1653.7148313696\\
0	1560.41500743329\\
0.2	1456.30636435587\\
0.4	1342.7164902584\\
0.6	1221.61477223603\\
0.8	1095.53267178257\\
1	967.414575810757\\
1.2	840.41183484764\\
1.4	717.643451627224\\
1.6	601.954191285905\\
1.8	495.702629055806\\
2	400.607093566798\\
2.2	317.667471870927\\
2.4	247.167698270227\\
2.6	188.750502105169\\
2.8	141.54555614231\\
3	104.326514392676\\
3.2	75.6721126751225\\
3.4	54.1107159070871\\
};

\addplot [color=red, dashed, line width=0.5pt, forget plot]
  table[row sep=newline]{%
-1	672.549405443449
-0.750000000000000	770.648708672899
-0.500000000000000	857.747579357578
-0.250000000000000	925.672681710499
0	967.020059620281
0.250000000000000	976.465113307418
0.500000000000000	951.872114412073
0.750000000000000	894.886270453746
1	810.799639455862
1.25000000000000	707.681284020389
1.50000000000000	594.981507459856
1.75000000000000	481.972112528274
2	376.408579466480
2.25000000000000	283.693537237558
2.50000000000000	206.638545234525
2.75000000000000	145.742253133937
3	99.7902589211141
3.25000000000000	66.5571759898304
3.50000000000000	43.4366261703315
};

\addplot [color=mycolor3, mark=square, mark size = 1pt, only marks, mark options={solid, mycolor3}, forget plot]
  table[row sep=crcr]{%
-0.5	32978.9773462783\\
0	25699.153674833\\
0.5	17032.8893333333\\
1	10999.5003367003\\
1.5	5130.07815066886\\
2	2066.19168679293\\
2.5	744.981566666667\\
3	238.250633333333\\
3.5	61.8018666666667\\
};
\addplot [color=red, mark=square, mark size = 1pt, only marks,  mark options={solid, red}, forget plot]
  table[row sep=crcr]{%
-0.5	1631.50522648084\\
0	1396.88346883469\\
0.5	1059.77663230241\\
1	760.878547105562\\
1.5	488.379912663755\\
2	260.31102079847\\
2.5	131.535829906379\\
3	58.2994333333333\\
3.5	25.7035333333333\\
};
\addplot [color=mycolor1, mark=square, mark size = 1pt, only marks, mark options={solid, mycolor1}, forget plot]
  table[row sep=crcr]{%
-0.5	383719.767567568\\
0	279996.164461248\\
0.5	154821.758032129\\
1	82575.2846003899\\
1.5	31864.9640533003\\
2	9314.2032\\
2.5	2743.6951\\
3	696.8093\\
3.5	107.263\\
};

\node[rotate=-16] at (axis cs: 2,700) {\footnotesize $m=2$};
\node[rotate=-14.5] at (axis cs: 1,18000) {\footnotesize $m=3$};
\node[rotate=-12] at (axis cs: 0,500000) {\footnotesize $m=4$};
\node[rotate=-45] at (axis cs: 1.5,2000000) {\footnotesize $m=k$};

\end{axis}
\end{tikzpicture}%
    	\vspace{-0.5em}
        \caption{\black{The average guesswork complexity of OSD in Algorithm \ref{Algo}} }
        \vspace{-0.25em}
    	\label{Fig:OSD_comp_SNR}
        
    \end{figure}

We \black{evaluate} Algorithm \ref{Algo} with $\lambda = 0.5$, \black{by comparing its guesswork complexity to bound} \eqref{equ::Newbound:XnYn:ordered:1k:order} and approximation \eqref{equ::Newbound:XnYn:ordered:1k:order:app:w1}\black{, as depicted} in Fig. \ref{Fig:OSD_comp_SNR}. \black{The decoding algorithm is tested on a} $(128,64)$ extended BCH code with various decoding orders $m$, and CRC-6 is used to provide additional parity checks for information bits. As shown, Algorithm \ref{Algo} \black{achieves the guesswork complexity predicted by} \eqref{equ::Newbound:XnYn:ordered:1k:order} and \eqref{equ::Newbound:XnYn:ordered:1k:order:app:w1}. \black{Moreover}, the decoding algorithm can maintain the near-ML decoding performance, achieving a BLER of $10^{-4}$ at the SNR of 3 dB with order-4 decoding via simulation. Detailed BLER results across SNRs are omitted for brevity.

We note that many OSD algorithms, \black{such as those in} \cite{yue2021probability,Chentao2019SDD, Wu2007OSDMRB,choi2021fast},  \black{can achieve} lower \black{guesswork} complexity than Algorithm \ref{Algo} at low-to-moderate SNRs. Besides \black{the early termination of decoding}, these methods \black{further reduce the number of guesses by discarding} unpromising TEPs. \black{However, since discarded TEPs can} occasionally \black{correspond to the correct codeword,  these techniques can degrade decoder BLER performance}. \black{Rather than analyzing such complexity-performance trade-offs, this paper only focuses on characterizing the fundamental guesswork complexity through \eqref{equ::Newbound:XnYn:ordered:1k:order} and \eqref{equ::Newbound:XnYn:ordered:1k:order:app:w1}, which represent the \textit{achievable guesswork complexity} when early termination is implemented without compromising BLER performance.}

\subsection{Efficient Design of \black{OSD-based} HARQ} \label{sec::HARQ}
OSD is regarded as a promising decoder for rate-compatible (RC) codes \cite{Mahyar2021primitive, yue2023efficient}. Codes are referred to as rate-compatible if they have the same information block length $k$ and their $k\times n$ generator matrices are nested, meaning that the generator matrix of a higher-rate code is a submatrix of the generator matrix of a lower-rate code. OSD \black{is particularly suitable for} RC codes in HARQ systems without compromising performance and latency, as it \black{operates directly on} the code generator matrix.

\begin{figure}  [t]
     \centering
        \input{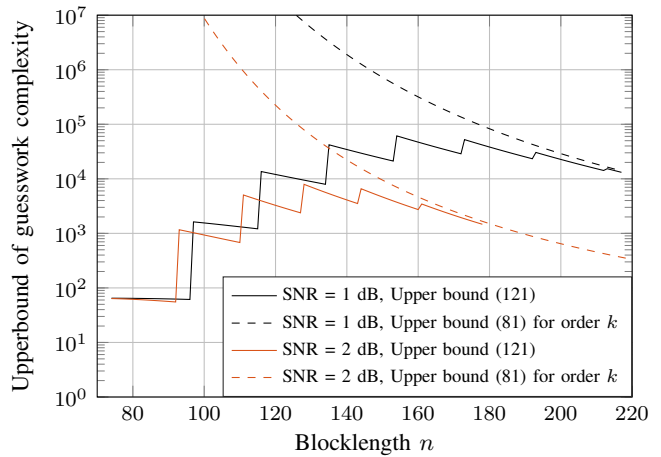}
    \vspace{-0.5em}
    \caption{The average guesswork complexity of OSD for achieving MLD. Results shown for RC codes in IR-HARQ with $k = 64$. }
    \vspace{-0.25em}
    \label{Fig:OSD_comp_HARQ}
    
\end{figure}

\begin{figure}  [t]
     \centering
        \begin{tikzpicture}

\definecolor{mycolor1}{rgb}{0.85000,0.32500,0.09800}%
\definecolor{mycolor2}{rgb}{0.92900,0.69400,0.12500}%
\definecolor{mycolor3}{rgb}{0.49400,0.18400,0.55600}%
\definecolor{mycolor4}{rgb}{0.46600,0.67400,0.18800}%
\definecolor{mycolor5}{rgb}{0.30100,0.74500,0.93300}%
\definecolor{mycolor6}{rgb}{0.00000,0.44700,0.74100}%

\begin{axis}[%
width=2.8in,
height=2in,
at={(1.01in,0.685in)},
scale only axis,
xmin=20,
xmax=128,
xlabel style={at={(0.5,1ex)},font=\color{white!15!black}, font=\small},
xlabel={Blocklength $n$},
ymode=log,
ymin=0.969125068344631,
ymax=1000,
ylabel style={at={(1.5ex,0.5)},font=\color{white!15!black}, font=\small},
ylabel={Upperbound of guesswork complexity},
axis background/.style={fill=white},
tick label style={font=\footnotesize},
xmajorgrids,
ymajorgrids,
legend style={at={(1,1)}, anchor=north east, fill opacity=0.5, text opacity=1, font=\scriptsize, legend cell align=left, align=left, draw=white!15!black}
]

\addplot [color=black]
  table[row sep=newline]{%
  18.0000    0.9326
   19.0000    0.9696
   20.0000    0.9865
   21.0000    0.9942
   22.0000    0.9975
   23.0000   12.9230
   24.0000   12.5941
   25.0000   12.2698
   26.0000   11.9543
   27.0000   11.6495
   28.0000   11.3563
   29.0000   11.0749
   30.0000   10.8053
   31.0000   10.5472
   32.0000   10.3002
   33.0000   10.0640
   34.0000    9.8382
   35.0000    9.6222
   36.0000    9.4156
   37.0000    9.2180
   38.0000    9.0289
   39.0000    8.8479
   40.0000    8.6745
   41.0000    8.5084
   42.0000    8.3491
   43.0000    8.1964
   44.0000    8.0498
   45.0000   21.2473
   46.0000   20.5356
   47.0000   19.8661
   48.0000   19.2355
   49.0000   18.6408
   50.0000   18.0793
   51.0000   17.5486
   52.0000   17.0464
   53.0000   16.5706
   54.0000   16.1194
   55.0000   15.6911
   56.0000   15.2842
   57.0000   14.8971
   58.0000   14.5287
   59.0000   14.1776
   60.0000   13.8428
   61.0000   13.5232
   62.0000   13.2179
   63.0000   12.9261
   64.0000   12.6469
   65.0000   12.3795
   66.0000   12.1234
   67.0000   11.8778
   68.0000   11.6421
   69.0000   11.4158
   70.0000   11.1984
   71.0000   13.6704
   72.0000   13.3590
   73.0000   13.0614
   74.0000   12.7766
   75.0000   12.5040
   76.0000   12.2429
   77.0000   11.9925
   78.0000   11.7523
   79.0000   11.5217
   80.0000   11.3002
   81.0000   11.0872
   82.0000   10.8824
   83.0000   10.6853
   84.0000   10.4955
   85.0000   10.3126
   86.0000   10.1363
   87.0000    9.9663
   88.0000    9.8021
   89.0000    9.6437
   90.0000    9.4906
   91.0000    9.3426
   92.0000    9.1996
   93.0000    9.0611
   94.0000    8.9272
   95.0000    8.7975
   96.0000    8.6718
   97.0000    8.5501
   98.0000    8.4320
   99.0000    8.3175
  100.0000    8.2065
  101.0000    8.2660
  102.0000    8.1551
  103.0000    8.0474
  104.0000    7.9430
  105.0000    7.8416
  106.0000    7.7430
  107.0000    7.6473
  108.0000    7.5543
  109.0000    7.4638
  110.0000    7.3757
  111.0000    7.2901
  112.0000    7.2067
  113.0000    7.1256
  114.0000    7.0465
  115.0000    6.9695
  116.0000    6.8944
  117.0000    6.8213
  118.0000    6.7499
  119.0000    6.6803
  120.0000    6.6124
  121.0000    6.5461
  122.0000    6.4815
  123.0000    6.4183
  124.0000    6.3566
  125.0000    6.2963
  126.0000    6.2375
  127.0000    6.1799
  128.0000    6.1236
};
\addlegendentry{SNR = -1 dB, Upper bound \eqref{equ::Newbound:XnYn:ordered:1k:order:app:w1}}

\addplot [color=black, dashed]
  table[row sep=newline]{%
   20.0000  447.2561
   22.0000  333.2177
   24.0000  250.5191
   26.0000  192.3388
   28.0000  151.0909
   30.0000  121.2961
   32.0000   99.3209
   34.0000   82.7805
   36.0000   70.0918
   38.0000   60.1858
   40.0000   52.3272
   42.0000   46.0009
   44.0000   40.8400
   46.0000   36.5783
   48.0000   33.0199
   50.0000   30.0189
   52.0000   27.4643
   54.0000   25.2714
   56.0000   23.3745
   58.0000   21.7217
   60.0000   20.2724
   62.0000   18.9937
   64.0000   17.8593
   66.0000   16.8476
   68.0000   15.9411
   70.0000   15.1252
   72.0000   14.3878
   74.0000   13.7188
   76.0000   13.1096
   78.0000   12.5529
   80.0000   12.0427
   82.0000   11.5737
   84.0000   11.1413
   86.0000   10.7416
   88.0000   10.3712
   90.0000   10.0272
   92.0000    9.7069
   94.0000    9.4082
   96.0000    9.1289
   98.0000    8.8673
  100.0000    8.6220
  102.0000    8.3913
  104.0000    8.1742
  106.0000    7.9696
  108.0000    7.7763
  110.0000    7.5935
  112.0000    7.4205
  114.0000    7.2564
  116.0000    7.1007
  118.0000    6.9527
  120.0000    6.8118
  122.0000    6.6777
  124.0000    6.5497
  126.0000    6.4277
  128.0000    6.3110
};
\addlegendentry{SNR = -1 dB, Upper bound \eqref{equ::Newbound:XnYn:ordered:1k} for order $k$}

\addplot [color=mycolor1]
  table[row sep=newline]{%
   18.0000    0.9332
   19.0000    0.9704
   20.0000    0.9872
   21.0000    0.9947
   22.0000   11.6471
   23.0000   11.1837
   24.0000   10.7346
   25.0000   10.3074
   26.0000    9.9046
   27.0000    9.5263
   28.0000    9.1717
   29.0000    8.8398
   30.0000    8.5290
   31.0000    8.2380
   32.0000    7.9654
   33.0000    7.7098
   34.0000    7.4701
   35.0000    7.2449
   36.0000    7.0332
   37.0000    6.8340
   38.0000    6.6463
   39.0000    6.4693
   40.0000    6.3023
   41.0000    6.1443
   42.0000    5.9949
   43.0000   12.1625
   44.0000   11.6597
   45.0000   11.1940
   46.0000   10.7618
   47.0000   10.3601
   48.0000    9.9859
   49.0000    9.6368
   50.0000    9.3106
   51.0000    9.0053
   52.0000    8.7191
   53.0000    8.4504
   54.0000    8.1977
   55.0000    7.9599
   56.0000    7.7357
   57.0000    7.5241
   58.0000    7.3242
   59.0000    7.1350
   60.0000    6.9557
   61.0000    6.7858
   62.0000    6.6245
   63.0000    6.4712
   64.0000    6.3253
   65.0000    6.1865
   66.0000    6.7161
   67.0000    6.5539
   68.0000    6.4000
   69.0000    6.2537
   70.0000    6.1147
   71.0000    5.9823
   72.0000    5.8562
   73.0000    5.7360
   74.0000    5.6212
   75.0000    5.5115
   76.0000    5.4066
   77.0000    5.3063
   78.0000    5.2102
   79.0000    5.1181
   80.0000    5.0297
   81.0000    4.9449
   82.0000    4.8635
   83.0000    4.7852
   84.0000    4.7100
   85.0000    4.6375
   86.0000    4.5678
   87.0000    4.5006
   88.0000    4.4358
   89.0000    4.3733
   90.0000    4.3129
   91.0000    4.2547
   92.0000    4.1984
   93.0000    4.1641
   94.0000    4.1105
   95.0000    4.0587
   96.0000    4.0086
   97.0000    3.9600
   98.0000    3.9130
   99.0000    3.8675
  100.0000    3.8233
  101.0000    3.7805
  102.0000    3.7389
  103.0000    3.6986
  104.0000    3.6595
  105.0000    3.6215
  106.0000    3.5845
  107.0000    3.5486
  108.0000    3.5137
  109.0000    3.4797
  110.0000    3.4467
  111.0000    3.4145
  112.0000    3.3832
  113.0000    3.3527
  114.0000    3.3229
  115.0000    3.2940
  116.0000    3.2657
  117.0000    3.2382
  118.0000    3.2113
  119.0000    3.1851
  120.0000    3.1595
  121.0000    3.1345
  122.0000    3.1101
  123.0000    3.0865
  124.0000    3.0632
  125.0000    3.0405
  126.0000    3.0182
  127.0000    2.9964
  128.0000    2.9752
};
\addlegendentry{SNR = 0 dB, Upper bound \eqref{equ::Newbound:XnYn:ordered:1k:order:app:w1}}

\addplot [color=mycolor1, dashed]
  table[row sep=newline]{%
   20.0000  257.4324
   22.0000  176.2703
   24.0000  123.3455
   26.0000   89.3736
   28.0000   67.1102
   30.0000   52.0573
   32.0000   41.5483
   34.0000   33.9894
   36.0000   28.4035
   38.0000   24.1748
   40.0000   20.9038
   42.0000   18.3248
   44.0000   16.2565
   46.0000   14.5724
   48.0000   13.1823
   50.0000   12.0209
   52.0000   11.0399
   54.0000   10.2030
   56.0000    9.4828
   58.0000    8.8579
   60.0000    8.3116
   62.0000    7.8309
   64.0000    7.4053
   66.0000    7.0264
   68.0000    6.6871
   70.0000    6.3821
   72.0000    6.1064
   74.0000    5.8564
   76.0000    5.6288
   78.0000    5.4207
   80.0000    5.2299
   82.0000    5.0545
   84.0000    4.8926
   86.0000    4.7429
   88.0000    4.6041
   90.0000    4.4750
   92.0000    4.3548
   94.0000    4.2425
   96.0000    4.1375
   98.0000    4.0390
  100.0000    3.9465
  102.0000    3.8595
  104.0000    3.7776
  106.0000    3.7002
  108.0000    3.6271
  110.0000    3.5578
  112.0000    3.4922
  114.0000    3.4300
  116.0000    3.3708
  118.0000    3.3145
  120.0000    3.2609
  122.0000    3.2098
  124.0000    3.1610
  126.0000    3.1144
  128.0000    3.0698
};
\addlegendentry{SNR = 0 dB, Upper bound \eqref{equ::Newbound:XnYn:ordered:1k} for order $k$}

\end{axis}
\end{tikzpicture}
    \vspace{-0.5em}
    \caption{\black{The average guesswork complexity of OSD for achieving MLD. Results shown for RC codes  in IR-HARQ with $k = 15$. }}
    \vspace{-0.25em}
    \label{Fig:OSD_comp_HARQ_16}
    
\end{figure}
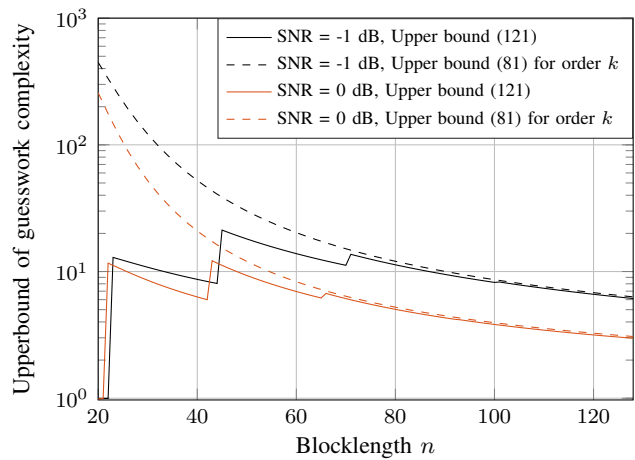

As demonstrated in Fig. \ref{Fig::comp_r_reach_MLD_fixn}, OSD \black{has relatively lower guesswork complexity for both} low-rate and high-rate codes. \black{This characteristic can be exploited to optimize HARQ system design by carefully selecting} code parameters for each retransmission to minimize the overall decoding latency. We show this by examining the guesswork complexity of OSD \black{when achieving MLD performance across various code rates and SNRs.} Given $k$, $n$, and SNR, \black{we first} determine the best achievable BLER $\epsilon_{\mathrm{ML}}$ for $(n,k)$ codes \black{using the NA bound}. \black{We then select the order $m = m_e$ of OSD  that ensure $\epsilon_m - \epsilon_{\mathrm{ML}} < \epsilon_{\mathrm{ML}}$, i.e., achieving the near MLD performance.} Then, \eqref{equ::Newbound:XnYn:ordered:1k:order:app:w1} is used to evaluate the \black{guesswork complexity of this OSD decoder}. \black{Figure \ref{Fig:OSD_comp_HARQ} shows results for information length $k = 64$ with code lengths ranging from 80 to 220, while Fig. \ref{Fig:OSD_comp_HARQ_16} presents results for $k = 15$ with lengths from 18 to 128.  In both cases, the guesswork complexity initially increases with blocklength but subsequently decreases.} \black{These results suggest an efficient OSD-based HARQ design strategy: beginning with low-rate transmissions and reserving higher rates for final transmission rounds.}

\black{Figures \ref{Fig:OSD_comp_HARQ} and \ref{Fig:OSD_comp_HARQ_16} also include the guesswork complexity bound \eqref{equ::Newbound:XnYn:ordered:1k:?} for order-$k$ OSD.} \black{This bound decreases monotonically with increasing blocklength, and then serves as an asymptotic limit} for the guesswork complexity of \black{ MLD-achieving OSD decoders} at larger block lengths (i.e., lower rates). This behavior can be explained through the complexity saturation threshold $m_s$. Specifically, \black{$m_s$ becomes smaller than the order $m_e$ required for MLD performance at low code rates, which effectively limits the guesswork complexity growth.}

It should be noted that the guesswork complexity curves in Figs. \ref{Fig:OSD_comp_HARQ} and \ref{Fig:OSD_comp_HARQ} \black{exhibit discrete jumps}. This occurs because general OSD settings permit only integer decoding orders. \black{Each discontinuous point corresponds a change in the required decoding order $m_e$ for achieving MLD, prompted by a change in the code rate.} \black{This behavior suggests a design strategy for OSD-based HARQ. Selecting the transmission blocklengths at the maximum values before the next order increment can optimize the trade-off between BLER performance and guesswork complexity.}

\black{Next, we compare two HARQ schemes with different rate selections. Both schemes employ the OSD decoder as described in Algorithm \ref{Algo}, which uses CRC-6 and threshold parameter $\lambda=0.75$. In the first transmission round, $n_1$ bits are transmitted and decoded. If decoding fails (i.e., no correct codeword is identified), additional bits are transmitted in the second round to reach a total length of $n_2$, followed by another decoding attempt. Both schemes use RC codes obtained by optimally puncturing a $(128,15)$ extended BCH code, with information length $k = 15$ and the following parameters:
\begin{itemize}
    \item Scheme 1: $n_1 = 48$ decoded by order-2 OSD, $n_2 = 72$ decoded by order-3 OSD.
    \item Scheme 2: $n_1 = 40$ decoded by order-1 OSD, $n_2 = 96$ decoded by order-4 OSD.
\end{itemize}
Here, the OSD decoding order for each transmission is selected to achieve the near MLD performance. Compared to scheme 1, scheme 2 employs a lower first-round rate but provides more redundancy in the second round.}

\black{
Figure \ref{Fig:OSD_FER_HARQ_16} compares the BLER achieved by these schemes across different SNRs. As depicted, Scheme 2 achieves better error performance, particularly at low SNRs. When SNR increases, the BLER of Scheme 2 experience an error floor, due to the false alert rate of short CRC used in Algorithm \ref{Algo}.
}

\begin{figure}  [t]
     \centering
        \begin{tikzpicture}

\definecolor{mycolor1}{rgb}{0.85000,0.32500,0.09800}%
\definecolor{mycolor2}{rgb}{0.92900,0.69400,0.12500}%
\definecolor{mycolor3}{rgb}{0.49400,0.18400,0.55600}%
\definecolor{mycolor4}{rgb}{0.46600,0.67400,0.18800}%
\definecolor{mycolor5}{rgb}{0.30100,0.74500,0.93300}%
\definecolor{mycolor6}{rgb}{0.00000,0.44700,0.74100}%

\begin{axis}[%
width=2.8in,
height=2in,
at={(1.01in,0.685in)},
scale only axis,
xmin=-4,
xmax=0,
xlabel style={at={(0.5,1ex)},font=\color{white!15!black}, font=\small},
xlabel={SNR (dB)},
ymode=log,
ymin=0,
ymax=1000,
ylabel style={at={(1.5ex,0.5)},font=\color{white!15!black}, font=\small},
ylabel={The averge number of TEPs (guesses)},
axis background/.style={fill=white},
tick label style={font=\footnotesize},
xmajorgrids,
ymajorgrids,
legend style={at={(1,1)}, anchor=north east, fill opacity=0.5, text opacity=1, font=\scriptsize, legend cell align=left, align=left, draw=white!15!black}
]

\addplot [color=black, mark=0, mark options={solid, black}]
  table[row sep=newline]{%
   -4.0000  320.8762
   -3.5000  269.6420
   -3.0000  223.7152
   -2.5000  172.5145
   -2.0000  142.2262
   -1.5000  108.8466
   -1.0000   82.0179
   -0.5000   61.2239
         0   43.6710
};
\addlegendentry{$n_1 = 48$, $n_2 = 72$, 2nd round}

\addplot [dashed, color=black, mark=o, mark options={solid, black}]
  table[row sep=newline]{%
   -4.0000   97.3516
   -3.5000   88.7617
   -3.0000   77.6495
   -2.5000   67.4184
   -2.0000   56.3337
   -1.5000   43.0013
   -1.0000   32.3698
   -0.5000   23.4152
         0   15.7304
};
\addlegendentry{$n_1 = 48$, $n_2 = 72$, 1st round}

\addplot [color=blue, mark=square, mark options={solid, blue}]
  table[row sep=newline]{%
   -4.0000  401.6363
   -3.5000  283.3627
   -3.0000  179.5441
   -2.5000  118.7653
   -2.0000   71.1055
   -1.5000   41.2527
   -1.0000   23.9194
   -0.5000   13.6219
         0    8.4395
};
\addlegendentry{$n_1 = 40$, $n_2 = 96$, 2nd round}

\addplot [dashed, color=blue, mark=square, mark options={solid, blue}]
  table[row sep=newline]{%
   -4.0000   13.6334
   -3.5000   12.9596
   -3.0000   12.0146
   -2.5000   11.1442
   -2.0000   10.0500
   -1.5000    8.8117
   -1.0000    7.5630
   -0.5000    6.3024
         0    5.1396
};
\addlegendentry{$n_1 = 40$, $n_2 = 96$, 1st round}

\end{axis}
\end{tikzpicture}
    \vspace{-0.5em}
    \caption{\black{The number of TEPs (guesses) comparison of two HARQ schemes with different blocklengths $n_1$ and $n_2$ for two transmission rounds. }}
    \vspace{-0.25em}
    \label{Fig:OSD_TEP_HARQ_16}
    
\end{figure}
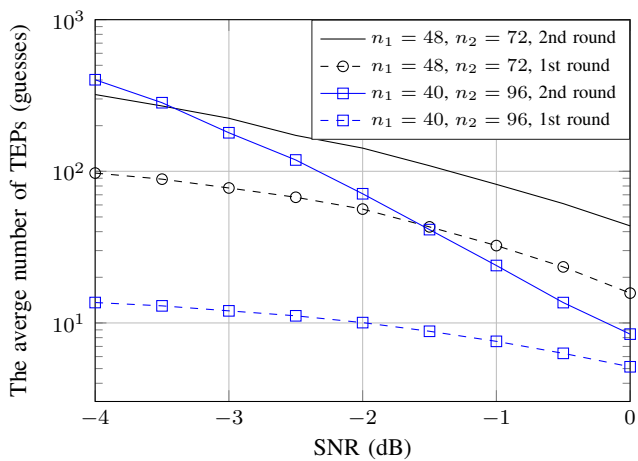

\black{
In terms of the guesswork complexity, as shown in Fig. \ref{Fig:OSD_TEP_HARQ_16}, Scheme 2 requires fewer TEPs (guesses) in both transmission rounds despite its better error performance. This reduction in guesswork complexity is particularly significant around SNR = 0 dB, where both schemes achieve BLER near $10^{-4}$. Moreover, our experiments show that both schemes can achieve similar throughput, with average effective blocklengths of 50 and 52 bits at SNR = 0 dB for Schemes 1 and 2, respectively.
}

\black{
These results validate our analysis based on Figs. \ref{Fig:OSD_comp_HARQ} and \ref{Fig:OSD_comp_HARQ}. That is, in OSD-based HARQ systems, starting with a lower rate transmission and using higher rates for final transmission rounds can simultaneously improve error performance and reduce guesswork complexity. Since both HARQ schemes use identical information length $k$ and comparable short blocklengths ($n_1<n_2<100$), we can approximate their relative complexity by comparing their guesswork complexity, as suggested by \ref{equ::OSD_complexity}.
}

\begin{figure}  [t]
     \centering
        \begin{tikzpicture}

\definecolor{mycolor1}{rgb}{0.85000,0.32500,0.09800}%
\definecolor{mycolor2}{rgb}{0.92900,0.69400,0.12500}%
\definecolor{mycolor3}{rgb}{0.49400,0.18400,0.55600}%
\definecolor{mycolor4}{rgb}{0.46600,0.67400,0.18800}%
\definecolor{mycolor5}{rgb}{0.30100,0.74500,0.93300}%
\definecolor{mycolor6}{rgb}{0.00000,0.44700,0.74100}%

\begin{axis}[%
width=2.8in,
height=2in,
at={(1.01in,0.685in)},
scale only axis,
xmin=-4,
xmax=0,
xlabel style={at={(0.5,1ex)},font=\color{white!15!black}, font=\small},
xlabel={Blocklength $n$},
ymode=log,
ymin=1e-4,
ymax=1,
ylabel style={at={(1.5ex,0.5)},font=\color{white!15!black}, font=\small},
ylabel={BLER},
axis background/.style={fill=white},
tick label style={font=\footnotesize},
xmajorgrids,
ymajorgrids,
legend style={at={(1,1)}, anchor=north east, fill opacity=0.5, text opacity=1, font=\scriptsize, legend cell align=left, align=left, draw=white!15!black}
]

\addplot [color=black, mark=o, mark options={solid, black}]
  table[row sep=newline]{%
-4	0.188501413760603
-3.50000000000000	0.126103404791929
-3	0.0734753857457752
-2.50000000000000	0.0374882849109653
-2	0.0183318056828598
-1.50000000000000	0.00749344323716748
-1	0.00272345988343592
-0.500000000000000	0.00092649380510989
0	0.000240000000000000
};
\addlegendentry{$n_1 = 48$, $n_2 = 72$}

\addplot [color=blue, mark=square, mark options={solid, blue}]
  table[row sep=newline]{%
-4	0.0428816466552316
-3.50000000000000	0.0209073803052478
-3	0.0101224820325944
-2.50000000000000	0.00447057245680309
-2	0.00182483416819497
-1.50000000000000	0.000840000000000000
-1	0.000470000000000000
-0.500000000000000	0.000280000000000000
0	0.000175000000000000
};
\addlegendentry{$n_1 = 40$, $n_2 = 96$}

\end{axis}
\end{tikzpicture}
    \vspace{-0.5em}
    \caption{\black{The BLER comparison of two HARQ schemes with different blocklengths $n_1$ and $n_2$ for two transmission rounds. }}
    \vspace{-0.25em}
    \label{Fig:OSD_FER_HARQ_16}
    
\end{figure}
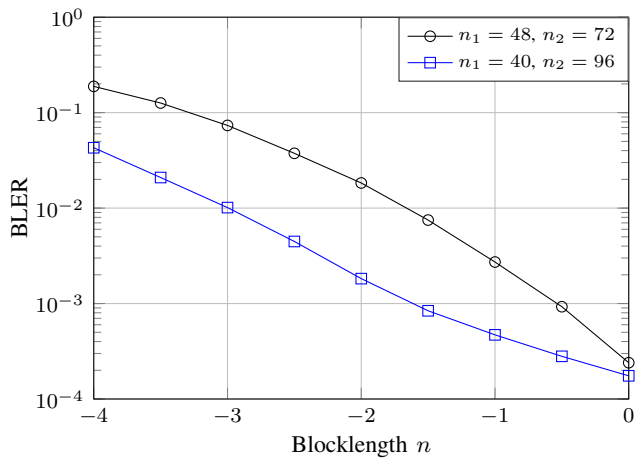

\subsection{Guesswork Complexity Cutoff Criterion for OSD}
\black{While our previous analysis in Sections \ref{sec::PB-OSD} and \ref{sec::HARQ} focus on the average guesswork complexity of OSD, controlling the variance in decoding complexity is also crucial for the decoder designs. In this section, we propose a guesswork complexity cutoff criterion (CCC) that utilizes second-order moment analysis to effectively limit worst-case guesswork complexity of OSD while maintaining near-optimal performance.}

\black{
For OSD with either optimal or Hamming weight-based processing of TEPs, the mode of the guesswork distribution is one. This property can be obtained from the characteristics of reliability-ordered bits in AWGN channels. Consider a BI-AWGN channel with BPSK modulation and noise power $\sigma^2$. The bit error probability for each unordered position is $Q(1/\sigma)<0.5$. After reliability ordering based on LLR magnitudes $|\ell_i|$, the first $k$ positions (MRB) exhibit even lower error probabilities than $Q(1/\sigma)$. Let $p_e(i)$ denote the error probability of the $i$-th most reliable bit, then we have $p_e(i) < Q(1/\sigma) <0.5$ for $i = 1,\ldots,k$. Consequently, the initial hard decision $\boldsymbol{b}_0$ of the MRB has a higher probability of being correct at each position than incorrect, and the zero error TEP (corresponding to guesswork value of 1) is the most probable case.
}

\black{
According to Gauss's inequality, for a unimodal random variable $X$ with mode zero and second-order moment $\tau^2$, the probability that $X$ deviates from its mode by more than $\zeta$ is bounded by:
\begin{equation}
    \mathbb{P}(|X|>\zeta) \leq 
    \begin{cases}
        \frac{4\tau^2}{9\zeta^2} & \text{for } \zeta\geq \frac{2\sqrt{\tau^2}}{\sqrt{3}}\\
        1- \frac{\zeta}{\sqrt{3\tau^2}} & \text{for } 0\leq \zeta \leq \frac{2\sqrt{\tau^2}}{\sqrt{3}}
    \end{cases}
\end{equation}
}

\black{
In the context of OSD, this inequality can be applied to bound the tail probability of guesswork. Let $X_{G}$ be the guesswork complexity of OSD. Then $X_{G}-1$ has the mode zero. Let $\mathbb{E}[(X_{G}\!-\!1)^2]$ denote the second-order moment of $X_{G}-1$. Then, the probability that the decoder requires more than $\zeta$ guesses to identify the correct error pattern is bounded by:
\begin{align}
    &\mathbb{P}(X_{G}>\zeta+1) \leq \\
    & \hspace{1cm}
    \begin{cases}
        \frac{4\mathbb{E}[(X_{G}\!-\!1)^2]}{9(\zeta)^2} & \text{for } \zeta\geq \frac{2\sqrt{\mathbb{E}[(X_{G}\!-\!1)^2]}}{\sqrt{3}}\\
        1- \frac{\zeta}{\sqrt{3\mathbb{E}[(X_{G}\!-\!1)^2]}} & \text{for } 0\leq \zeta \leq \frac{2\sqrt{\mathbb{E}[(X_{G}\!-\!1)^2]}}{\sqrt{3}}
    \end{cases}
\end{align}
}

\black{
For a target BLER $\epsilon_b$, setting $\mathbb{P}(X_{G}>\zeta +1) \leq \epsilon_b$ yields the following cutoff number of guesses.
\begin{equation} \label{equ::CCC}
    \zeta_{\mathrm{cut}} =  
    \begin{cases}
        \zeta_1 +1, & \ \ \text{if } \ \zeta_1 \geq \frac{2\sqrt{\mathbb{E}[(X_{G}\!-\!1)^2]}}{\sqrt{3}}, \\
        \zeta_2 +1, & \ \ \text{if } \  0 \leq \zeta_2 \leq \frac{2\sqrt{\mathbb{E}[(X_{G}\!-\!1)^2]}}{\sqrt{3}},
    \end{cases}
\end{equation}
where $\zeta_1 = \sqrt{\frac{4\mathbb{E}[(X_{G}\!-\!1)^2]}{9\epsilon_b}} $ and $\zeta_2 = \sqrt{3\mathbb{E}[(X_{G}\!-\!1)^2]}(1\!-\!\epsilon_b)$.
}

\black{Since $X_{G}$ is generally very large as shown in Section \ref{sec::OSD:bound}, we take $ \mathbb{E}[(X_{G}-1)^2] \approx \mathbb{E}[X_{G}^2] $.
For an order-$m$ OSD, the second-order moment $\mathbb{E}[X_{G}^2]$ can be fast evaluated using our derived bounds \eqref{equ::Newbound:XnYn:ordered:1k:order:app:w1} prior to decoding. According to \eqref{equ::pe}, $p_e$ in \eqref{equ::Newbound:XnYn:ordered:1k:order:app:w1} represents the average error probability of the first $k$ most reliable bits, which therefore is estimated as
\begin{equation}
    p_e \approx \frac{1}{k} \sum_{i =1}^{k} \frac{1}{1+\exp(|\tilde{\ell}_i|)}
\end{equation}
where $|\tilde{\ell}_i|$ represents the magnitude of ordered LLRs from received symbols. This $p_e$ can then be substituted into \eqref{equ::Newbound:XnYn:ordered:1k:order:app} with $\omega=2$ to estimate $ \mathbb{E}[(X_{G}-1)^2] \approx \mathbb{E}[X_{G}^2] $.
}

\black{
To obtain $\zeta_{\mathrm{cut}}$, we set the target BLER for CCC as $\epsilon_b = (\epsilon_m)^\alpha$ for an order-$m$ OSD with BLER $\epsilon_m$ given in \eqref{equ::OSD:error:rate}, where $\alpha$ controls the trade-off between complexity reduction and performance loss. Our experiments suggest $\alpha = 0.8$ provides an effective balance.
}

\black{
We summarize the procedure of OSD combining CCC in Algorithm \ref{Algo:CCC}. The algorithm first calculates the second-order moment $\mathbb{E}[X_{G}^2]$ and corresponding cutoff threshold $\zeta_\mathrm{cut}$ based on received signal reliability. Then it processes TEPs following Hamming weight ordering until either reaching the maximum number $\xi_{\max}$ or the complexity cutoff number $\zeta_\mathrm{cut}$. We note that CCC can be combined with any OSD variants, since it estimates $\zeta_\mathrm{cut}$ prior to the decoding.
}

\begin{algorithm} 
    \caption{OSD with CCC}
    \label{Algo:CCC}
    \begin{algorithmic} [1]
        \REQUIRE Received signal $\mathbf{y}$, Target BLER $\epsilon_b$, Parameter $\alpha$
        \ENSURE Decoded result $\hat{\mathbf{c}}$

        \STATE Preprocessing: reliability ordering and GE
        \STATE Evaluate $\mathbb{E}[X_{G}^2]$ using (119) with $\omega=2$
        \STATE Find cutoff threshold $\zeta_{\mathrm{cut}}$ using \eqref{equ::CCC} with $\epsilon_b = (\epsilon_m)^\alpha$.
        \STATE Get hard decision $\boldsymbol{b}_0$ from MRB
        \FOR{$\xi = 1$ to $\min\{\xi_{\max}, \zeta_{\mathrm{cut}}\}$}
            \STATE Generate TEP $\mathbf{e}_{\xi}$ following Hamming weight order
            \STATE Obtain guess $\boldsymbol{b}_{\xi} = \boldsymbol{b}_0 \oplus \mathbf{e}_{\xi}$
            \STATE Generate codeword candidate $\boldsymbol{v}_{\xi} = \boldsymbol{b}_{\xi}\widetilde{\mathbf{G}}$
        \ENDFOR
        \RETURN The codeword candidate with minimum Euclidean distance to $\mathbf{y}$
    \end{algorithmic}
\end{algorithm}

\black{
We evaluate the effectiveness of the proposed CCC through simulation of a $(128,64)$ extended BCH code with order-4 OSD. The baseline decoder is the PB-OSD proposed in \cite{yue2021probability}. Then, PB-OSD is further enhanced by the CCC technique in the approach as outlined in Algorithm \ref{Algo:CCC}.
}

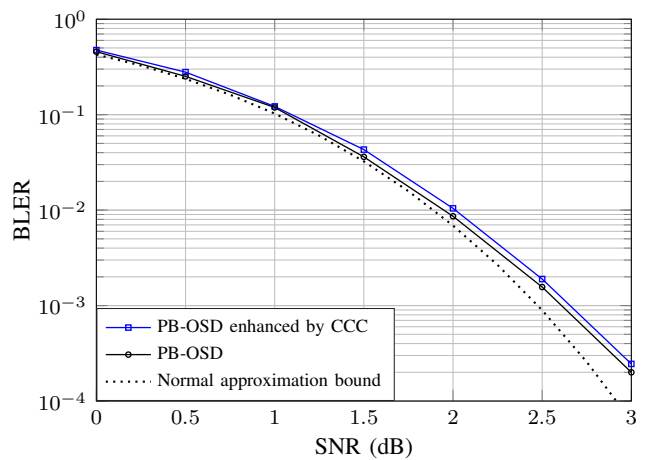
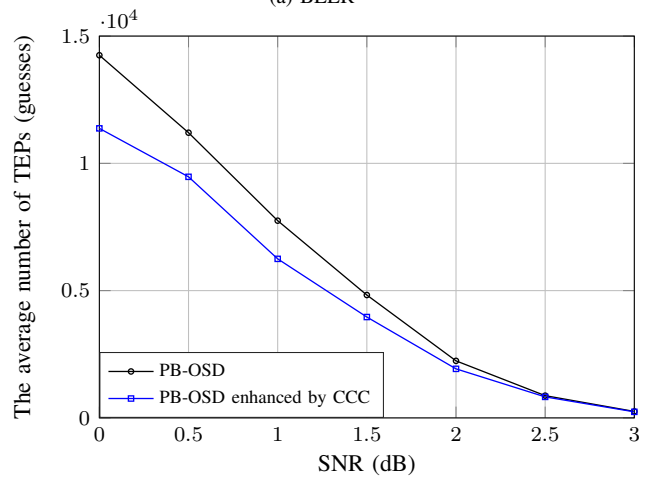
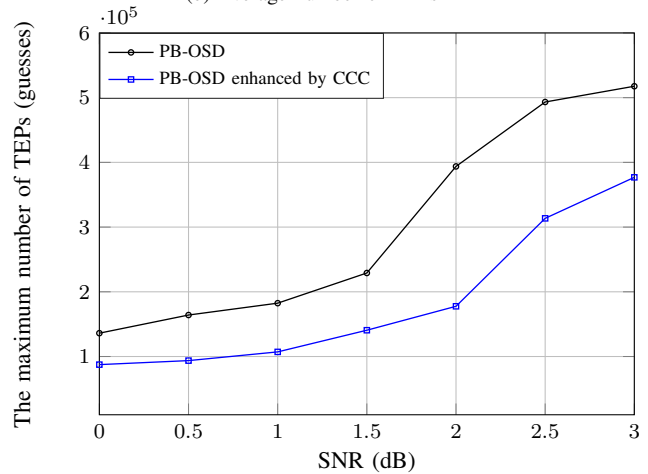
\begin{figure}[!htb]
    \centering
    \begin{subfigure}[b]{0.45\textwidth}
        \centering
\definecolor{mycolor1}{rgb}{0.85000,0.32500,0.09800}%
\definecolor{mycolor2}{rgb}{0.92900,0.69400,0.12500}%
\definecolor{mycolor3}{rgb}{0.00000,0.44706,0.74118}%
\definecolor{mycolor4}{rgb}{0.14902,0.14902,0.14902}%
\begin{tikzpicture}

\begin{axis}[%
width=2.8in,
height=2in,
at={(1.01in,0.685in)},
scale only axis,
xmin=0,
xmax=3,
xlabel style={at={(0.5,1ex)},font=\color{white!15!black}, font=\small},
xlabel={SNR (dB)},
ymode=log,
ymin=1e-4,
ymax=1,
yminorticks=true,
ylabel style={at={(1.5ex,0.5)},font=\color{white!15!black}, font=\small},
ylabel={BLER},
axis background/.style={fill=white},
tick label style={font=\footnotesize},
xmajorgrids,
ymajorgrids,
yminorgrids,
legend style={at={(0,0)}, anchor=south west , fill opacity=0.5, text opacity=1, font = \scriptsize	 , legend cell align=left, align=left, draw=white!15!black}
]

\addplot [color= blue, mark=square, mark size = 1pt, line width=0.5pt, mark options={solid}]
 table[row sep=crcr]{%
0	0.475\\
0.5	0.279\\
1	0.122\\
1.5	0.0430809792843691\\
2	0.0104403386340236\\
2.5	0.00189673863114378\\
3	0.000245\\
};
\addlegendentry{PB-OSD enhanced by CCC}

\addplot [color= black, mark=o, mark size = 1pt, line width=0.5pt, mark options={solid}]
  table[row sep=crcr]{%
0	0.457\\
0.5	0.252\\
1	0.119\\
1.5	0.0361532899493854\\
2	0.00862311327946333\\
2.5	0.0015633305193384\\
3	0.0002\\
};
\addlegendentry{PB-OSD}

\addplot [thick, dotted, color=black]
  table[row sep=crcr]{%
    0	0.426577286637981\\
    0.250000000000000	0.327324788777251\\
    0.500000000000000	0.237462723440728\\
    0.750000000000000	0.161754562076805\\
    1	0.102689677840520\\
    1.25000000000000	0.0602633528716957\\
    1.50000000000000	0.0323950953629263\\
    1.75000000000000	0.0157885686103981\\
    2	0.00689536709410716\\
    2.25000000000000	0.00266238435534207\\
    2.50000000000000	0.000894735320003527\\
    2.75000000000000	0.000256979776118879\\
    3	6.17391099148114e-05\\
    3.25000000000000	1.20962528375712e-05\\
    3.50000000000000	1.87521478750642e-06\\
};
\addlegendentry{Normal approximation bound}

\end{axis}
\end{tikzpicture}%
        \caption{BLER}
        \label{fig:CCC::BLER}
    \end{subfigure}

    \begin{subfigure}[b]{0.45\textwidth}
        \centering
\definecolor{mycolor1}{rgb}{0.85000,0.32500,0.09800}%
\definecolor{mycolor2}{rgb}{0.92900,0.69400,0.12500}%
\definecolor{mycolor3}{rgb}{0.00000,0.44706,0.74118}%
\definecolor{mycolor4}{rgb}{0.14902,0.14902,0.14902}%
\begin{tikzpicture}

\begin{axis}[%
width=2.8in,
height=2in,
at={(1.01in,0.685in)},
scale only axis,
xmin=0,
xmax=3,
xlabel style={at={(0.5,1ex)},font=\color{white!15!black}, font=\small},
xlabel={SNR (dB)},
%ymode=log,
ymin=0,
ymax=15000,
yminorticks=true,
ylabel style={at={(1.5ex,0.5)},font=\color{white!15!black}, font=\small},
ylabel={The average number of TEPs (guesses)},
axis background/.style={fill=white},
tick label style={font=\footnotesize},
xmajorgrids,
ymajorgrids,
yminorgrids,
legend style={at={(0,0)}, anchor=south west , fill opacity=0.5, text opacity=1, font = \scriptsize	 , legend cell align=left, align=left, draw=white!15!black}
]

\addplot [color= black, mark=o, mark size = 1pt, line width=0.5pt, mark options={solid}]
 table[row sep=crcr]{%
0	14246.654\\
0.5	11204.293\\
1	7748.042\\
1.5	4826.63195950832\\
2	2241.11892056716\\
2.5	874.274427039365\\
3	253.909805\\
};
\addlegendentry{PB-OSD }

\addplot [color= blue, mark=square, mark size = 1pt, line width=0.5pt, mark options={solid}]
  table[row sep=crcr]{%
0	11374.195\\
0.5	9471.112\\
1	6248.127\\
1.5	3962.50784655623\\
2	1926.95039946738\\
2.5	825.554945054945\\
3	238.133955\\
};
\addlegendentry{PB-OSD enhanced by CCC}

\end{axis}
\end{tikzpicture}%
        \caption{Average number of TEPs}
        \label{fig:CCC::ave}
    \end{subfigure}

    \begin{subfigure}[b]{0.45\textwidth}
        \centering
\definecolor{mycolor1}{rgb}{0.85000,0.32500,0.09800}%
\definecolor{mycolor2}{rgb}{0.92900,0.69400,0.12500}%
\definecolor{mycolor3}{rgb}{0.00000,0.44706,0.74118}%
\definecolor{mycolor4}{rgb}{0.14902,0.14902,0.14902}%
\begin{tikzpicture}

\begin{axis}[%
width=2.8in,
height=2in,
at={(1.01in,0.685in)},
scale only axis,
xmin=0,
xmax=3,
xlabel style={at={(0.5,1ex)},font=\color{white!15!black}, font=\small},
xlabel={SNR (dB)},
%ymode=log,
ymin=10000,
ymax=600000,
yminorticks=true,
ylabel style={at={(1.5ex,0.5)},font=\color{white!15!black}, font=\small},
ylabel={The maximum number of TEPs (guesses)},
axis background/.style={fill=white},
tick label style={font=\footnotesize},
xmajorgrids,
ymajorgrids,
yminorgrids,
legend style={at={(0,1)}, anchor=north west , fill opacity=0.5, text opacity=1, font = \scriptsize	 , legend cell align=left, align=left, draw=white!15!black}
]

\addplot [color= black, mark=o, mark size = 1pt, line width=0.5pt, mark options={solid}]
 table[row sep=crcr]{%
0	136068\\
0.5	164100\\
1	182433\\
1.5	229072\\
2	393730\\
2.5	493282\\
3	517640\\
};
\addlegendentry{PB-OSD }

\addplot [color= blue, mark=square, mark size = 1pt, line width=0.5pt, mark options={solid}]
  table[row sep=crcr]{%
0	87480\\
0.5	93670\\
1	107147\\
1.5	140628\\
2	177607\\
2.5	313496\\
3	376876\\
};
\addlegendentry{PB-OSD enhanced by CCC}

\end{axis}
\end{tikzpicture}%
        \caption{Worst-case number of TEPs}
        \label{fig:CCC::max}
    \end{subfigure}

    \caption{\black{The performance of enhancing OSD with the proposed CCC.}}
    \label{fig:CCC}
\end{figure}

\black{
Figure \ref{fig:CCC::BLER} compares the BLER performance of both decoders against the normal approximation bound. As shown, incorporating CCC results in only a minor degradation in error performance, with approximately 0.1 dB loss at BLER of $10^{-4}$. One can apply a higher $\alpha$ to close this small performance gap. The average guesswork complexity comparison is presented in Fig. \eqref{fig:CCC::ave}. The CCC-enhanced decoder demonstrates consistently lower guesswork complexity across all SNRs. This reduction is particularly significant at low SNRs; for instance, at SNR = 0 dB, CCC reduces the average number of TEPs from 14,246 to 11,374. Most notably, CCC provides substantial improvement in controlling worst-case guesswork complexity, as illustrated in Fig. \ref{fig:CCC::max}. The maximum number of TEPs required by PB-OSD increases significantly with SNR, reaching over 510,000 at SNR = 3 dB. In contrast, the CCC-enhanced decoder maintains a more stable worst-case complexity performance, with the maximum number of TEPs staying below 370,000 at high SNRs.
}

\section{Conclusion} \label{sec::conclusion}

This paper presents a comprehensive analysis of the achievable \black{guesswork} complexity of ordered statistics decoding (OSD) in binary additive white Gaussian noise (AWGN) channels. \black{The achievable guesswork complexity is defined as the number of guesses (test error patterns) required by OSD upon identifying the correct codeword estimate, which is characterized through new tight upper bounds developed for guesswork over ordered-statistic sequences.} 

\black{By applying Hölder's inequality to analyze ordered statistics, we establish that} the achievable \black{guesswork}  complexity of order-$k$ OSD \black{(where $k$ is the information length)} is tightly approximated by a modified Bessel function, which increases near-exponentially with code blocklength. \black{For an order-$m$ OSD with $m<k$, we derive bounds and approximations to conveniently estimate its guesswork complexity.} Furthermore, \black{our analysis reveals a fundamental guesswork complexity saturation threshold for OSD}. Increasing the OSD decoding order \black{beyond this threshold can} improve error performance without raising the average guesswork complexity.  

The theoretical results provide valuable insights for the design \black{of OSD decoders}. The derived achievable \black{guesswork} complexity enables a rapid assessment of the performance-complexity trade-offs \black{across different OSD orders}. \black{We demonstrate the application of these theoretical results through two scenarios. First, we discussed the optimized design of HARQ systems with OSD, which significantly improves the trade-off between BLER and the average decoding complexity. Second, we developed a guesswork complexity cutoff criterion (CCC), which effectively constrains worst-case guesswork complexity in existing OSD algorithms while maintaining near-optimal error performance.}

%
%Another way: Observe that
%\begin{align}
%    \mathbb{P}(y|0) + \mathbb{P}(y|1) &= \frac{1}{\sqrt{2\pi\sigma^2}}\exp\left(\frac{-y^2-1}{2\sigma^2}\right)\left(\exp{\frac{x}{\sigma^2}}+\exp{\frac{-x}{\sigma^2}}\right)\\
%    &=  \frac{1}{\sqrt{2\pi\sigma^2}}\exp\left(\frac{-y^2-1}{2\sigma^2}\right)\cosh{\frac{y}{\sigma^2}}
%\end{align}
%Therefore,
%\begin{align}
%     \mathbb{P}_{X|Y}(x|y)^q \, \mathbb{P}_{Y}(y) &=\frac{1}{2} \frac{[\mathbb{P}(y|x)]^q}{[\mathbb{P}(y|0) + \mathbb{P}(y|1)]^{q-1}}\\
%     &= \frac{1}{2} \frac{(\frac{1}{\sqrt{2\pi\sigma^2}})^q\exp\left(\frac{-q(y-x)^2}{2\sigma^2}\right) }{ (\frac{2}{\sqrt{2\pi\sigma^2}})^{q-1}\exp\left(\frac{-(q-1)(y^2+1)}{2\sigma^2}\right)(\cosh{\frac{x}{\sigma^2}})^{q-1}}\\
%     &= \frac{1}{2^q}\frac{1}{\sqrt{2\pi\sigma^2}} \frac{\exp\left(\frac{-q(y-x)^2}{2\sigma^2}\right) }{ \exp\left(\frac{-(q-1)(y^2+1)}{2\sigma^2}\right)}\left(\text{sech}{\frac{x}{\sigma^2}}\right)^{q-1}\\
%     & = \frac{2}{\sqrt{8\pi\sigma^2}} \exp\left(-\frac{y^2-2qxy+qx^2-q+1}{2\sigma^2}\right) \frac{1}{\cosh{\frac{y}{\sigma^2}})^{q-1}}\\
%\end{align}

\ifCLASSOPTIONcaptionsoff
  \newpage
\fi



%\bibliographystyle{IEEEtran}

\bibliography{reference/IEEEabrv, reference/OSDAbrv, reference/SurveyAbrv, reference/ClassicAbrv, reference/LearningAbrv, reference/MathAbrv, reference/GrandAbrv, reference/NOMAAbrv, reference/PolarAbrv}

\begin{thebibliography}{10}
\providecommand{\url}[1]{#1}
\csname url@samestyle\endcsname
\providecommand{\newblock}{\relax}
\providecommand{\bibinfo}[2]{#2}
\providecommand{\BIBentrySTDinterwordspacing}{\spaceskip=0pt\relax}
\providecommand{\BIBentryALTinterwordstretchfactor}{4}
\providecommand{\BIBentryALTinterwordspacing}{\spaceskip=\fontdimen2\font plus
\BIBentryALTinterwordstretchfactor\fontdimen3\font minus \fontdimen4\font\relax}
\providecommand{\BIBforeignlanguage}[2]{{%
\expandafter\ifx\csname l@#1\endcsname\relax
\typeout{** WARNING: IEEEtran.bst: No hyphenation pattern has been}%
\typeout{** loaded for the language `#1'. Using the pattern for}%
\typeout{** the default language instead.}%
\else
\language=\csname l@#1\endcsname
\fi
#2}}
\providecommand{\BIBdecl}{\relax}
\BIBdecl

\bibitem{Changyang2023xURLLC}
C.~She, C.~Pan, T.~Q. Duong, T.~Q.~S. Quek, R.~Schober, M.~Simsek, and P.~Zhu, ``Guest editorial {xURLLC} in {6G}: Next generation ultra-reliable and low-latency communications,'' \emph{{IEEE} J. Sel. Areas Commun.}, vol.~41, no.~7, pp. 1963--1968, 2023.

\bibitem{tataria20216g}
H.~Tataria, M.~Shafi, A.~F. Molisch, M.~Dohler, H.~Sj{\"o}land, and F.~Tufvesson, ``{6G} wireless systems: Vision, requirements, challenges, insights, and opportunities,'' \emph{Proc. {IEEE}}, vol. 109, no.~7, pp. 1166--1199, 2021.

\bibitem{erseghe2016coding}
T.~Erseghe, ``Coding in the finite-blocklength regime: Bounds based on {L}aplace integrals and their asymptotic approximations,'' \emph{{IEEE} Trans. Inf. Theory}, vol.~62, no.~12, pp. 6854--6883, 2016.

\bibitem{Mahyar2019ShortCode}
M.~{Shirvanimoghaddam}, M.~S. {Mohammadi}, R.~{Abbas}, A.~{Minja}, C.~{Yue}, B.~{Matuz}, G.~{Han}, Z.~{Lin}, W.~{Liu}, Y.~{Li}, S.~{Johnson}, and B.~{Vucetic}, ``Short block-length codes for ultra-reliable low latency communications,'' \emph{{IEEE} Commun. Mag.}, vol.~57, no.~2, pp. 130--137, February 2019.

\bibitem{3GPPRel16Coding}
``{{5G NR} Multiplexing and channel coding},'' {GPP TS 38.212 version 16.2.0 Release 16}, Tech. Rep., Jul. 2020.

\bibitem{yue2023efficient}
C.~Yue, V.~Miloslavskaya, M.~Shirvanimoghaddam, B.~Vucetic, and Y.~Li, ``Efficient decoders for short block length codes in {6G URLLC},'' \emph{{IEEE} Commun. Mag.}, vol.~61, no.~4, pp. 84--90, 2023.

\bibitem{Grasslcodetables}
M.~Grassl, ``{Bounds on the minimum distance of linear codes and quantum codes},'' Online available at \url{http://www.codetables.de} (Accessed: Dec. 22, 2022).

\bibitem{cavarec2020learning}
B.~Cavarec, H.~B. Celebi, M.~Bengtsson, and M.~Skoglund, ``A learning-based approach to address complexity-reliability tradeoff in {OS} decoders,'' in \emph{2020 54th Asilomar Conference on Signals, Systems, and Computers}.\hskip 1em plus 0.5em minus 0.4em\relax IEEE, 2020, pp. 689--692.

\bibitem{larue2022neural}
G.~Larue, L.-A. Dufrene, Q.~Lampin, H.~Ghauch, and G.~R.-B. Othman, ``Neural belief propagation auto-encoder for linear block code design,'' \emph{IEEE Transactions on Communications}, vol.~70, no.~11, pp. 7250--7264, 2022.

\bibitem{Fossorier1995OSD}
M.~P.~C. Fossorier and S.~Lin, ``Soft-decision decoding of linear block codes based on ordered statistics,'' \emph{{IEEE} Trans. Inf. Theory}, vol.~41, no.~5, pp. 1379--1396, Sep 1995.

\bibitem{duffy2021guessing}
K.~R. Duffy, M.~M{\'e}dard, and W.~An, ``Guessing random additive noise decoding with symbol reliability information ({SRGRAND}),'' \emph{{IEEE} Trans. Commun.}, 2021.

\bibitem{duffy2018guessing}
K.~R. Duffy, J.~Li, and M.~M{\'e}dard, ``Guessing noise, not code-words,'' in \emph{2018 IEEE International Symposium on Information Theory (ISIT)}.\hskip 1em plus 0.5em minus 0.4em\relax IEEE, 2018, pp. 671--675.

\bibitem{abbas2023guessing}
S.~M. Abbas, M.~Jalaleddine, and W.~J. Gross, ``Guessing random additive noise decoding ({GRAND}),'' in \emph{Guessing Random Additive Noise Decoding: A Hardware Perspective}.\hskip 1em plus 0.5em minus 0.4em\relax Springer, 2023, pp. 3--17.

\bibitem{riaz2023sub}
A.~Riaz, A.~Yasar, F.~Ercan, W.~An, J.~Ngo, K.~Galligan, M.~Medard, K.~R. Duffy, and R.~T. Yazicigil, ``A sub-0.8 {p}j/b 16.3 {G}bps/mm 2 universal soft-detection decoder using {ORBGRAND} in 40nm {CMOS},'' in \emph{2023 IEEE International Solid-State Circuits Conference (ISSCC)}.\hskip 1em plus 0.5em minus 0.4em\relax IEEE, 2023, pp. 432--434.

\bibitem{yue2021revisit}
C.~Yue, M.~Shirvanimoghaddam, B.~Vucetic, and Y.~Li, ``A revisit to ordered statistics decoding: Distance distribution and decoding rules,'' \emph{{IEEE} Trans. Inf. Theory}, vol.~67, no.~7, pp. 4288--4337, 2021.

\bibitem{yue2021probability}
C.~Yue, M.~Shirvanimoghaddam, G.~Park, O.-S. Park, B.~Vucetic, and Y.~Li, ``Probability-based ordered-statistics decoding for short block codes,'' \emph{{IEEE} Commun. Lett.}, vol.~25, no.~6, pp. 1791--1795, 2021.

\bibitem{yue2022ordered}
C.~Yue, M.~Shirvanimoghaddam, B.~Vucetic, and Y.~Li, ``Ordered-statistics decoding with adaptive {Gaussian} elimination reduction for short codes,'' in \emph{2022 IEEE Globecom Workshops (GC Wkshps)}.\hskip 1em plus 0.5em minus 0.4em\relax IEEE, 2022, pp. 492--497.

\bibitem{choi2019fast}
C.~Choi and J.~Jeong, ``Fast and scalable soft decision decoding of linear block codes,'' \emph{{IEEE} Commun. Lett.}, vol.~23, no.~10, pp. 1753--1756, 2019.

\bibitem{wang2021efficient}
F.~Wang, J.~Jiao, K.~Zhang, S.~Wu, Y.~Li, and Q.~Zhang, ``Efficient ordered statistics decoder for ultra-reliable low latency communications,'' in \emph{ICC 2021-IEEE International Conference on Communications}.\hskip 1em plus 0.5em minus 0.4em\relax IEEE, 2021, pp. 1--6.

\bibitem{yue2022linear}
C.~Yue, M.~Shirvanimoghaddam, G.~Park, O.-S. Park, B.~Vucetic, and Y.~Li, ``Linear-equation ordered-statistics decoding,'' \emph{{IEEE} Trans. Commun.}, vol.~70, no.~11, pp. 7105--7123, 2022.

\bibitem{LCOSD2022}
Y.~Wang, J.~Liang, and X.~Ma, ``Local constraint-based ordered statistics decoding for short block codes,'' in \emph{2022 IEEE Information Theory Workshop (ITW)}, 2022, pp. 107--112.

\bibitem{dhakal2016error}
P.~Dhakal, R.~Garello, S.~K. Sharma, S.~Chatzinotas, and B.~Ottersten, ``On the error performance bound of ordered statistics decoding of linear block codes,'' in \emph{2016 IEEE International Conference on Communications (ICC)}.\hskip 1em plus 0.5em minus 0.4em\relax IEEE, 2016, pp. 1--6.

\bibitem{scholl2014hardware}
S.~Scholl and N.~Wehn, ``Hardware implementation of a reed-solomon soft decoder based on information set decoding,'' in \emph{2014 Design, Automation \& Test in Europe Conference \& Exhibition (DATE)}.\hskip 1em plus 0.5em minus 0.4em\relax IEEE, 2014, pp. 1--6.

\bibitem{kim2021fpga}
C.~Kim, D.~Rim, J.~Choe, D.~Kam, G.~Park, S.~Kim, and Y.~Lee, ``Fpga-based ordered statistic decoding architecture for b5g/6g urllc iiot networks,'' in \emph{2021 IEEE Asian Solid-State Circuits Conference (A-SSCC)}.\hskip 1em plus 0.5em minus 0.4em\relax IEEE, 2021, pp. 1--3.

\bibitem{wu2007preprocessing_and_diversification}
Y.~Wu and C.~N. Hadjicostis, ``Soft-decision decoding of linear block codes using preprocessing and diversification,'' \emph{{IEEE} Trans. Inf. Theory}, vol.~53, no.~1, pp. 378--393, 2007.

\bibitem{jin2006probabilisticConditions}
W.~{Jin} and M.~{Fossorier}, ``Probabilistic sufficient conditions on optimality for reliability based decoding of linear block codes,'' in \emph{2006 IEEE International Symposium on Information Theory}, 2006, pp. 2235--2239.

\bibitem{Chentao2019SDD}
C.~Yue, M.~Shirvanimoghaddam, Y.~Li, and B.~Vucetic, ``Segmentation-discarding ordered-statistic decoding for linear block codes,'' in \emph{2019 IEEE Global Communications Conference (GLOBECOM)}.\hskip 1em plus 0.5em minus 0.4em\relax IEEE, 2019, pp. 1--6.

\bibitem{wang2021self}
F.~Wang, J.~Jiao, K.~Zhang, S.~Wu, Y.~Li, and Q.~Zhang, ``Self-adaptive ordered statistics decoder for finite block length raptor codes toward {URLLC},'' \emph{{IEEE} Internet Things J.}, vol.~9, no.~5, pp. 3282--3297, 2021.

\bibitem{arikan1996inequality}
E.~Arikan, ``An inequality on guessing and its application to sequential decoding,'' \emph{{IEEE} Trans. Inf. Theory}, vol.~42, no.~1, pp. 99--105, 1996.

\bibitem{malone2004guesswork}
D.~Malone and W.~G. Sullivan, ``Guesswork and entropy,'' \emph{{IEEE} Trans. Inf. Theory}, vol.~50, no.~3, pp. 525--526, 2004.

\bibitem{hanawal2010guessing}
M.~K. Hanawal and R.~Sundaresan, ``Guessing revisited: A large deviations approach,'' \emph{{IEEE} Trans. Inf. Theory}, vol.~57, no.~1, pp. 70--78, 2010.

\bibitem{christiansen2012guesswork}
M.~M. Christiansen and K.~R. Duffy, ``Guesswork, large deviations, and shannon entropy,'' \emph{{IEEE} Trans. Inf. Theory}, vol.~59, no.~2, pp. 796--802, 2012.

\bibitem{balakrishnan2014order}
N.~Balakrishnan and A.~C. Cohen, \emph{Order statistics \& inference: estimation methods}.\hskip 1em plus 0.5em minus 0.4em\relax Elsevier, 2014.

\bibitem{fossorier1996first}
M.~P. Fossorier and S.~Lin, ``First-order approximation of the ordered binary-symmetric channel,'' \emph{{IEEE} Trans. Inf. Theory}, vol.~42, no.~5, pp. 1381--1387, 1996.

\bibitem{feller1968}
W.~Feller, \emph{An Introduction to Probability Theory and Its Applications}, 3rd~ed.\hskip 1em plus 0.5em minus 0.4em\relax John Wiley \& Sons, 1968, vol.~1.

\bibitem{Wu2007OSDMRB}
Y.~Wu and C.~N. Hadjicostis, ``Soft-decision decoding using ordered recodings on the most reliable basis,'' \emph{{IEEE} Trans. Inf. Theory}, vol.~53, no.~2, pp. 829--836, 2007.

\bibitem{choi2021fast}
C.~Choi and J.~Jeong, ``Fast soft decision decoding algorithm for linear block codes using permuted generator matrices,'' \emph{{IEEE} Commun. Lett.}, vol.~25, no.~12, pp. 3775--3779, 2021.

\bibitem{Mahyar2021primitive}
M.~Shirvanimoghaddam, ``Primitive rateless codes,'' \emph{{IEEE} Trans. Commun.}, vol.~69, no.~10, pp. 6395--6408, 2021.

\end{thebibliography}
\bibliographystyle{IEEEtran}  


\iffalse
\begin{IEEEbiography}{Michael Shell}
Biography text here.
\end{IEEEbiography}

\begin{IEEEbiographynophoto}{John Doe}
Biography text here.
\end{IEEEbiographynophoto}

%\newpage

\begin{IEEEbiographynophoto}{Jane Doe}
Biography text here.
\end{IEEEbiographynophoto}

\fi


%\vfill


\end{document}